\documentclass[preprint,aps,prc,showpacs,nofootinbib,secnumarabic,floatfix]{revtex4-1}

\usepackage{graphicx}% Include figure files
\usepackage{bm}% bold math
\usepackage{color}
\usepackage{gensymb}
\usepackage{amsmath}
\usepackage{slashed}

\newcommand{\ltsim}{\protect\raisebox{-0.5ex}{$\:\stackrel{\textstyle <}
	{\sim}\:$}}
\newcommand{\gtsim}{\protect\raisebox{-0.5ex}{$\:\stackrel{\textstyle >}
	{\sim}\:$}}

\newcommand{\bvec}[1]{\ensuremath{\boldsymbol{#1}}}

\begin{document}

\title{Slow neutron production as a probe of hadron formation in high-energy $\gamma^*A$ reactions}

\author{A.B. Larionov$^{1,2}$\footnote{E-mail address: larionov@fias.uni-frankfurt.de},
        M. Strikman$^3$\footnote{E-mail address: mxs43@psu.edu}}

\affiliation{$^1$Frankfurt Institute for Advanced Studies (FIAS), 
  D-60438 Frankfurt am Main, Germany\\
  $^2$Institut f\"ur Theoretische Physik, Universit\"at Giessen, D-35392 Giessen, Germany\\
  $^3$Pennsylvania State University, University Park, PA 16802, USA}

\date{\today}

\begin{abstract}
  Deep Inelastic Scattering (DIS) experiments at the planned Electron-Ion Collider will be affected by details of the 
  hadron formation inside the nuclear volume. Besides semi-inclusive particle production experiments decays of the target nucleus
  via emission of neutrons provide an additional opportunity to probe this domain.
  This paper reports on the hybrid dynamical+statistical calculations of low-energy neutron production in
  muon- and virtual photon-induced collisions with nuclei.
  We confirm the conclusion that the E665 data on neutron production in $\mu^-$ + Pb DIS
  at 470 GeV indicate a strong suppression of the final state interaction for hadrons with momenta above
  $\sim 1$ GeV/c. Ultraperipheral heavy-ion collisions at the Large Hadron Collider (LHC)
  and the Relativistic Heavy Ion Collider (RHIC) can be used to test this suppression. 
  The calculations of the neutron multiplicity distributions and $p_t$-spectra
  in photon - nucleus collisions at the energies accessible at the LHC and RHIC are presented for several models
  of hadron formation. We argue that studies of neutron production in ultraperipheral heavy ion collisions
  open a new window on the small-$x$ dynamics and hadron component of the photon wave function.
\end{abstract}

%\pacs{24.85.+p;~ 	%Quarks, gluons, and QCD in nuclear reactions
%      25.30.Mr;~ 	%Muon-induced reactions (including the EMC effect)
%      25.75.Bh;~ 	%Hard scattering in relativistic heavy ion collisions}

%\keywords{hadron formation; color transparency; transport model; deep inelastic scattering; ultraperipheral heavy-ion collisions;
%          slow neutrons}   

\maketitle

\section{Introduction}
\label{intro}

The planned Electron-Ion Collider (EIC) will give us access to the quark- and gluon-substructure of bound nucleons in energy regimes that extend
the earlier experiments at Hadron-Electron Ring Accelerator (HERA) and Thomas Jefferson National Accelerator Facility (JLab) into a new domain \cite{Accardi:2012qut}.
Experiments at the EIC will use the nucleus both as a target and a laboratory for quantum chromodynamics (QCD) where, e.g., the quark hadronization can be studied.
In all these experiments one tacitly assumes an approximate factorization of the initial DIS process and the final state interactions (FSI) of the particles produced.
All experimental observables will necessarily be affected by both. To gain any knowledge of the microscopic quark-gluon structure of bound nucleons it is, therefore,
mandatory to have the FSI under quantitative control. 

In the DIS processes, the production of a leading hadron involves a multistep fragmentation of a leading quark.
The attenuation of leading hadrons has been measured by the European Muon Collaboration (EMC) \cite{Ashman:1991cx} and HERMES Collaboration \cite{Airapetian:2000ks}.
In the models of refs. \cite{Accardi:2002tv,Akopov:2002nc,Gallmeister:2005ad,Gallmeister:2007an}, which are based on the string fragmentation,
the attenuation is explained by the pre-hadronic absorption on surrounding nucleons.
In the model of ref. \cite{Kopeliovich:2003py}, both the gluon bremsstrahlung and the pre-hadronic absorption are included to describe
the experimental data. In the models of refs. \cite{Wang:2002ri,Arleo:2003jz}, the same data are described by modifications
of the quark fragmentation function due to the medium-induced gluon radiation.

So far, the only theoretical analysis of both EMC \cite{Ashman:1991cx} and HERMES \cite{Airapetian:2000ks} experiments is that of Gallmeister et al. in ref. \cite{Gallmeister:2007an}.
These authors used a transport model to follow the initial fragmentation all the way to the asymptotic, final state of all ejected particles. This is necessary since experimental acceptances
can overshadow the underlying microscopic mechanism as was illustrated by Falter et al. in ref. \cite{Falter:2004uc}.

At high energies and hard scattering processes color transparency, which reduces the FSI of initially produced fast particles, 
is expected to be present and can be studied in semi-inclusive lepton-induced particle production spectra.
An alternative is the study of hadron production in the nuclear fragmentation region.
If the produced fast particles interact with the target nucleus only weakly the excitation of the target remnant should be small.
It is, therefore, promising to study the production of low-energy hadrons emitted from the target remnant.
Such studies were performed so far only by the E665 Collaboration at Fermilab. 
In ref. \cite{Adams:1995nu}, the production of low energy ($E < 10$ MeV) neutrons in $\mu^-$+Pb DIS at 470 GeV have been studied.
The first theoretical analysis of these data has been performed in ref. \cite{Strikman:1998cc}. These authors have demonstrated that
the energy dependence of the neutron spectrum can be reproduced  for a range of assumptions about the formation time.
However, in order to reproduce  the total neutron multiplicity of about five neutrons per event, it was necessary to assume that only nucleons produced
in DIS with momenta $\le $ 1 GeV/c could interact. This surprisingly low value is well below the momentum range where the 
formation time and color transparency effects become important. This hints at {\it  a new dynamics} for the hadron formation in the nucleus fragmentation region.

In another E665 experiment \cite{Adams:1994ri} on $\mu^-$+Xe DIS at 490 GeV, the multiplicities of so-called gray tracks, ascribed to protons
in the momentum range 200-600 MeV/c, have been measured. The theoretical analysis of this experiment in ref. \cite{CiofidegliAtti:2004pv} has shown
that the data on the $Q^2$- and $x$-dependence of the average number of gray tracks can be described  when one assumes the color transparency only for the leading hadrons,
in line with the assumptions of ref. \cite{Strikman:1998cc}. In this model only slow recoiling nucleons from elementary lepton-nucleon DIS re-interact with the nucleus.

In this paper, we suggest a  way to study the formation of hadrons in the nucleus fragmentation region  at the collider energies
before the advent of the EIC. The proposed strategy is  based on the experience gained from the studies of the ultraperipheral collisions
(UPCs) of heavy ions at the LHC and direct photon interactions at HERA. In analogy with DIS, one can select a hard interaction of a quasireal
photon coherently emitted by one of the nuclei with a parton of the other nucleus by triggering on the production of two high $p_T$ jets which together
carry practically all energy of the photon. They have large invariant mass, are mostly produced at central rapidities and can be  easily detected
by the CMS and Atlas detectors \cite{Angerami:2319206}. The hard photon interaction thus initiates a cascade of particles. If the nucleus is more 
transparent for the energetic products, counting slow neutrons with energies up to about 10 MeV then uses the nucleus as a {\it "micro calorimeter'' }

In the present study we simulate the initial $\gamma^* N$ event on the bound nucleon via the PYTHIA model
\footnote{We will mostly discuss the PYTHIA results in terms of photon-gluon interactions which are predominant
  in the studied region of Bjorken $x \sim 10^{-2}-10^{-1}$,
  although also a moderate admixture of photon-quark interactions can be present in PYTHIA events as well.}.
The slow neutrons are mostly produced in the interaction of the struck nucleon debris with surrounding nucleons and we keep track of these
interactions by means of quantum-kinetic transport theory as encoded in the  Giessen Boltzmann-Uehling-Uhlenbeck (GiBUU) model \cite{Buss:2011mx}. 
For the low-energy particles we supplement this description with the statistical multifragmentation model (SMM) \cite{Botvina:1987jp,Bondorf:1995ua}
that describes the evaporation of low-energy neutrons by an excited nuclear residue.

The paper is organized as follows. Section \ref{model} gives an overview of the theoretical framework containing the $\gamma^* N$ event simulation
by the PYTHIA model included in the GiBUU package, the propagation of produced particles through the nucleus which leads to the energy
deposition in the form of hole excitations in the nuclear residue, and the statistical deexcitation of the nuclear residue.
Numerical results are presented in sec. \ref{results}.
First, we benchmark the model against experimental data on slow neutron production in $\mu^-+ A$ collisions at 470 GeV.
Next, we perform calculations for the LHC and RHIC kinematics of ultraperipheral heavy-ion collisions
in which one of the nuclei serves as a source of hard quasireal photons interacting with another nucleus.
We show that the multiplicity distribution and $p_t$-spectrum of soft neutrons are directly related to pre-hadronic interactions in the nucleus. 
In sec. \ref{directions}, we discuss directions for further studies, in particular, the expectations for enhanced neutron production
in the resolved photon kinematics. Finally, in sec. \ref{summary} we summarize our results and draw some conclusions.

Appendix \ref{results_pA} contains the numerical results of simulations of proton-nucleus reactions at $p_{\rm lab} = 1-2$ GeV/c 
as a test of the interface between the dynamical and statistical parts of our hybrid theoretical approach.
We determine the time window when the nuclear residue reaches thermal equilibrium and thus the SMM becomes
applicable. The energy spectra of neutrons from $pA$ collisions are well described by our hybrid calculations
in a quite broad range of neutron energy, up to 160 MeV. Remarkably, the neutron yields below 10 MeV are governed
by statistical evaporation and agree very well with experiment.

\section{Theoretical framework}
\label{model}

We are addressing slow particle production in the interaction  of a highly-energetic (virtual) photon with a nuclear target.
The physical picture of this process is quite complicated and, in-fact, is not fully covered by any existing theoretical approach.
Hence, largely guided by experience gained from previous studies of the particle- and nucleus-nucleus interactions
\cite{Barashenkov:1974qj,Bondorf:1995ua,Botvina:1995,Strikman:1998cc,Buss:2011mx}
we rely on the hybrid description combining three steps: (1) hard interaction of the incoming $\gamma^*$ with a bound nucleon,
(2) propagation of produced particles in the target nucleus, (3) statistical evaporation of slow neutrons from a nuclear residue.
For each step we use a dedicated and well tested in its energy range theoretical model.
Below in this section, we describe all three model steps and -- in some more detail -- the interfaces between models,
i.e. the hadron formation process and the determination of the parameters of the nuclear residue.

\subsection{Hard interaction}
\label{HardInt}
In the case of high energy lepton-nucleus interaction the struck nucleon is chosen randomly neglecting nuclear shadowing
effects, see discussions below in secs. \ref{E665} and \ref{gammaA_LHC}.
This corresponds to the probability distribution of the interaction with protons and neutrons
\begin{equation}
  dP_i =\frac{\rho_i(\bvec{r}) d^3r}{A}~,        \label{dP_i}
\end{equation}
where $\rho_i(\bvec{r})~(i=p,n)$ are the densities of protons and neutrons.
The collision of virtual photon with the struck nucleon is simulated via PYTHIA version 6.4 \cite{Sjostrand:2006za}.

The kinematics of the lepton scattering depends on experimental conditions. In the case of DIS the elementary lepton-nucleon cross section
can be written as follows (cf. \cite{Christy:2007ve}):
\begin{equation}
  \frac{d\sigma}{d\Omega dE^\prime} = \Gamma[\sigma_T(W^2,Q^2)+\epsilon \sigma_L(W^2,Q^2)]~,       \label{dsigma_dOmegadE^prime}
\end{equation}
where $\Omega$ and $E^\prime$ are, respectively, the solid angle and the energy of the scattered lepton in the laboratory frame,
\begin{equation}
   \Gamma=\frac{\alpha E^\prime (W^2-m_N^2)}{(2\pi)^2 Q^2 m_N E (1-\epsilon)}       \label{Gamma}
\end{equation}
is the flux of virtual photons, $\sigma_T (\sigma_L)$ is the photoabsorption cross section for purely transverse (longitudinal)
photons, and
\begin{eqnarray}
  \epsilon &=& \left[1+2(1+\frac{\nu^2}{Q^2})\tan^2\frac{\theta}{2}\right]^{-1} \nonumber \\
           &=& \frac{2(1-y-Q^2/4E^2)}{2(1-y-Q^2/4E^2)+y^2(1+Q^2/\nu^2)}      \label{epsilon}
\end{eqnarray}
is the relative flux of longitudinal virtual photons, $y=\nu/E$ is the
fraction of the beam energy $E$ carried off by a virtual photon.
The second equality of Eq.(\ref{epsilon}), which is actually implemented in GiBUU,
is satisfied in the limit of the zero lepton mass exactly.

In actual simulations, Eq.(\ref{dsigma_dOmegadE^prime}) has been applied in the rest frame of the struck nucleon
(which has a finite momentum due to the Fermi motion) by using the relation
\begin{equation}
  \frac{d\sigma}{dydQ^2} = \frac{\pi}{E^\prime} \frac{d\sigma}{d\Omega dE^\prime}~,     \label{dsigma_dydQ^2}
\end{equation}
where, in the Lorentz-invariant form,  $y=pq/pk$ with $p, k$, and $q$ being the four-momenta of the struck nucleon,
initial lepton, and the virtual photon, respectively.  Note that in the present GiBUU calculations the reduced cross section
$\sigma^*=\sigma_T+\epsilon \sigma_L$ is extracted from PYTHIA with default settings (MSTP(14)=30), i.e. it includes the mixture
of all possible processes (point-like photon, VMD etc.).

\subsection{Hadron formation}
\label{HadrForm}

Particles produced in a hard interaction are not immediately formed and thus interact with the target nucleons
not like ordinary hadrons with a fully developed internal structure. It is well accepted that the interaction of such {\it pre-hadrons}
with target nucleons is significantly reduced. A pre-hadron converts to the ordinary hadron within the distance called {\it formation length}
or, equivalently, within the time interval called {\it formation time}.

In ref. \cite{Gallmeister:2007an}, the consistent description of the HERMES and EMC data has been reached within the GiBUU transport model
with the pre-hadron production and formation times of the yo-yo picture. Following this picture, the difference between
the formation and production proper times for a hadron with mass $m_h$ should be of the order of $m_h/2\kappa$
(1/4 of the $q \bar q$ string oscillation period \cite{Andersson:1983ia}) where $\kappa=1$ GeV/fm is the string tension.

This estimate can be compared with that of the color transparency model for the hadron attenuation in exclusive processes
like $e A \to e + \mbox{meson} + A^*$ at large enough $Q^2$ and momenta of the produced meson, see \cite{Dutta:2012ii} for review and references.
In this case, the color-neutral $q\bar q$ dipole of the transverse size $r_{\perp} \sim 1/Q$ is formed and interacts with the nucleon with
the cross section $\propto r_{\perp}^2$ as follows from perturbative QCD. Such a ``point-like'' configuration is not static. However, it can be decomposed
in the hadronic basis of the states with the same momentum but different energies. Therefore, the decomposition looses its coherence on the time
scale $\sim 2p_h/\Delta M^2$ where $\Delta M^2$ is of the order of the difference between the squared masses of the particle to be formed and
its nearest radially excited state. This time scale can be regarded as a formation time of the hadron, or the expansion time of the point-like
configuration to the normal hadronic size. It is expected from the analyses of pion electroproduction at JLab
\cite{Larson:2006ge,Larionov:2016phv}, that $\Delta M^2 \simeq 0.7$ GeV$^2$. This corresponds to almost the same value of the hadron formation
proper time as predicted by the yo-yo picture.

Thus, the hadron formation length is similar in DIS and exclusive processes and can be written as 
\begin{equation}
  l_{\rm form}= a p_h~,
                   \label{form}
\end{equation}
where $p_h$ is a hadron momentum, and $a=(0.4-0.6)$ fm\,c/GeV is a constant factor.
Eq.(\ref{form}) provides a guideline for our present-day understanding of the hadron formation
length. This equation suggests that the hadron will be formed inside a heavy nucleus, i.e. the formation length
is less than a nuclear radius, if $p_h \ltsim 10$ GeV/c, while at higher momenta the pronounced transparency should be observed.

To describe the hadron formation effects we applied the three different prescriptions: 

(i) Default formation procedure of the GiBUU model \cite{Gallmeister:2005ad,Gallmeister:2007an,Buss:2011mx} based on the production
and formation space-time points extracted from the JETSET part of the PYTHIA. The ratio of the effective pre-hadron-nucleon cross section
$\sigma_{\rm eff}$ to the usual total hadron-nucleon cross section $\sigma_0$ varies linearly with time between production
and formation times as
\begin{equation}
  \sigma_{\rm eff}(t)/\sigma_0=X_0+(1-X_0)\frac{t-t_{\rm prod}}{t_{\rm form}-t_{\rm prod}}~,          \label{sigma_eff_GiBUU}
\end{equation}
where $X_0=r_{\rm lead} a/Q^2$ is the pedestal value of the ratio, $r_{\rm lead}$ is the ratio of the number of leading quarks to the
total number of quarks in a hadron, and $a=1$ GeV$^2$ is a constant factor.

(ii) Quantum diffusion model (QDM) of ref. \cite{Farrar:1988me}.
This model has been applied for exclusive hard processes where, in the interaction point, the hadrons are in small-size configurations.
In this case, the expression for the effective cross section is formally the same as Eq.(\ref{sigma_eff_GiBUU}).
However, the production time $t_{\rm prod}$ is set to the time when the hard interaction
happens, while the formation time is defined as $t_{\rm form}=t_{\rm prod}+l_{\rm form}/c$,
where the formation length $l_{\rm form}$ is given by Eq.(\ref{form}). In numerical calculations we set $a=0.56$ fm\,c/GeV in Eq.(\ref{form})
as concluded from the analyses of pion electroproduction at JLab (see sec. \ref{intro}).
As far as the size of the quark-gluon configuration in the hard interaction point is concerned, in the case of DIS
it is more difficult to match it to other observations.
Since the model has a tendency to overestimate the neutron rate pretty significantly
we will make a simplifying assumption that the interaction in the production point is strongly suppressed, that is $X_0=0$.
As we will see below, even with this assumption, the model predicts a significantly higher neutron rate
than observed experimentally.

(iii) A simple cutoff 
\begin{equation}
  \sigma_{\rm eff}/\sigma_0=\Theta(p_{\rm cut}-p_h)~,     \label{sigma_eff_cut}
\end{equation}
where $\Theta(x)$ is a Heaviside step function ($\Theta(x)=0$ for $x<0$ and $\Theta(x)=1$ for $x>0$). The cutoff momentum
$p_{\rm cut}$ should be of the order of few GeV/c. It can be chosen from comparison with experimental data.

\subsection{Particle propagation in the nucleus}
\label{PartProp}

The propagation of pre-hadrons and hadrons is described in the framework of the GiBUU model \cite{Buss:2011mx}.
A pre-hadron is converted to an ordinary hadron at the time moment $t_{\rm form}$ which is determined for every produced pre-hadron.
The mass and quantum numbers (spin, isospin, etc.) are not changed in that conversion. 
During formation time period, the pre-hadrons differ from respective hadrons only in the total interaction cross sections with the target nucleons
and decay width (that disappears for the pre-hadrons). The branching ratios of different outgoing channels are the same for pre-hadronic
and hadronic interactions. 
The nuclear mean field and Coulomb potentials are also supposed to be the same for pre-hadrons and hadrons
\footnote{It is of course questionable whether the pre-hadrons feel the same potentials. However, since the kinetic energy of a pre-hadron
  is much larger than the potential energy, the action of the mean field potential does not practically influence the pre-hadron propagation.}.
Bearing all that in mind we call 'hadron' both hadronic and pre-hadronic states below in this subsection.

The GiBUU model explicitly solves the coupled set of kinetic equations for the system of hadrons
in time and six-dimensional phase space.
Between collisions and resonance decays the particles propagate according to the Hamiltonian equations of motion
with the mean field potentials. The nuclear potential is described by the relativistic mean field model NL3 of ref. \cite{Lalazissis:1996rd}.
The Coulomb potential acting on charge particles is also included. Particle collisions are simulated in the geometrical minimum distance scheme.
The two particles approaching their minimal distance $d$ during a given time step will collide during this time step if
$d < \sqrt{\sigma_{\rm tot}/\pi}$ where $\sigma_{\rm tot}$ is the total interaction cross section of these particles. Collisions are simulated by
a Monte Carlo algorithm taking into account elastic and inelastic scattering channels. If the invariant energy $\sqrt{s}$ is larger than some threshold
value (2.2 GeV for meson-baryon collision, 3.4 GeV for baryon-baryon collision, and 2.38 GeV for antibaryon-baryon collision), the hadron-hadron collision
is simulated via the PYTHIA and (for antibaryon-baryon collisions only) the FRITIOF version 7.02 \cite{Andersson:1992iq}
models.
The threshold values of $\sqrt{s}$ are selected to ensure the smooth behaviour of the hadronic cross sections over the transition
from the resonance regime at low energies to the string-dominated regime at high energies (see sec. 3.3.2 of ref.\cite{Buss:2011mx}
and refs. \cite{Larionov:2011fs,WeilPhD} for detail).
The proton and neutron density profiles are chosen in a Woods-Saxon form with geometrical parameters taken from the Skyrme-Hartree-Fock
systematics \cite{Lenske_priv}. The Fermi motion is taken into account in the local Fermi approximation. The Pauli blocking factors are included for
the nucleons in the final state of scattering events. 
The nuclear mean field and Pauli blocking factors are supposed to be constant during the time evolution of the hadronic system.
This allows to save CPU time for the elementary-particle-induced reactions on nuclei without significant accuracy lost.

There is also one important technical aspect which deserves to be mentioned. Most of the previous GiBUU calculations for the lepton-nucleus
interactions were done with the so-called perturbative particles to increase statistics and save CPU time, see e.g. Appendix D.2 of ref. \cite{Buss:2011mx}.
This method works very well for fast particles but not for slow ones, since the nucleus stays intact in calculations with perturbative particles.
In particular, using the perturbative particles would lead to the drastic overestimation of the number of holes in the nucleus, see discussion below
in sec. \ref{DetRes}. Therefore, to avoid such problems, we apply in calculations the real-particles method which takes into account the trailing of the nucleus
in the course of the cascade and thus provides much better description of the energy deposition in the nuclear residue.
The importance of the trailing effect in proton-nucleus collisions at the energies of several GeV has already been found in early intranuclear cascade
(INC) studies (cf. \cite{Barashenkov:1974qj} and refs. therein).

\subsection{Determination of an excited nuclear residue}
\label{DetRes}

The outgoing hadrons from a DIS event on the bound nucleon propagate through the nucleus inducing the cascade of interactions.
This eventually leads to the emission of slow neutrons which can be produced either directly or by statistical decay
of the excited residual nucleus.

The determination of the parameters of the residual nucleus can be done by various methods. In $pA$ collisions below pion production threshold,
it is possible to apply the energy balance for the determination of the residue excitation energy \cite{Bertini:1963zzc}.
However, this requires the utmost precision in the calculation of the total kinetic energy
of emitted particles which renders this method impractical at the proton beam energy $\gtsim 1$ GeV.
In the case of strongly violent nuclear dynamics, like central $pA$ collisions at GeV energies and central $AA$ collisions above Fermi energy,
the nuclear excitation energy may reach several MeV/nucleon. The residual nuclear system strongly deviates from the ground state due to thermal excitation and may,
for example, have a low density and/or a collective flow. In such situations, the only way is to use theoretical models with well defined ground state
which allows to calculate directly the energy of the residual nuclear system and subtract its ground state energy by using the corresponding many-body
Hamiltonian (cf. refs. \cite{Sangster:1992qr,Botvina:1995}) or relativistic energy density functional (cf. ref. \cite{Gaitanos:2007mm}). However, if the system
looses only relatively few ($\sim 10$) nucleons, the accuracy of such direct methods ($\sim 1$ MeV/nucleon) is not enough, while the old-fashioned
residue determination in terms of particle-hole excitations works significantly better (cf. Eqs.(62),(63) in ref. \cite{Bondorf:1995ua}).

Thus, we start adding up hole excitations in the nucleus neglecting, for simplicity,
particle excitations. The corrections for the latter will be introduced at the last step.
The mass number $A_{\rm res}$, charge number $Z_{\rm res}$, excitation energy $E^*_{\rm res}$,
and momentum $\bvec{p}_{\rm res}$ of the residual nucleus can be determined in every parallel ensemble\footnote{Parallel ensembles
can be regarded as ``events'' in GiBUU simulations.} by counting the hole excitations as
\begin{eqnarray}
  A_{\rm res}   &=& A - n_{\rm h}~,                      \label{A_res}\\
  Z_{\rm res}   &=& Z - \sum_{i=1}^{n_{\rm h}} Q_i~,        \label{Z_res}\\
  E^*_{\rm res} &=& \sum_{i=1}^{n_{\rm h}} (E_{F,i}-E_i)~,       \label{E_res}\\
  \bvec{p}_{\rm res} &=& - \sum_{i=1}^{n_{\rm h}} \bvec{p}_i~,    \label{p_res}
\end{eqnarray}
where $A$($Z$) is the target nucleus mass (charge) number; $n_{\rm h}$ is the total number of holes; $Q_i=1(0)$ for the proton (neutron)
hole; $E_i$ and $\bvec{p}_i$ are, respectively, the single particle energy and momentum; $E_{F,i}$ is the Fermi energy.
The dependence of the Fermi energy on the hole index ``i'' is mostly due to the difference of the Fermi energies for protons
and neutrons\footnote{Since we use the empirical Woods-Saxon density profiles in combination with relativistic mean field potentials,
the Fermi energies of protons and neutrons have also some spurious coordinate dependence. Comparison with the calculation using
the relativistic Thomas-Fermi method for the nuclear ground states \cite{Gaitanos:2010fd} shows, however, that the resulting excitation energy
is practically insensitive to the details of the nuclear ground state density profile.}. 
A hole is added to the nuclear residue for every two-body collision involving the target nucleon which did not collide yet,
starting from the initial DIS event.

Let us discuss now the contribution of particle excitations which is neglected  in Eqs.(\ref{A_res})-(\ref{p_res}).
The particle excitations are the bound nucleons with energies above Fermi level. A particle excitation is created if the recoil
nucleon is not energetic enough and thus gets captured in the nuclear potential well. Hence, not every hole excitation leads to
the reduction of the nucleus mass number. We have, therefore, redefined the quantities $A_{\rm res}$, $Z_{\rm res}$, and $\bvec{p}_{\rm res}$  
using the position space and counting all bound nucleons which are left in the target after the emission of the direct ones.
The directly emitted particles have been determined by the requirement of their separation in the position space by at least 3 fm
from all other nucleons.
The choice of the minimum separation distance does not influence the result once this distance
is larger than the internucleon distance of 1-2 fm and the time of the system evolution is large enough (see discussion in ref. \cite{Larionov:2011fs}).
If not specifically mentioned, the results reported below are obtained with the GiBUU calculation stopped at the time $t=100$ fm/c.
The mass and charge numbers of the residue and its momentum are then replaced, respectively, by  
the total mass and charge numbers and momentum of the bound nucleons. The residue excitation energy of Eq.(\ref{E_res}) is, however, not modified.
This is because, in contrast to the INC models \cite{Barashenkov:1972id}, we allow for collisions of the bound nucleons
above Fermi level with other nucleons. As a result, the number of particle excitations grows, but they are approaching
the Fermi level relaxing their kinetic energies to the low-lying particle-hole excitations.

\subsection{Evaporation from the excited nuclear residue}
\label{evapor}

The decay of excited nuclear residues has been described with a help of SMM \cite{Botvina:1987jp,Bondorf:1995ua}.
This model is focused on the description of the multiple simultaneous breakup of 
highly-excited nuclear systems (with excitation energy of the order of few MeV/nucleon). However, the SMM includes also the sequential decays of excited fragments
which are treated similar to the Weisskopf model \cite{Weisskopf:1937zz}.\footnote{In some contrast to the Weisskopf model, SMM includes
also the decay width of an excited compound nucleus with respect to the emission of fragments heavier than $\alpha$-particle taking into
account that the emitted fragment may be in an excited state stable to nucleon emission (see Eq. (34) in ref. \cite{Botvina:1987jp}).}
To save CPU time,
we have applied the SMM in the evaporation mode, switching-off the simultaneous multifragment breakup, since in the considered
reactions the yield of nuclear residues with excitation energies exceeding $1-2$ MeV/nucleon is small
(cf. Fig.~\ref{fig:ResProp} below).
In some selected cases, we performed the full SMM calculation, including multifragmentation, and
found out that this does not lead to noticeable changes  of the neutron multiplicities and spectra.

\section{Results}
\label{results}

\subsection{Muon-induced DIS on nuclei}
\label{E665}

In this section we analyze the E665 data on slow neutron production in inclusive $\mu^- + ^{208}$Pb and $\mu^- + ^{40}$Ca DIS processes
at the beam energy of 470 GeV \cite{Adams:1995nu}. The scattered muon was sampled as described in sec. \ref{HardInt} 
with cuts $\nu > 20$ GeV, $Q^2 > 0.8$ GeV$^2$. In the studied kinematics, a rather modest, on the scale of 20\%, nuclear shadowing
dominated by the resolved photon interactions with two nucleons is present for $R_A=\sigma(\mu Pb)/A\sigma(\mu N)$.
In the setup of the E665 experiment, both inelastic and diffractive events were included.
Applying Abramovsky-Gribov-Kancheli cutting rules \cite{Abramovsky:1973fm} we  observe that
the events where projectile interacted with two nucleons and diffractive events contribute equal cross sections to the final state.
While in the diffractive events no neutrons are produced, in the interactions where two nucleons are wounded roughly twice as many neutrons
are produced. Overall, this results in the disappearance of shadowing for inclusive cross section (similar to the cancellation for central
rapidities in ref. \cite{Abramovsky:1973fm}) and in the enhancement of the average number of wounded nucleons per event:
\begin{equation} 
  n_{\rm wound}= 1/R_A \sim 1.25,
\end{equation}
(see e.g. ref. \cite{Bertocchi:1976bq}) increasing the average neutron multiplicity by roughly the same factor. 

This implies that the constraints on the secondary interactions are somewhat stronger than
those which we find in our analysis.  For now we will neglect this effect.

\begin{figure}
  \includegraphics[scale = 0.50]{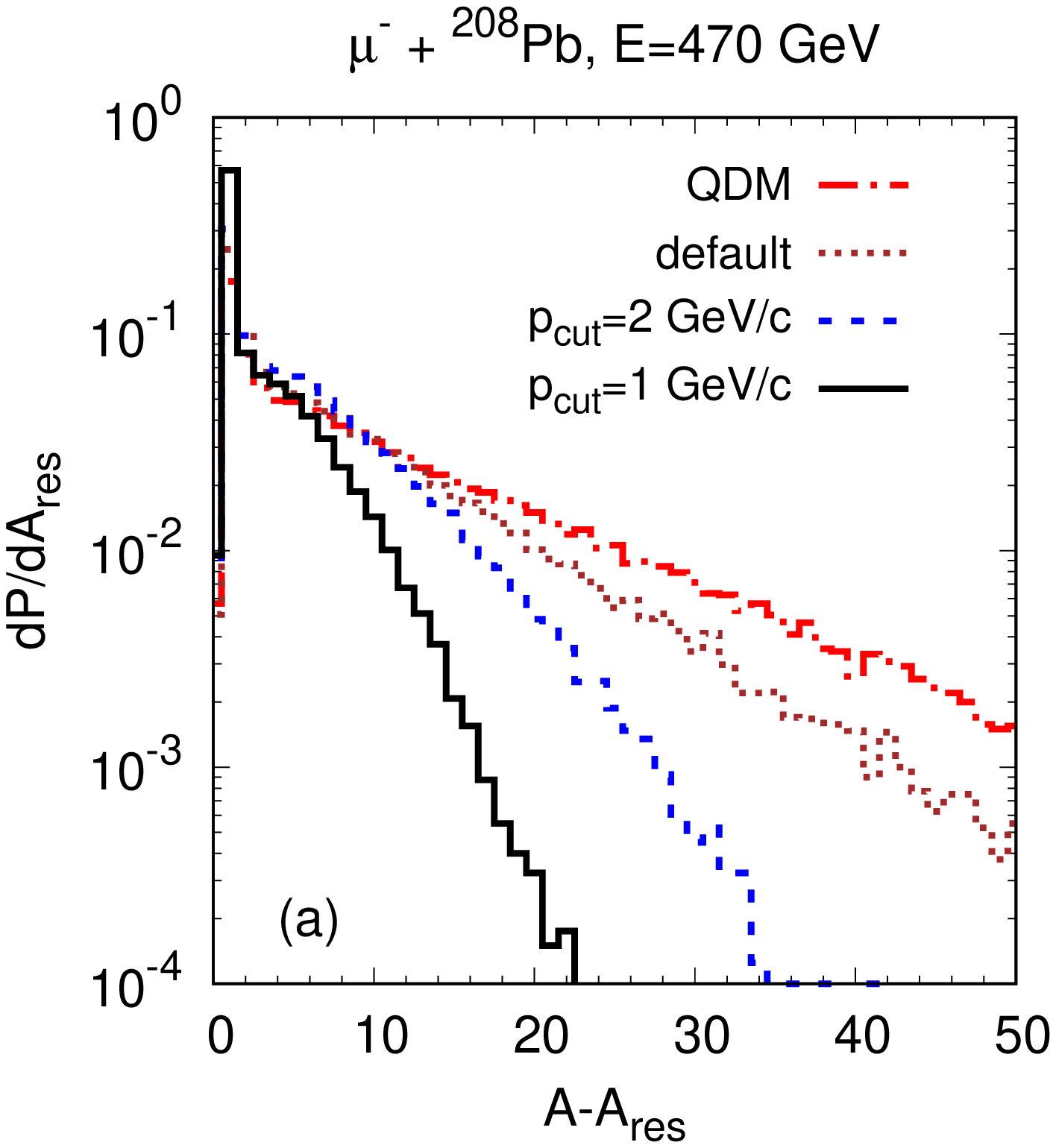}
  \includegraphics[scale = 0.50]{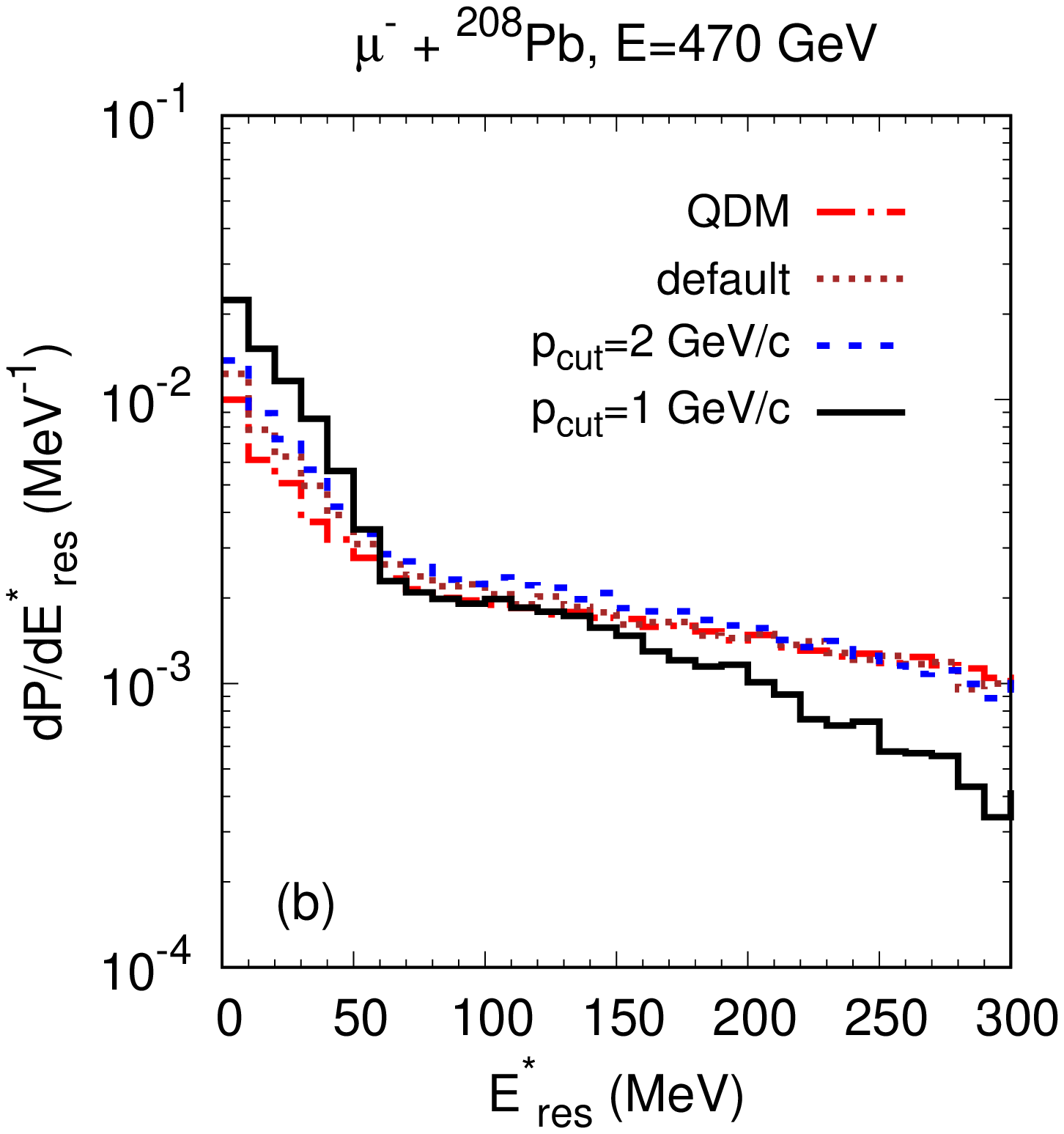}
  \includegraphics[scale = 0.50]{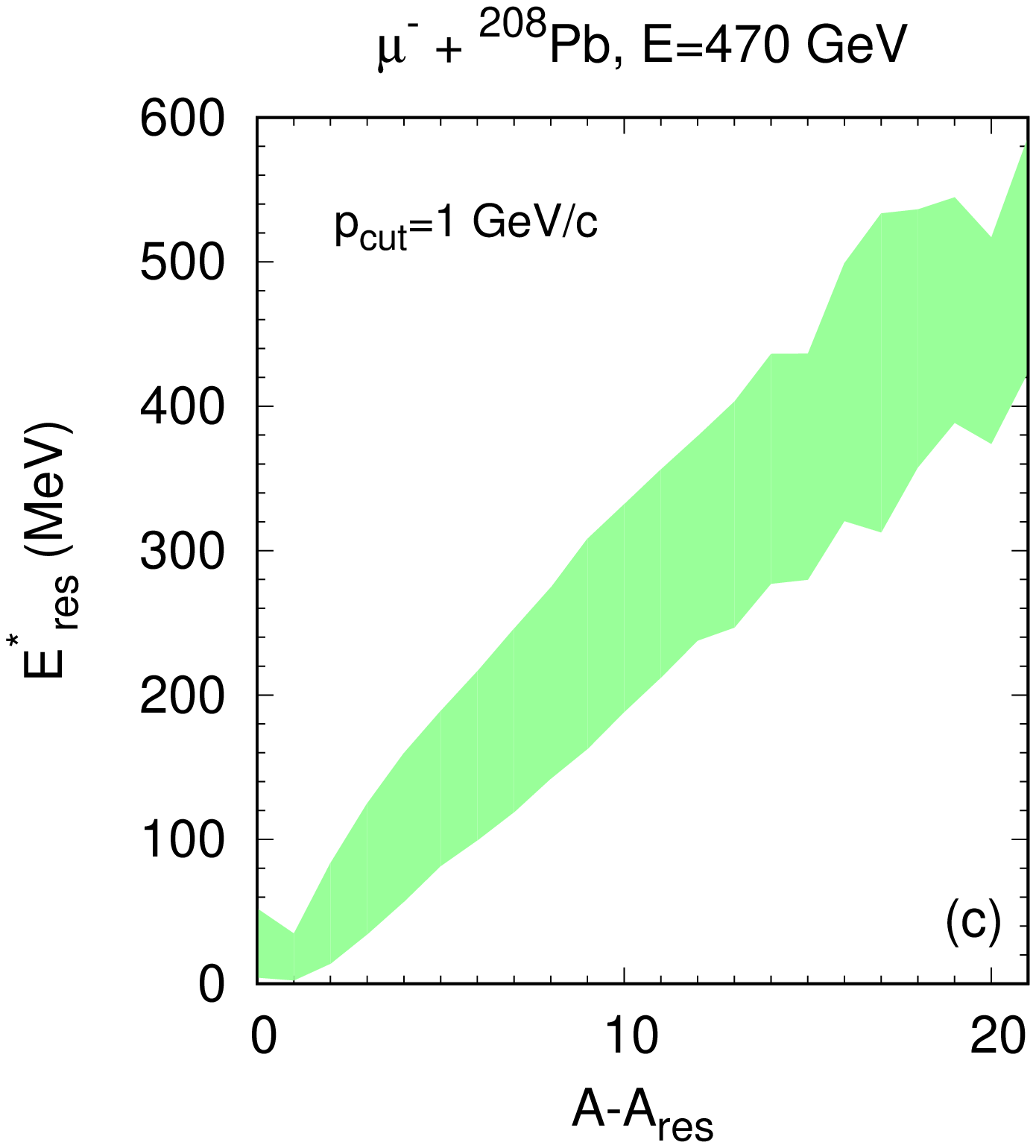}
  \caption{\label{fig:ResProp} Characteristics of the residual nucleus in $\mu^- + ^{208}$Pb DIS at 470 GeV.
  (a) Probability distribution of the mass number loss $A-A_{\rm res}$.
  (b) Probability distribution of the excitation energy.
    Different lines correspond to different prescriptions for the hadron formation:
    dash-dotted (red) line -- QDM calculation,
    dotted (brown) line -- GiBUU default,
    dashed (blue) line -- cutoff momentum 2 GeV/c,
    solid (black) line -- cutoff momentum 1 GeV/c.
  (c) Excitation energy as a function of the mass number loss obtained in GiBUU calculation
      with cutoff momentum $p_{\rm cut}=1$ GeV/c.
      The upper (lower) boundary of the band corresponds to the average value
      $\langle E^* \rangle$ plus (minus) the standard deviation.}
\end{figure}
Fig.~\ref{fig:ResProp} displays the mass number loss distribution (a), the excitation energy distribution (b)
and the correlation between the excitation energy and  the mass number loss (c) of the residual nucleus 
for the $^{208}$Pb target. Calculations in all models produce
a sharp peak in the mass number loss distribution for the removal of only one nucleon and 
a broad tail towards larger mass number loss. For example, in the calculation with $p_{\rm cut}=1$ GeV/c,
$57\%$ of residual nuclei have $A_{\rm res}=207$, while in the QDM calculation -- $18\%$. In most of such events,
a hole is produced due to initial hard DIS and the produced particles do not interact with the residue.
Thus, the calculation with the strongest restriction on the momentum of interacting particle produces the largest fraction of
one-hole events. The excitation energy distributions have a two-slope structure with a kink at $E^*_{\rm res} = 50-70$ MeV.
The region $E^*_{\rm res} < E_F \simeq 37$ MeV is saturated by one-hole excitations mostly. Higher excitation energies are only possible for
the multiple (2,3,...) hole excitations. The dependence of the excitation energy on the number of removed nucleons is close to a linear one,
$\langle E^*_{\rm res} \rangle \simeq 25~\mbox{MeV} (A-A_{\rm res})$.\footnote{Note that without contribution of particle excitations,
  this dependence would be $\langle E^*_{\rm res} \rangle \simeq 10~\mbox{MeV} (A-A_{\rm res})$. The difference arises due to the change of the number
  of removed nucleons, see discussion in sec. \ref{DetRes}.}
Due to the Fermi motion, however, the distribution in the excitation energy for the fixed number of removed nucleons is quite broad.
We checked that this distribution practically does not depend on the treatment of the pattern of hadron formation 
in calculations.

\begin{figure}
  \includegraphics[scale = 0.53]{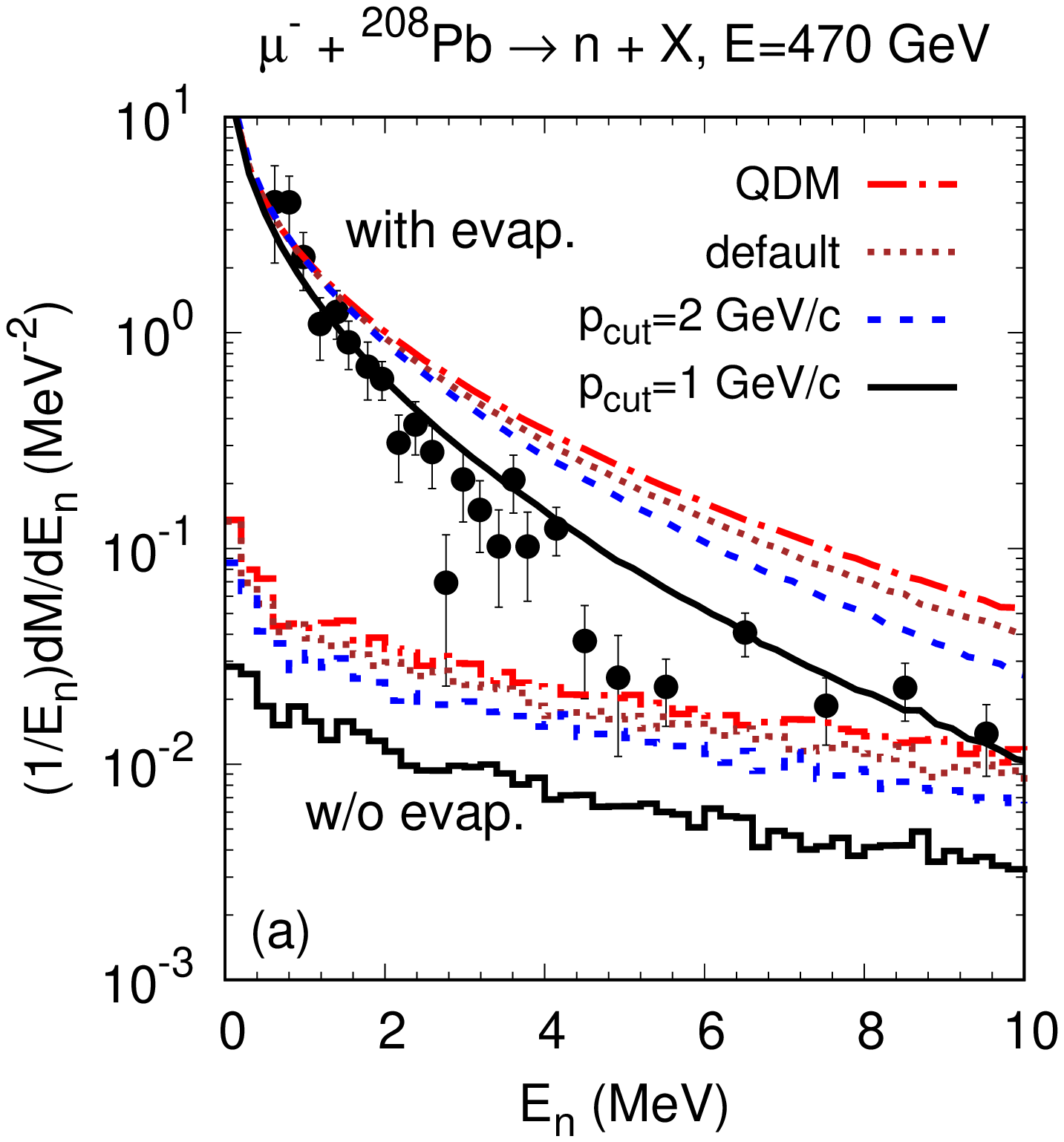}
  \includegraphics[scale = 0.53]{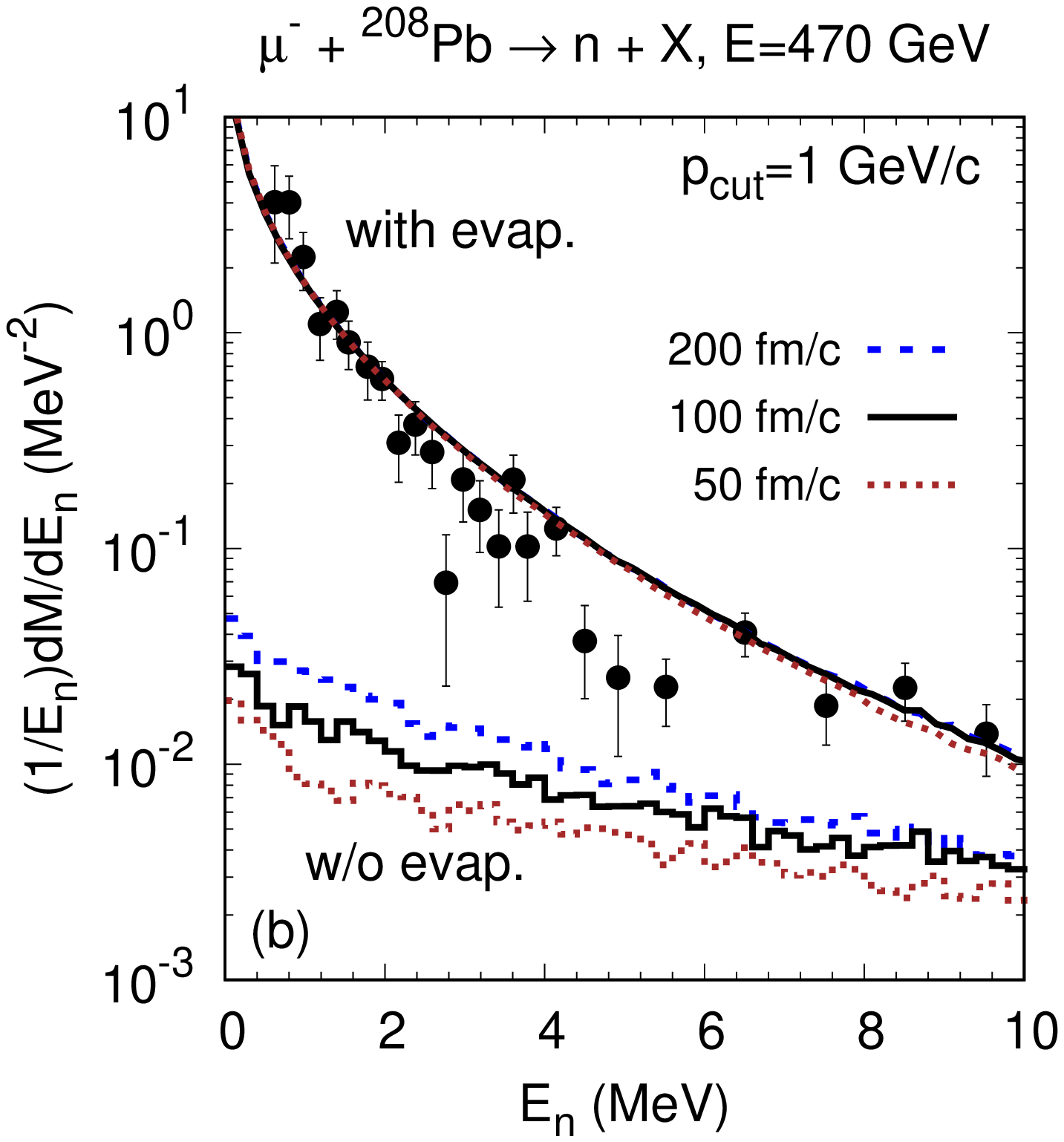}
  \caption{\label{fig:Pb_E665} Energy spectrum of emitted neutrons in $\mu^- + ^{208}$Pb DIS at 470 GeV.
    (a) GiBUU results with fixed maximum time (100 fm/c) for the different treatments
    of hadron formation (line notations are the same as in Fig.~\ref{fig:ResProp}).
    (b) GiBUU results with cutoff momentum 1 GeV/c for the different maximum times:
    dashed (blue) line -- 200 fm/c,
    solid (black) line -- 100 fm/c,
    dotted (brown) line -- 50 fm/c.
    Upper (lower) lines
    are calculated with (without) adding evaporated neutrons from the nuclear
    residue. Experimental data are from ref. \cite{Adams:1995nu}.} 
\end{figure}
The spectrum of emitted neutrons is shown in Fig.~\ref{fig:Pb_E665}a. Below 1 MeV, the spectrum is almost entirely due to
statistical evaporation from the excited nuclear residue. In this low energy region, the number of neutrons depends weakly on
the excitation energy of the nuclear residue (and thus on the treatment of hadron formation), since with decreasing
$E^*_{\rm res}$ the energy spectrum of evaporated neutrons becomes more steep. The sensitivity of the spectrum to 
various treatments of hadron formation increases with neutron energy. The QDM and default calculations
overestimate neutron yields. Only applying a very strong restriction on the momenta of interacting
particles ($p < p_{\rm cut}=1$ GeV/c) allows us to describe the E665 data.
Using the in-medium reduced low-energy $NN$ cross sections of refs. \cite{Li:1993rwa,Li:1993ef} leads
to $\sim 20\%$ reduction in the neutron multiplicity. This, however, would bring the calculation with $p_{\rm cut}=1$ GeV/c
to even closer agreement with the E665 data for the lead target.

In Fig.~\ref{fig:Pb_E665}b, the neutron spectra for different maximum times of the GiBUU calculation are shown.
The dynamically emitted component of the spectra at low energies ($E_n \ltsim 6$ MeV) is moderately sensitive
to variation of the maximum time, however, the full spectrum is practically stable.

We have also performed the calculation by setting $p_{\rm cut}$ to a very large value (1 TeV) which can be considered
as no formation at all. 
This calculation produces almost indistinguishable with QDM yields
of neutrons below 10 MeV (of course, at higher neutron energies the QDM calculation gives smaller yields than
the calculation with instantly formed hadrons experiencing more final state interactions).
However, the default GiBUU calculation gives smaller yields of slow
neutrons as compared to the QDM one.
It appears that the difference originates mainly from much larger expansion time,
i.e. the string proper time between the hard interaction and first string breaking space-time points,
in the GiBUU  model \cite{Gallmeister:2005ad}.
In the nucleus rest frame, the expansion time corresponds to the time period $t < t_{\rm prod}$ when
the effective cross section of Eq.(\ref{sigma_eff_GiBUU}) is set to zero in the default GiBUU calculation.

\begin{figure}
  \includegraphics[scale = 0.50]{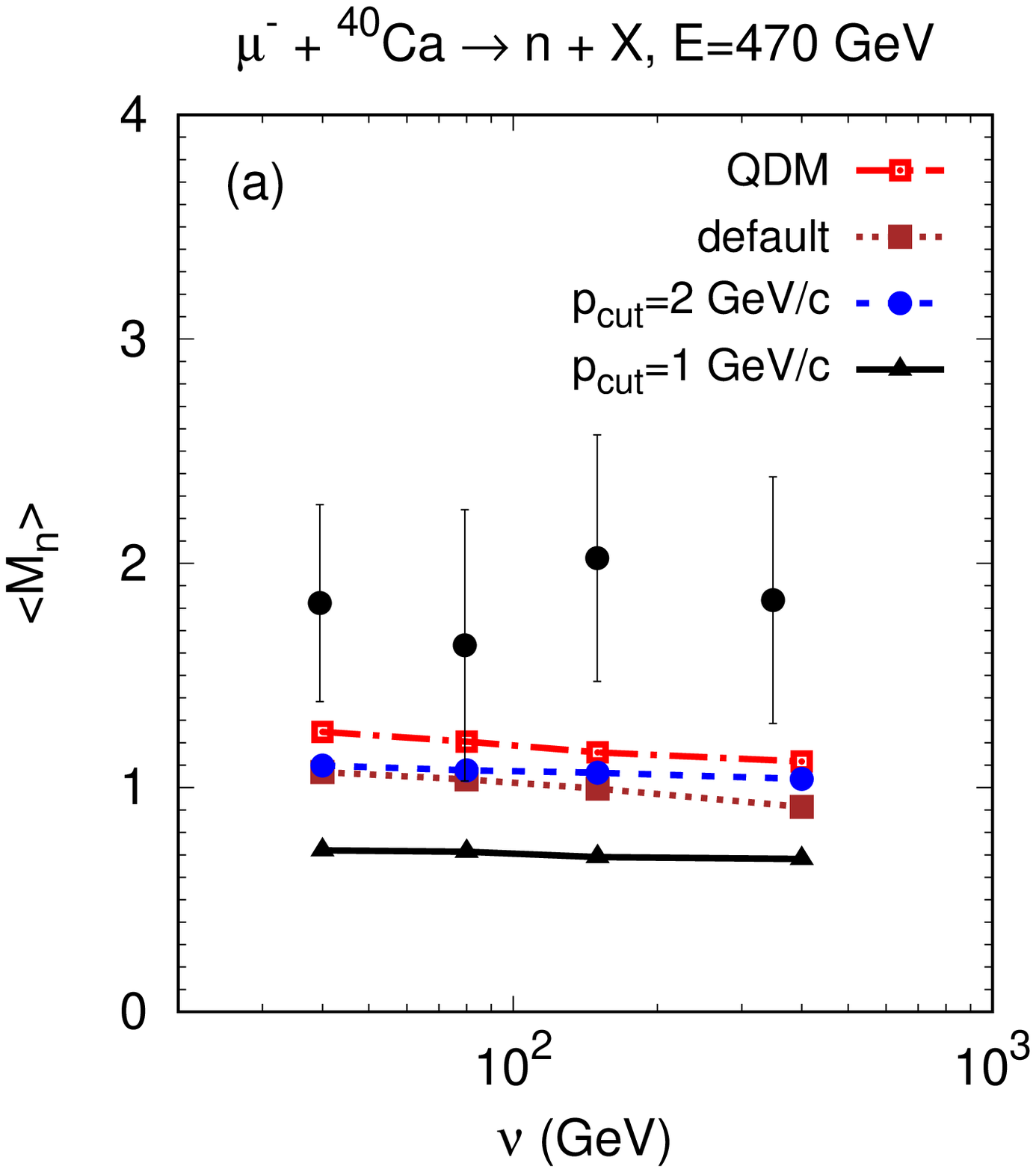}
  \includegraphics[scale = 0.50]{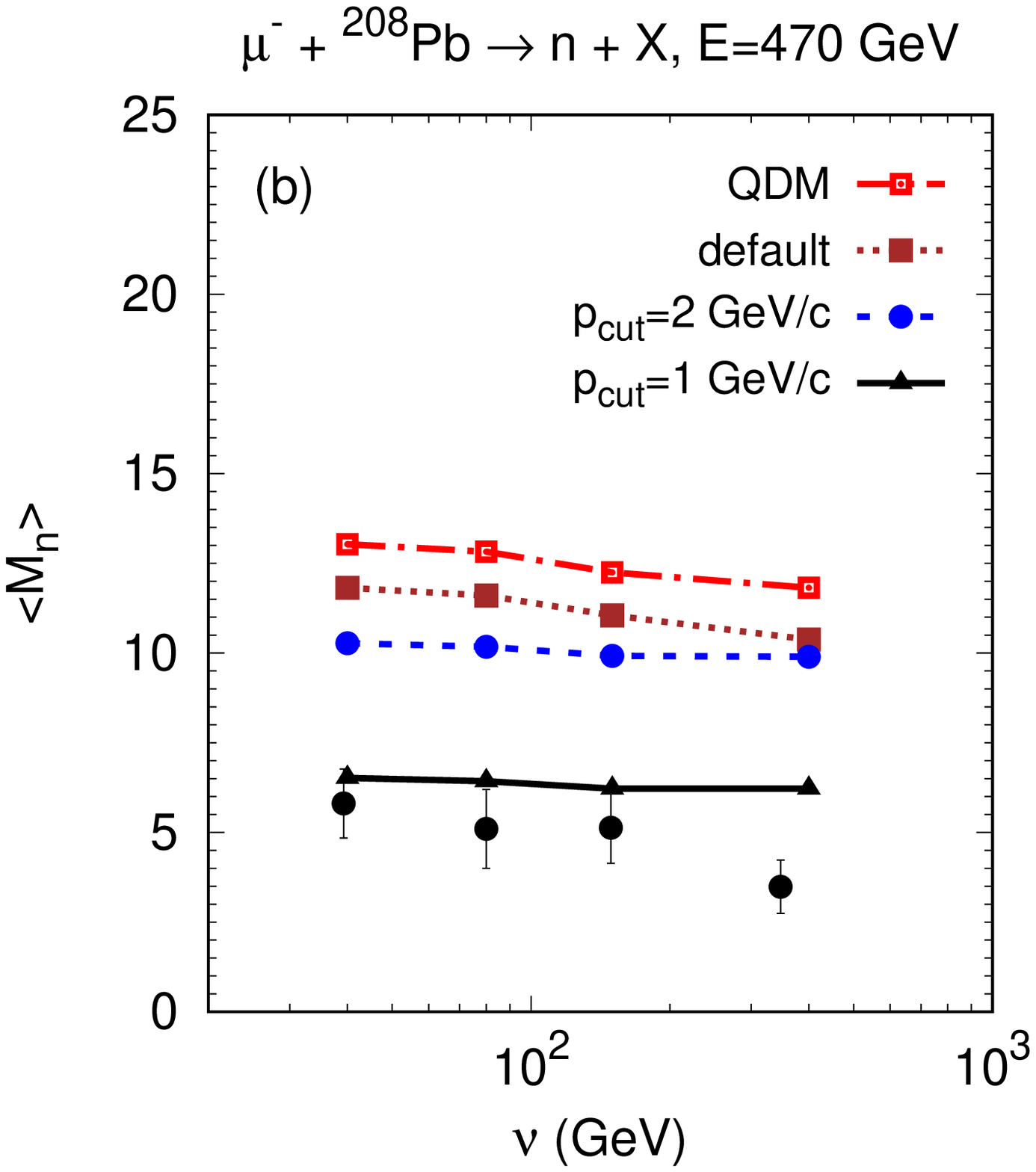}
  \caption{\label{fig:Mn_vs_nu} Multiplicity of neutrons with energy $E_n < 10$ MeV
    as a function of a virtual photon energy for
    (a) $\mu^- + ^{40}$Ca and (b) $\mu^- + ^{208}$Pb DIS at 470 GeV.
    Lines with symbols show the GiBUU results with different prescriptions for hadron formation:
    dash-dotted (red) with boxes -- QDM calculation, dotted (brown) with boxes -- GiBUU default,
    dashed (blue) with circles -- cutoff momentum  2 GeV/c,
    solid (black) with triangles --  cutoff momentum  1 GeV/c.
    Experimental data are from ref. \cite{Adams:1995nu}.}
\end{figure}
Having successfully described the E665 data for the lead target with $p_{\rm cut}=1$ GeV/c, we would
expect that the description of the data for a lighter target nucleus would require even a lower value
of the cutoff momentum. This is because a hadron likely passes through the nucleus without 
interaction, if its formation length is comparable with the nuclear size. Thus, we conclude from Eq.(\ref{form})
that the cutoff momentum should be proportional to the nuclear radius. This naive picture is, however,
not supported by the experimental data. Fig.~\ref{fig:Mn_vs_nu} shows the average neutron multiplicity
vs the photon energy $\nu$ for the calcium and lead targets.
In the case of the $^{40}$Ca target, the neutron multiplicities are underestimated by a factor of 2.5
in the calculation with $p_{\rm cut}=1$ GeV/c and by 30\% -- in the QDM calculation.
Thus, we fail to describe the E665 data for the calcium target based on the present-day formation length concept.
This  suggests (if the  E665 neutron data are confirmed) a substantially more complicated mechanism of formation
of slow hadrons probably involving a significant degree of coherence between the produced and spectator  partons
which ultimately form slow hadrons.

\subsection{Quasireal photon interactions with nuclei at the LHC and RHIC}
\label{gammaA_LHC}

Nowadays, the feasibility of the studies of the heavy ion UPCs at colliders is widely realized.  
In such processes, one of the ions serves as a source of a quasireal photon which interacts with the second ion
leaving the first ion intact. The UPCs are now intensively studied at the LHC.
The most recent review of the UPC physics at the LHC can be found in ref. \cite{Baltz:2007kq}.
For a summary of the recent results and references we refer the reader to ref. \cite{Angerami:2319206}.

The maximum photon longitudinal momentum can be estimated in the c.m. frame of colliding nuclei (collider
laboratory frame) from the condition that the photon wave length is equal to the radius of the Lorentz-contracted emitting nucleus
\cite{Baltz:2007kq}, i.e. $k_L^{\rm max} \simeq \gamma_L/R_A$, where $\gamma_L$ is the Lorentz factor.
Assuming symmetric colliding system, we get $k^{\rm max} = \gamma_L 2 k_L^{\rm max} \simeq 2 \gamma_L^2/R_A$
in the rest frame of another colliding nucleus.
The estimated values of the maximum photon momentum $k^{\rm max}$ and of the $\gamma^* N$ c.m. energy $W$
are listed in Table~\ref{tab:kmax} (see also Table 1 in ref. \cite{Baltz:2007kq})
\begin{table}[htb]
  \caption{\label{tab:kmax}Parameters of UPC Au+Au at RHIC and Pb+Pb at LHC.}
  \begin{center}
    \begin{tabular}{|c|c|c|c|c|}
    \hline
    ~~~~& $\sqrt{s_{NN}}$ (TeV)  & $\gamma_L$ &  $k^{\rm max}$ (TeV/c)  & $W$ (GeV) \\
    \hline
  RHIC  & 0.2                   & 106        &  0.642                &   34.7 \\
  LHC   & 5.5                   & 2931       &  477                  &   946  \\
    \hline
    \end{tabular}
  \end{center}
\end{table}

In this paper we are discussing the simplest case when the photon experiences a hard interaction with one nucleon only
and model it using the PYTHIA. 
In this model, the fragmentation of a nucleon weakly depends on
the light cone momentum fraction $x_N$ of the nucleon carried by the removed parton  
for moderate $x_N$ and on the type of the removed parton.
In principle, due to the color effects, the cases of the removal of a gluon (which is the dominating process in $AA$ UPCs)   
and a quark from the nucleus may differ. In particular, in the Lund model, the removal of a gluon, the $\gamma g \to \mbox{two jets}$ process,
corresponds to the attachment of two strings to the nucleon leading to somewhat softer jets as compared to the quark case,    
$\gamma q \to q^\prime$. (This effect would be possible to test in $pA$ UPCs.)
For now we neglect this difference and rely on the inclusive set of the PYTHIA events under certain kinematic conditions. 

Since the PYTHIA model describes a $\gamma^*$-nucleon collision with the photon emitted by the scattered lepton,
it is impossible to initialize an exactly real photon. However, by varying the kinematics of the lepton scattering
one can change the photon virtuality $Q^2$ (or Bjorken $x=Q^2/2 \nu m_N$) and the relative flux of the longitudinal
photons $\epsilon$. We have checked that the variation of $\epsilon$ in the interval $0.02-0.10$ has practically
no effect on the results. In the calculations below we set $\epsilon=0.05$.
The $z$-axis is directed along the photon momentum.

In the dijet events measured at the LHC the jets are typically required to have $p_t > 20$ GeV/c (cf. \cite{Aad:2015xis}).
For the back-to-back jets, this roughly corresponds to setting the minimum dijet invariant mass $M^{(0)}_{\rm dijet}=40$ GeV.
We have chosen $W=100$ GeV and $W=500$ GeV as representative values of the $\gamma^* N$ c.m. energy at the LHC.
The Bjorken $x$ has been set to $M^{(0)2}_{\rm dijet}/W^2$
which gives $x=0.16$ for $W=100$ GeV and $x=6.4 \times 10^{-3}$ for $W=500$ GeV.
This is equivalent to modeling the dijet production with invariant mass $M_{\rm dijet} > M^{(0)}_{\rm dijet}$
by the direct photon on a gluon carrying the light cone momentum fraction of the nucleon
$x_N=M^2_{\rm dijet}/W^2$.     
The lower boundary imposed on the dijet invariant mass
guaranties the smallness of the shadowing effects.

We have also performed calculations at RHIC kinematics with $W=30$ GeV, $x=0.16$.
This corresponds to setting the threshold dijet invariant mass $M^{(0)}_{\rm dijet}=12$ GeV.

\begin{figure}
  \includegraphics[scale = 0.53]{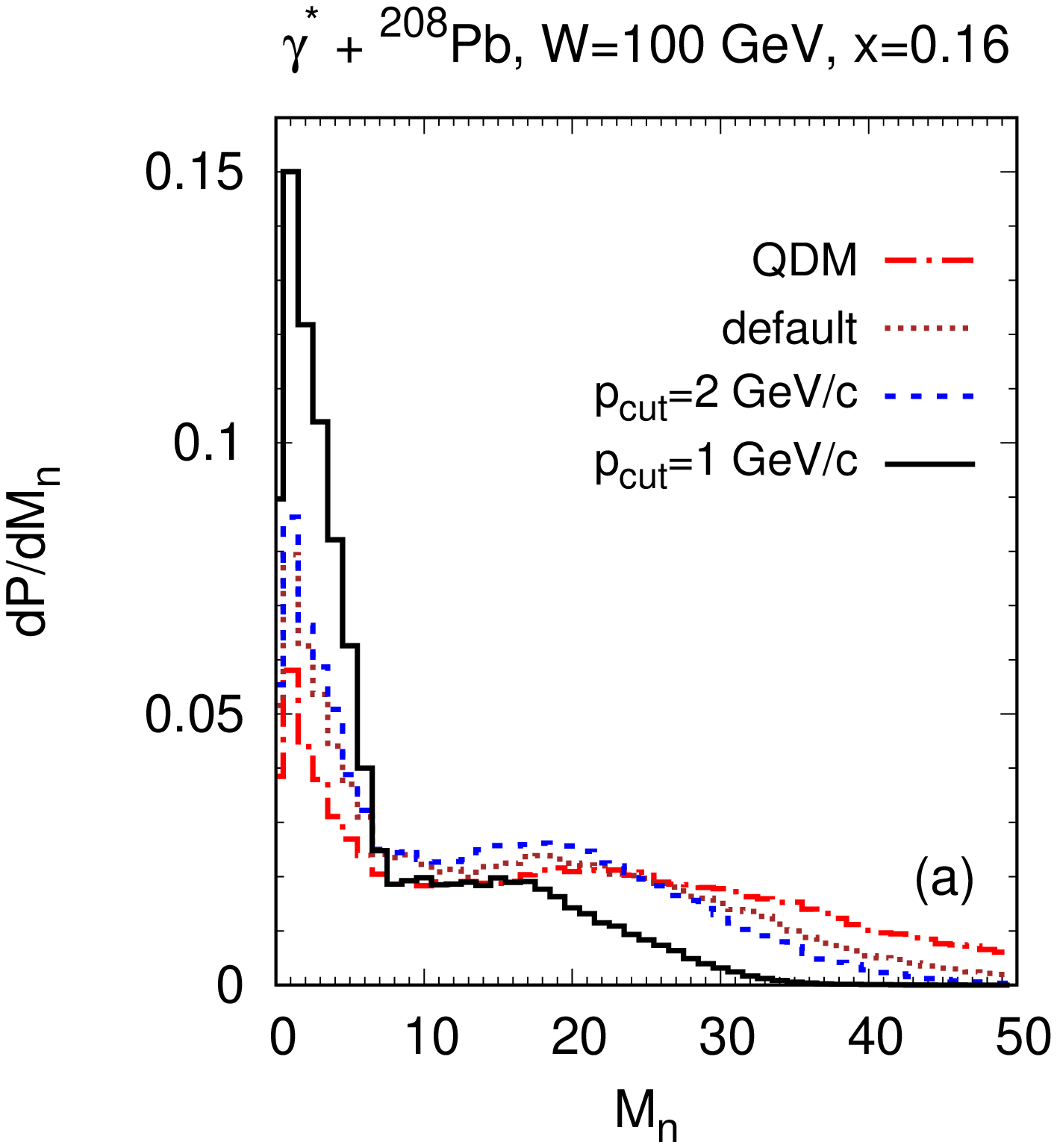}
  \includegraphics[scale = 0.53]{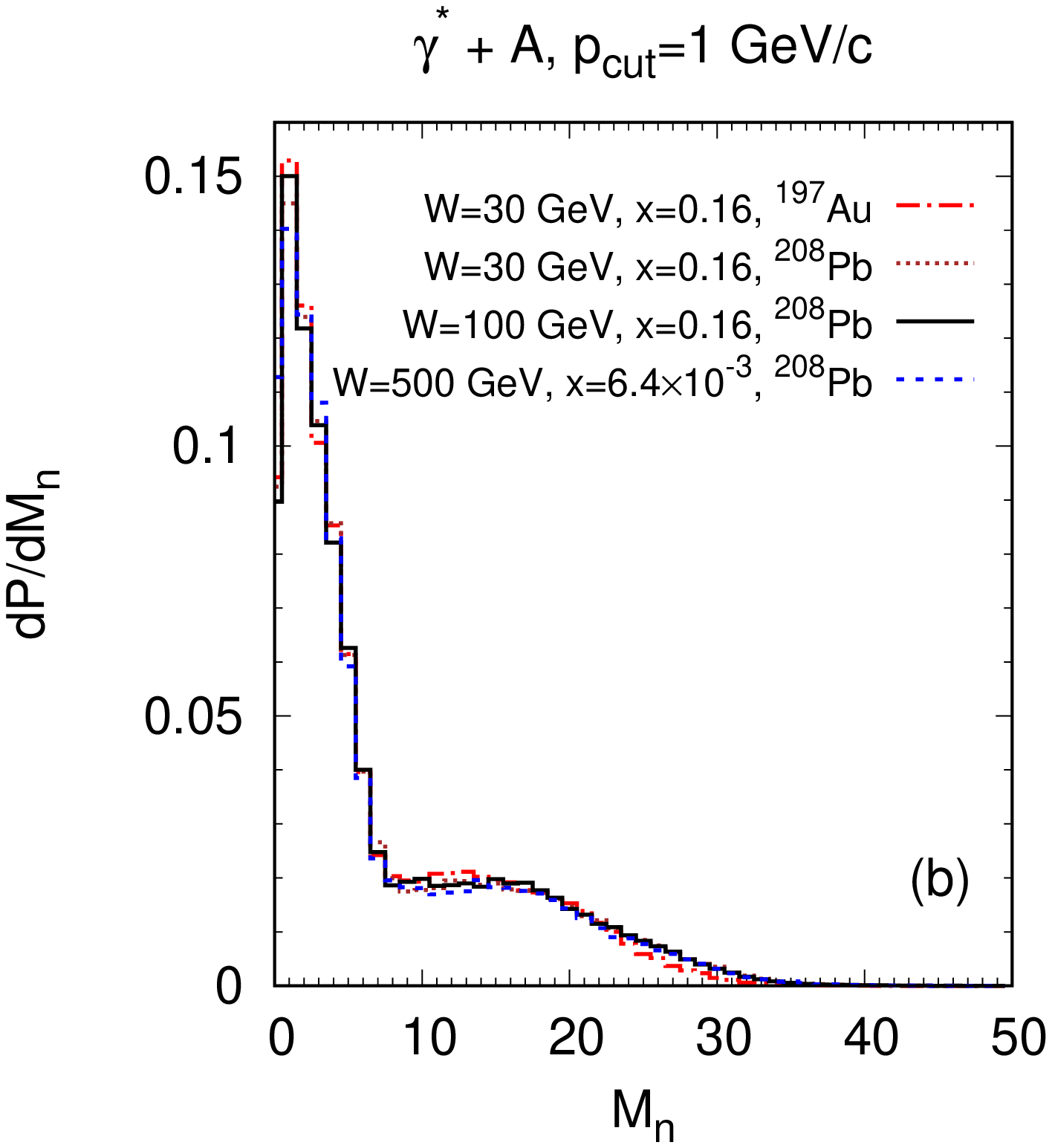}
  \caption{\label{fig:dP_dMn} Neutron multiplicity distributions for $\gamma^*$ + nucleus deep inelastic
    collisions. Panel (a) shows the results with fixed photon kinematics $W=100$ GeV, $x=0.16$ on the $^{208}$Pb target for  
    different prescriptions for hadron formation 
   (line notations are the same as in Fig.~\ref{fig:ResProp}).
     The average neutron multiplicities (standard deviations) are:
    $17.8(14.1)$ for QDM, $14.7(12.3)$ for default, $13.1(10.9)$ for $p_{\rm cut}=2$ GeV/c, and $7.2(7.7)$ for $p_{\rm cut}=1$ GeV/c.
     Panel (b) compares the results calculated with $p_{\rm cut}=1$ GeV/c for the different photon kinematics and nuclear targets
     as indicated. The condition $x_F>0.1$ is applied for neutron selection (see Eq.(\ref{x_F}) and text below it).}
\end{figure}

The mass number loss
and excitation energy distributions of residual nuclei and the correlation between $A-A_{\rm res}$
and $E^*_{\rm res}$ in the LHC kinematics are almost identical to those for the E665 kinematics shown in Fig.~\ref{fig:ResProp}.
(Thus, the respective plots for the LHC are not shown.)
The shape of the neutron multiplicity distributions shown in Fig.~\ref{fig:dP_dMn}(a) for the LHC kinematics reflects
the shape of the excitation energy distributions of the nuclear residues (Fig.~\ref{fig:ResProp}b)
with a broad maximum near $E^*_{\rm res}=0$ and a long tail of high energy excitations.
The variation of the photon kinematics and of the target nucleus has only a little effect on the neutron multiplicity distribution
as demonstrated in Fig.~\ref{fig:dP_dMn}(b) for $p_{\rm cut}=1$ GeV/c.
This holds true also for other prescriptions for the hadron formation (not shown).
Some excess of events with low neutron multiplicities is visible in the case of $^{197}$Au target with smaller $N/Z$ ratio. 

In the ultraperipheral collision of heavy ions, it is convenient to study the transverse momentum spectrum of neutrons from
the target nucleus since this spectrum is longitudinal-boost-invariant.

Experimentally, the neutrons are supposed to be detected
in the zero degree calorimeter (ZDC) in the direction of the target nucleus. In the present exploratory work we do not intend to simulate
all details of the experimental setups at
the LHC and RHIC, which can be, in principle, done via the GiBUU+SMM framework.

The separation of neutrons in the direction of the target nucleus from those in the direction of the dijet 
can be done in terms of the longitudinal-boost-invariant Feynman variable $x_F$ defined as 
\begin{equation}
   x_F=\frac{E-p^z}{(E_A-p^z_A)/A}~,    \label{x_F_def}
\end{equation}
where $E$ ($E_A$) and $p^z$ ($p^z_A$) are the particle (target nucleus)  energy and the longitudinal component of momentum, respectively.
In the target nucleus rest frame the Feynman variable is expressed as
\begin{equation}
   x_F=\frac{E-p^z}{m_N}~.    \label{x_F}
\end{equation}
The condition that the particle is emitted in the direction of the target nucleus in the collider laboratory frame is thus
$x_F > x_F^0$, where $x_F^0=m_t/\sqrt{s_{NN}}$ with $m_t=\sqrt{m^2+p_t^2}$ being the particle transverse mass. 
In the LHC and RHIC kinematics the boundary values $x_F^0$ for the neutron with zero transverse momentum are $1.7\times10^{-4}$ and
$4.7\times10^{-3}$, respectively. These values are still larger than those of $x_F^{\rm min}=m_N^2/W^2$ as obtained from
the extreme kinematics of the $\gamma^* N \to N$ transition.

In concrete calculations we choose the cut $x_F > 0.1$ relevant, for example, for the ZDC of ATLAS detector \cite{White:2011tq}.
\begin{figure}
  \includegraphics[scale = 0.50]{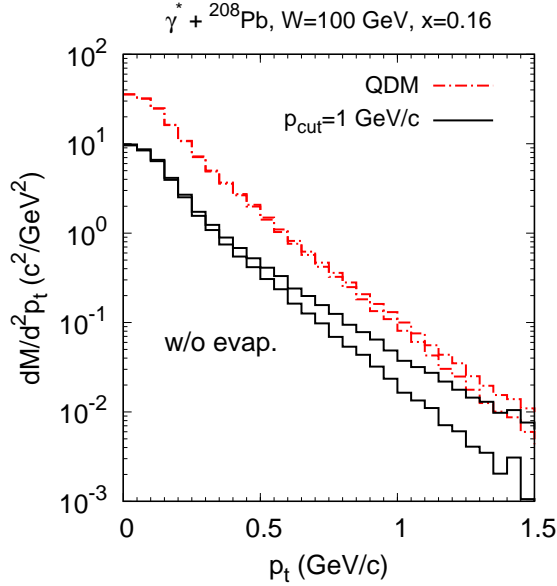}
  \caption{\label{fig:dM_d2pt_xF} Neutron transverse momentum spectra for $\gamma^*+^{208}$Pb deep inelastic collisions
    for fixed photon kinematics $W=100$ GeV, $x=0.16$. Solid (black) lines -- calculation with $p_{\rm cut}=1$ GeV/c.
    Dash-dotted (red) lines -- QDM calculation. Upper and lower lines display the spectra without and with condition $x_F > 0.1$,
    respectively. Statistically evaporated neutrons are not included.} 
\end{figure}
As we see from Fig.~\ref{fig:dM_d2pt_xF}, the chosen $x_F$-cut practically does not influence the low-$p_t$
neutron spectra originating mostly from target fragmentation.

At the intermediate transverse momenta, $p_t \sim 1$ GeV/c, the $x_F$-cut reduces the neutron yield somewhat.
The reduction effect, however, depends on the scenario of hadron formation: The neutrons with transverse momenta
of a few GeV/c originate mostly from the nucleon fragmentation in the reaction $\gamma +N\to \mbox{two jets} + n + X$
on the bound nucleon. In the case of QDM calculation, such neutrons may still interact with surrounding nucleons and, therefore,
increase their $x_F$ values due to deceleration. In the case of calculation with $p_{\rm cut}=1$ GeV/c, however, these neutrons
do not interact with nucleus, and thus a part of them remains in the 'invisible' region $x_F < 0.1$. 

The measurement of the neutron production at $p_t \gtsim 1$ GeV/c offers an additional test of the hadron formation scenario. Namely,
if only nucleons with momenta below 1 GeV/c interact with the residual nucleus,
the inclusive cross section in this kinematics should be linear in $A$
reflecting the direct mechanism of the neutron production
\footnote{Observing this effect would require a highly segmented ZDC.
  The ZDCs used in the current collider experiments  detect with a very high efficiency neutrons produced with momenta below the Fermi momentum in the nucleus rest frame.
  To detect neutrons originating from the $\gamma N$ reaction on the bound nucleon one would need a different detector like the one which is employed by
the LHCf \cite{Adriani:2018ess}.}.

\begin{figure}
  \includegraphics[scale = 0.50]{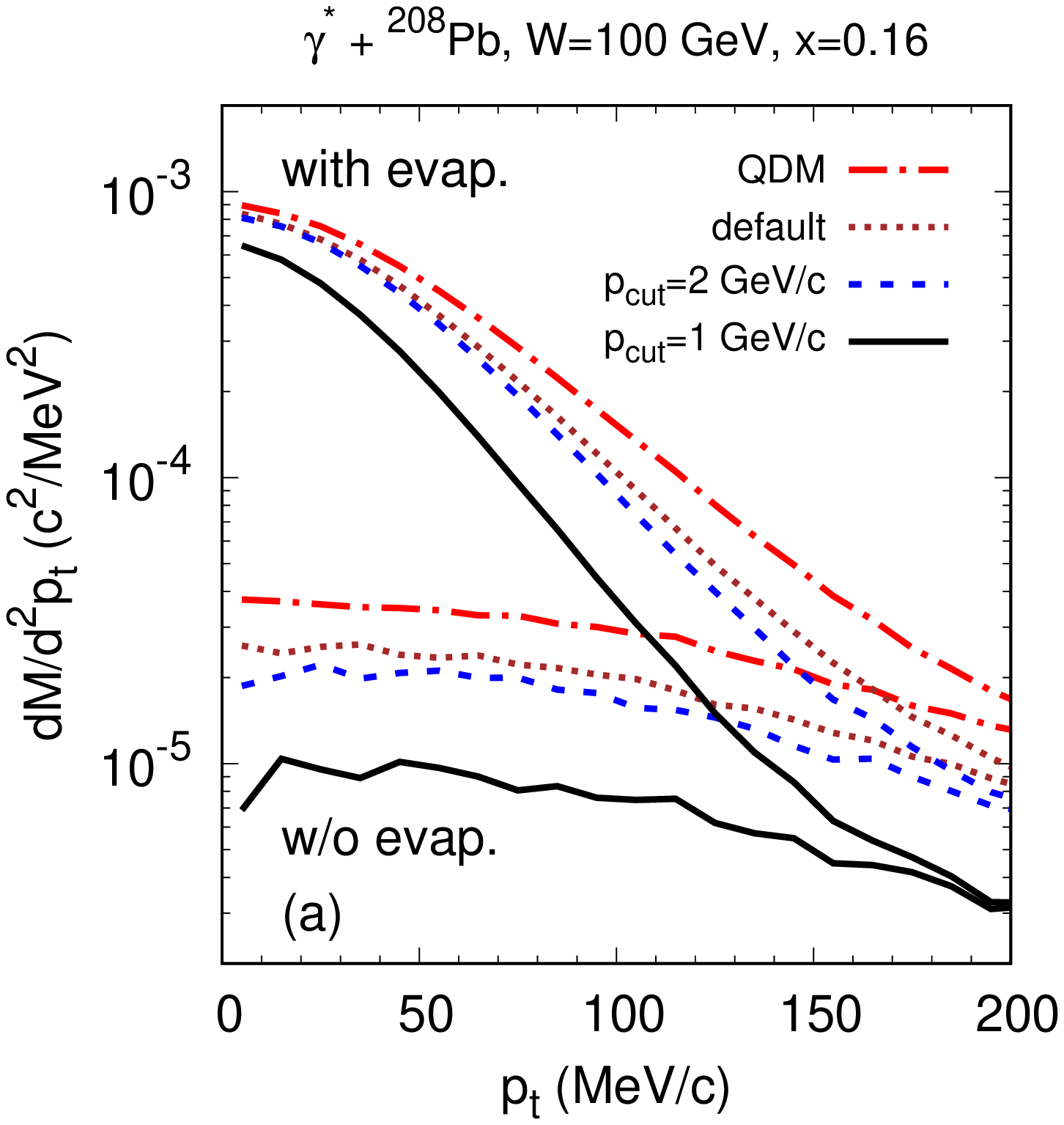}
  \includegraphics[scale = 0.50]{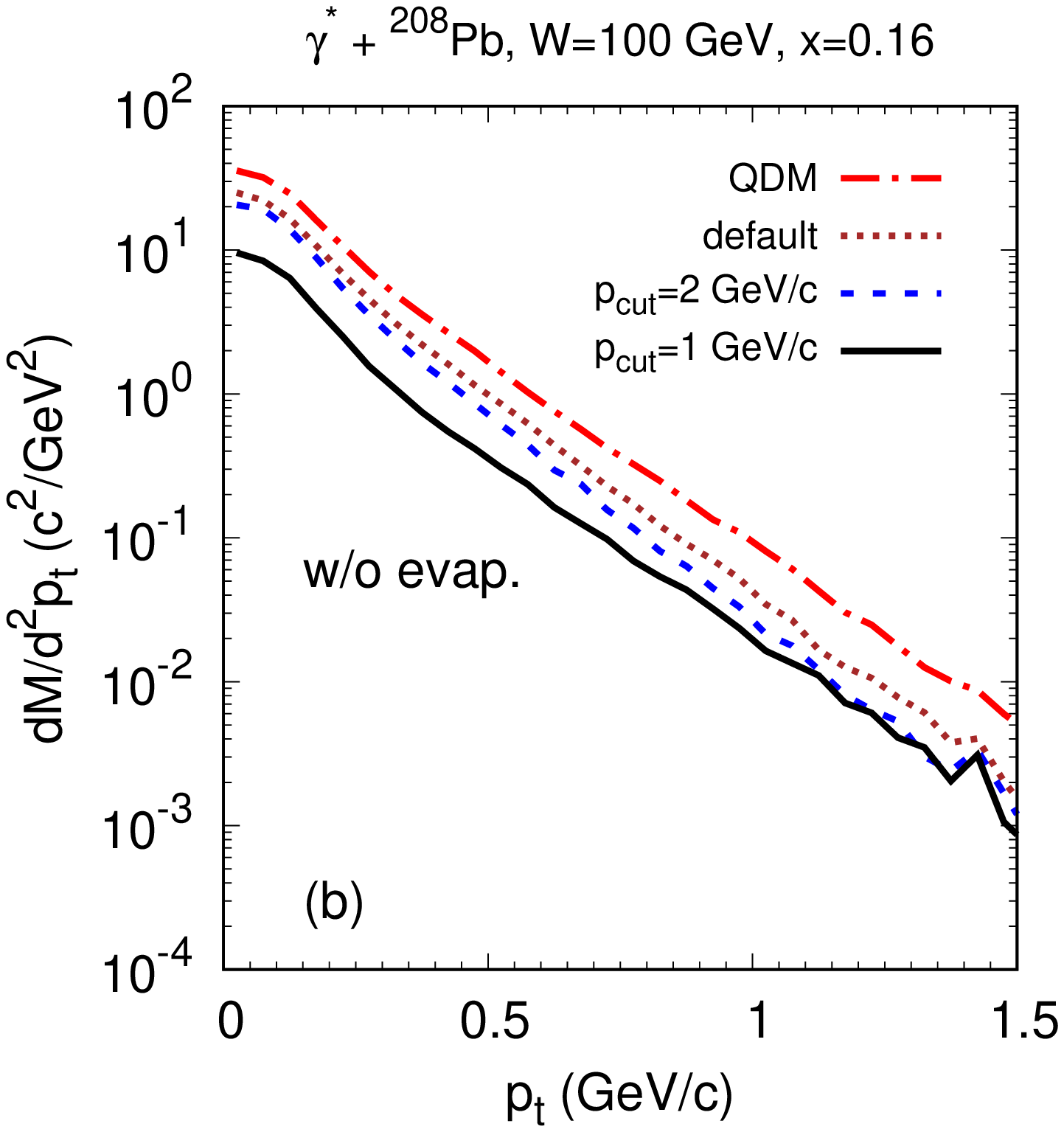}
  \includegraphics[scale = 0.50]{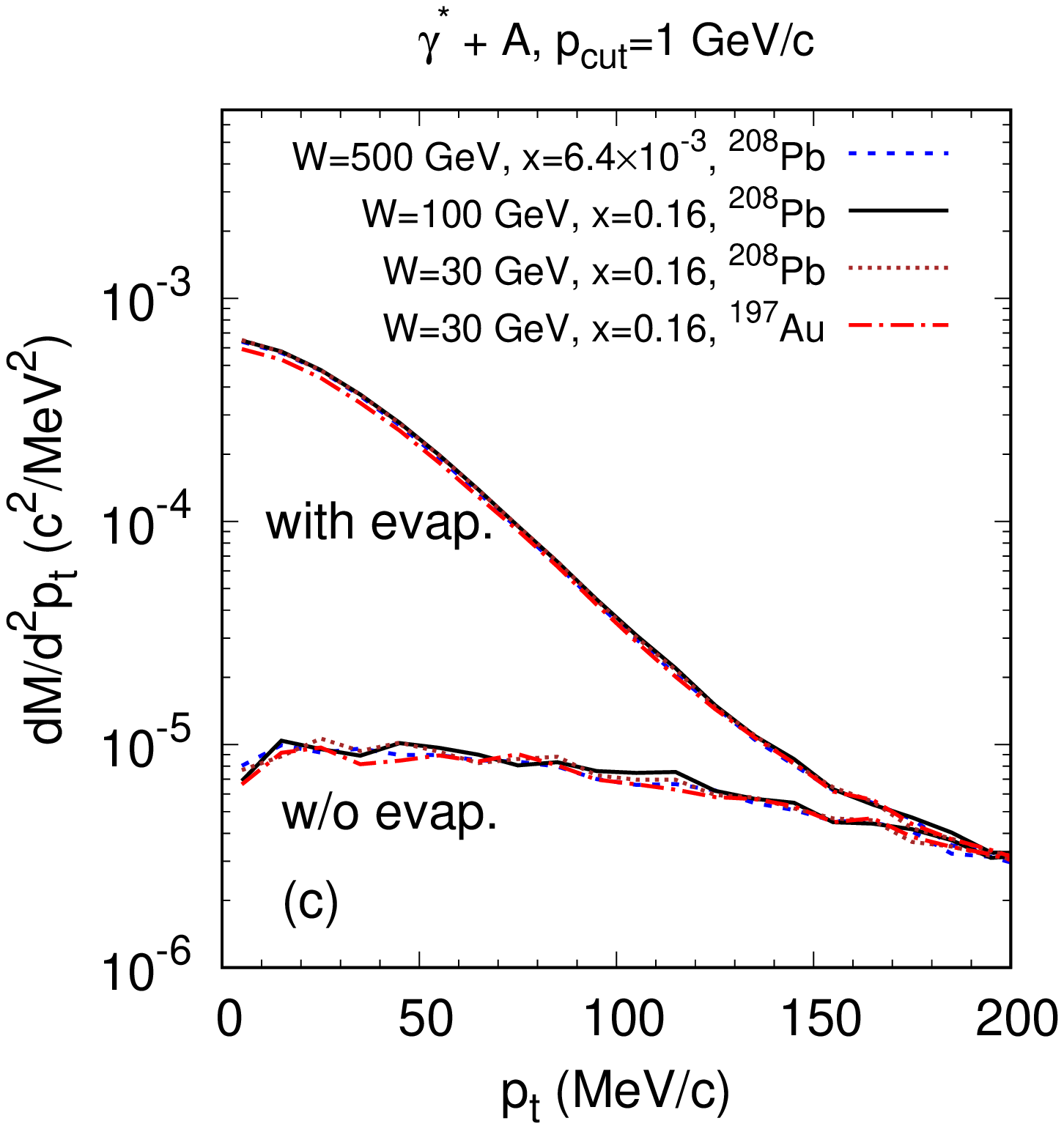}
  \includegraphics[scale = 0.50]{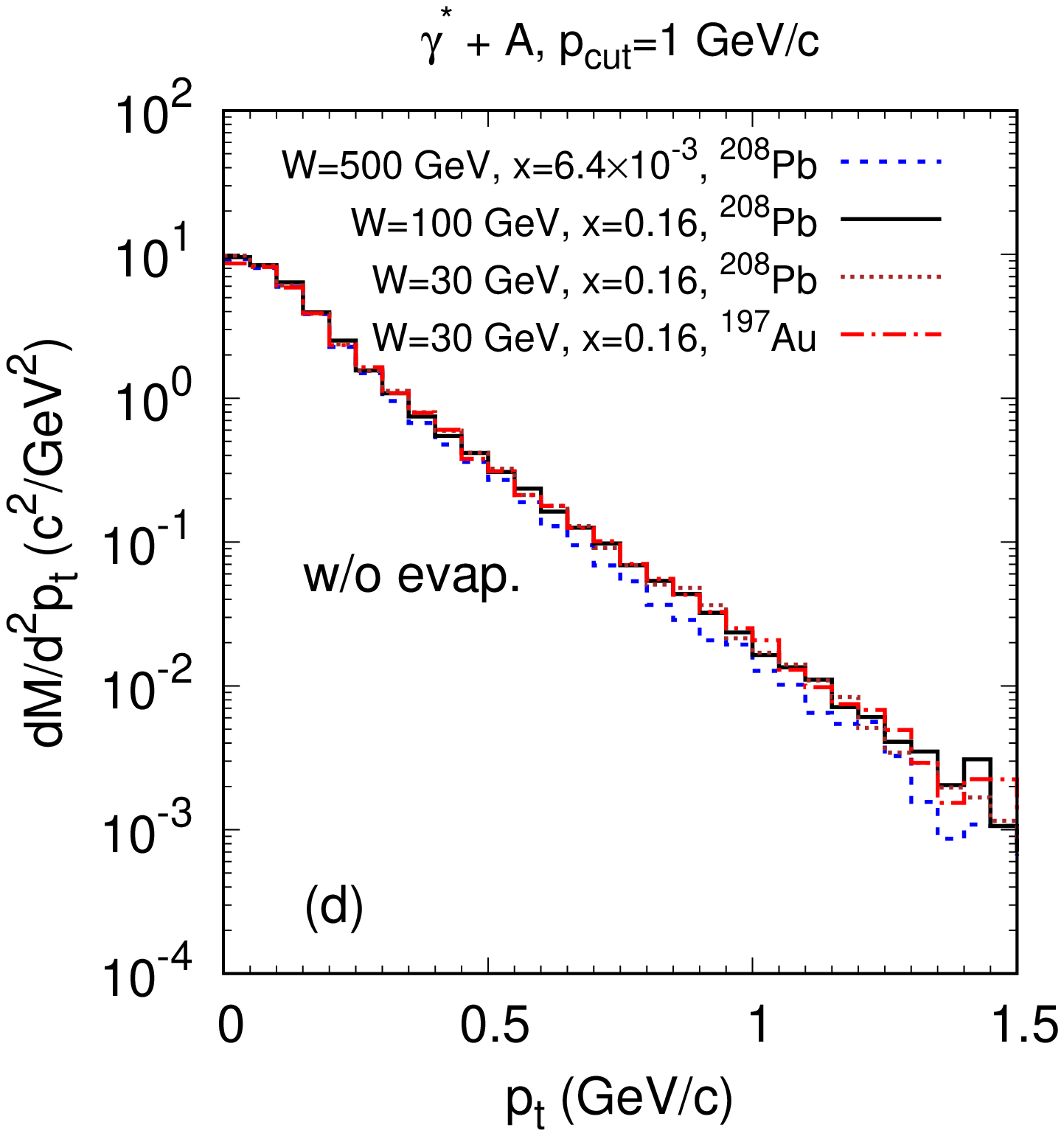}
  \caption{\label{fig:dM_d2pt_Pb} Neutron transverse momentum spectra for $\gamma^*$+nucleus deep inelastic collisions.
    (a) Spectra for fixed photon kinematics $W=100$ GeV, $x=0.16$ on the $^{208}$Pb target with different prescriptions
    for hadron formation (line notations are the same as in Fig.~\ref{fig:ResProp}).
    Upper (lower) lines show calculations with (without) statistical evaporation.
    (b) Same as (a) without evaporation, but for the large range of neutron transverse momentum.
    (c) Spectra for the different photon kinematics and nuclear targets as indicated calculated with $p_{\rm cut}=1$ GeV/c.
    (d) Same as (c) without evaporation, but for the large range of neutron transverse momentum.
    The spectra are calculated with condition $x_F>0.1$.}
\end{figure}

Fig.~\ref{fig:dM_d2pt_Pb}(a) shows the neutron low-$p_t$ spectra for different prescriptions for hadron formation.
At very small transverse momenta, statistical evaporation hides the effect of hadron formation.
However, already for $p_t$ above $\sim100$ MeV/c we observe a much  stronger dependence of the neutron emission
on the pattern of hadron formation. 
As shown in Fig.~\ref{fig:dM_d2pt_Pb}(c), the spectra at low $p_t$'s
are almost independent on $W$, $x$ and on the choice of the nuclear target.
This means that the convolution with the photon flux should not influence the low-$p_t$ spectra of neutrons.

The above behaviour of the neutron $p_t$ spectra continues also towards higher transverse momenta as shown in Figs.~\ref{fig:dM_d2pt_Pb}b,d,
if the cutting condition $x_F>0.1$ is applied. However, the $p_t$ region above 1 GeV/c is significantly influenced by the $x_F$ cut.
In particular, as we can see from Fig.~\ref{fig:dM_d2pt_xF}, in full spectra, the dependence on the hadron formation scenario tends to disappear with
increasing $p_t$. On the other hand, the full neutron spectra at $p_t \gtsim 1$ GeV/c become influenced by photon kinematics (not shown).
The detailed discussion of this transverse momentum range goes beyond the scope of our work. We would only mention one feature which exists
both in the full and $x_F$-cut spectra at $p_t$ above 1 GeV/c. Namely, the per-event spectra on the proton and nucleus are similar
at high $p_t$'s, however, the latter spectra are significantly enhanced at low $p_t$'s (cf. Fig.~\ref{fig:dM_d2pt_neut}a and Fig.~\ref{fig:dM_d2pt_Pb}d). 
This is a clear indication of the production mechanism of low-$p_t$ neutrons in secondary interactions.
Another indication is the sensitivity of the low-transverse-momentum neutron yield to the hadron formation.

\subsection{Discussion}
\label{discussion}

\begin{figure}
  \includegraphics[scale = 0.50]{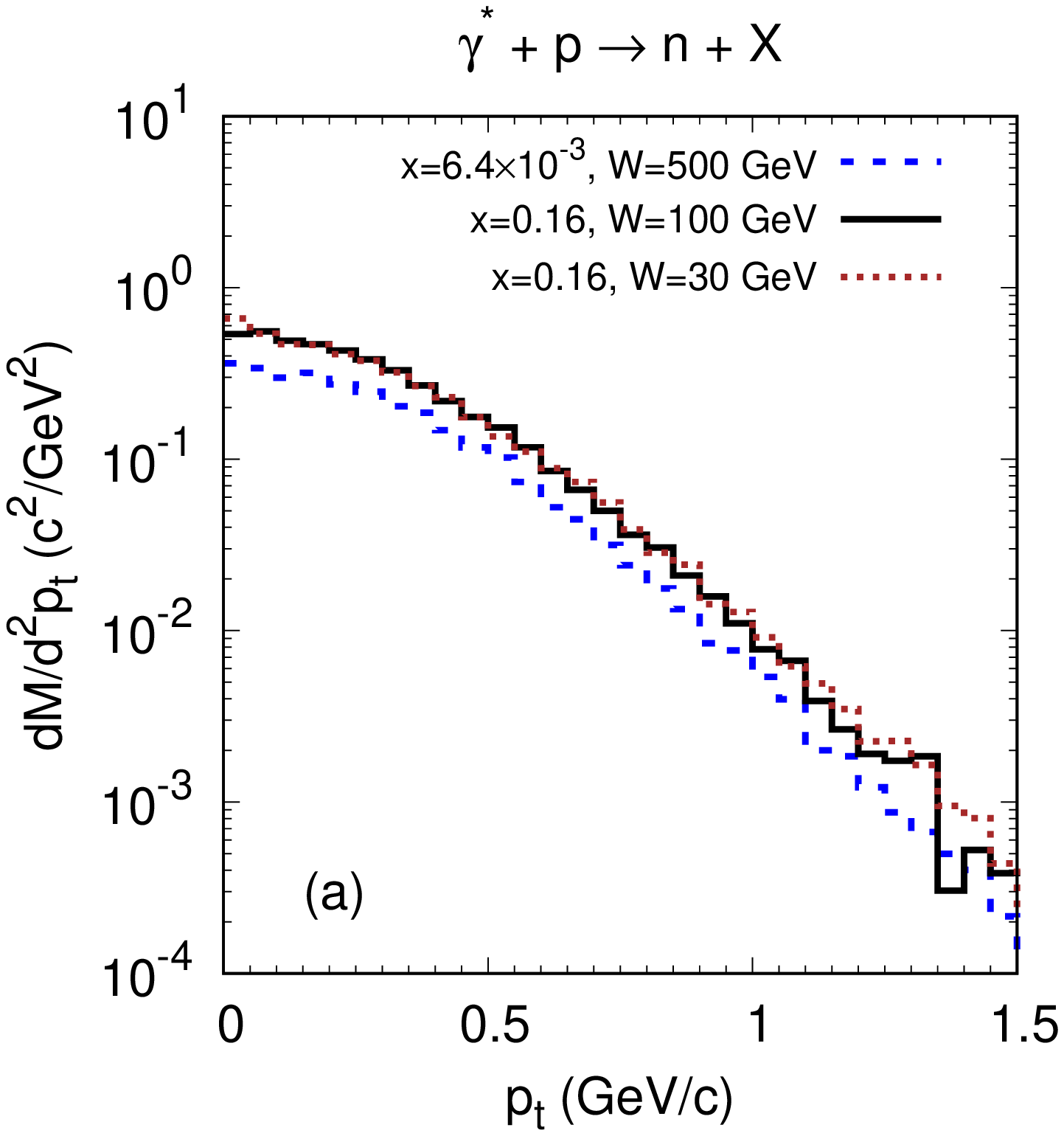}
  \includegraphics[scale = 0.50]{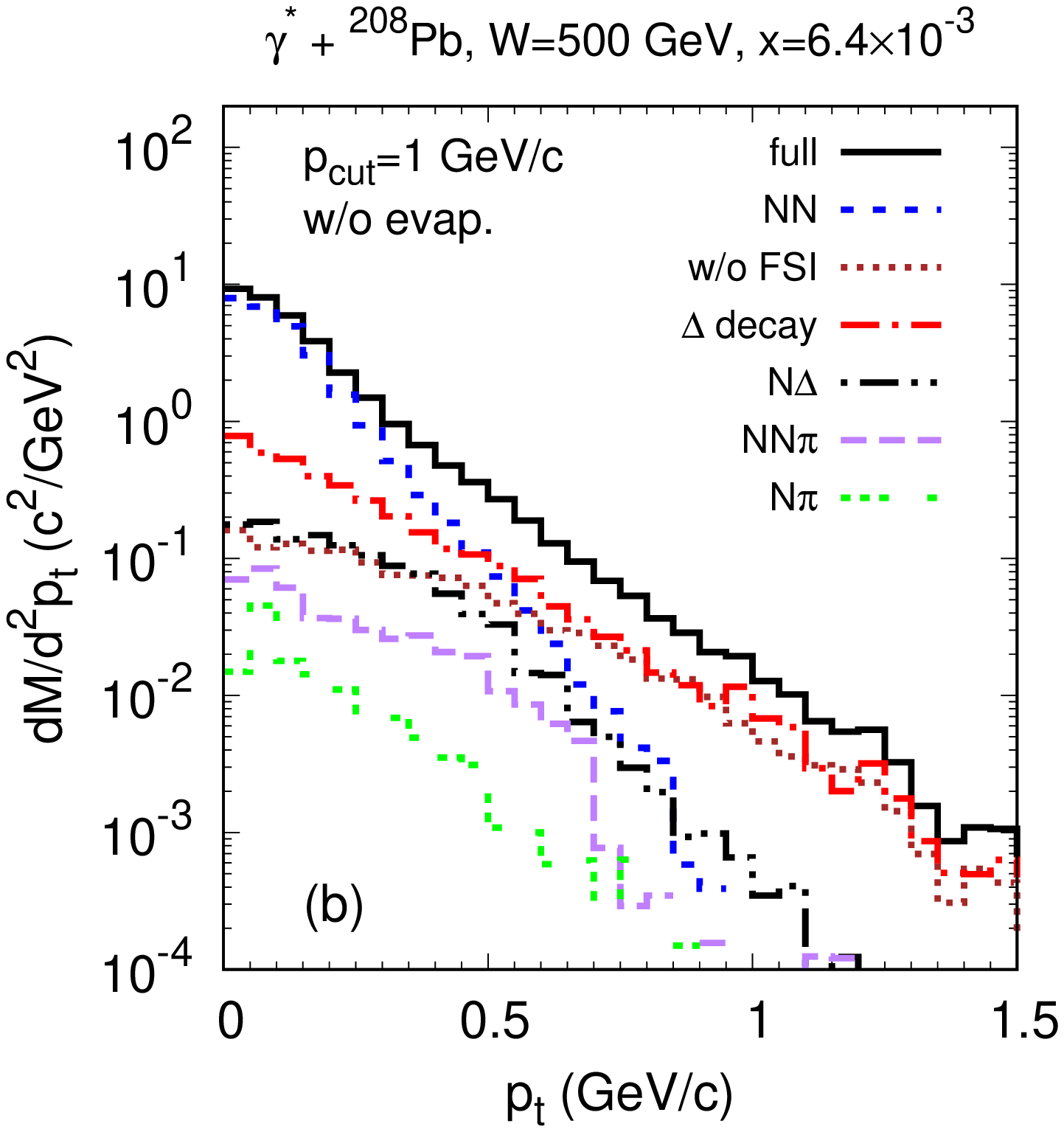}  
  \caption{\label{fig:dM_d2pt_neut}
    (a) Neutron transverse momentum spectra for $\gamma^* + p$ deep inelastic collisions.
    Different lines correspond to different
    photon kinematics as indicated.
    (b) Neutron transverse momentum spectra for $\gamma^* + ^{208}$Pb deep inelastic collisions
    for photon kinematics $W=500$ GeV, $x=6.4\times10^{-3}$ calculated with $p_{\rm cut}=1$ GeV/c without evaporation.
    Shown are, as indicated, the full spectrum without evaporation and the partial contributions of various neutron production
    channels. The spectra are calculated with condition $x_F>0.1$.}
\end{figure}

To understand better the sources of slow neutrons we present in  Fig.~\ref{fig:dM_d2pt_neut}b the decomposition of the neutron
$p_t$-spectrum in $\gamma^* + ^{208}$Pb deep inelastic collisions in partial components.
The direct production channel (denoted 'w/o FSI') includes neutrons produced in a primary $\gamma^* N$ collision on a bound nucleon
and emitted without final state interactions.
The partial components of the secondary interactions are defined in the same way as in $pA$ collisions (see Fig.~\ref{fig:pPb}c).
The high-$p_t$ part is dominated by the direct neutrons and by the neutrons produced in $\Delta$ decays, while the low-$p_t$ part
-- by the neutrons scattered elastically on a nucleon at the end of their interaction chain.
Similar to the case of $pA$ collisions, the neutrons produced in inelastic channels and resonance decays and then scattered elastically constitute
$\sim15\%$ of the latter component.
Therefore, most of the low-$p_t$ neutrons are produced in the multiple
$NN$ elastic scattering processes.

\begin{figure}
  \includegraphics[scale = 0.40]{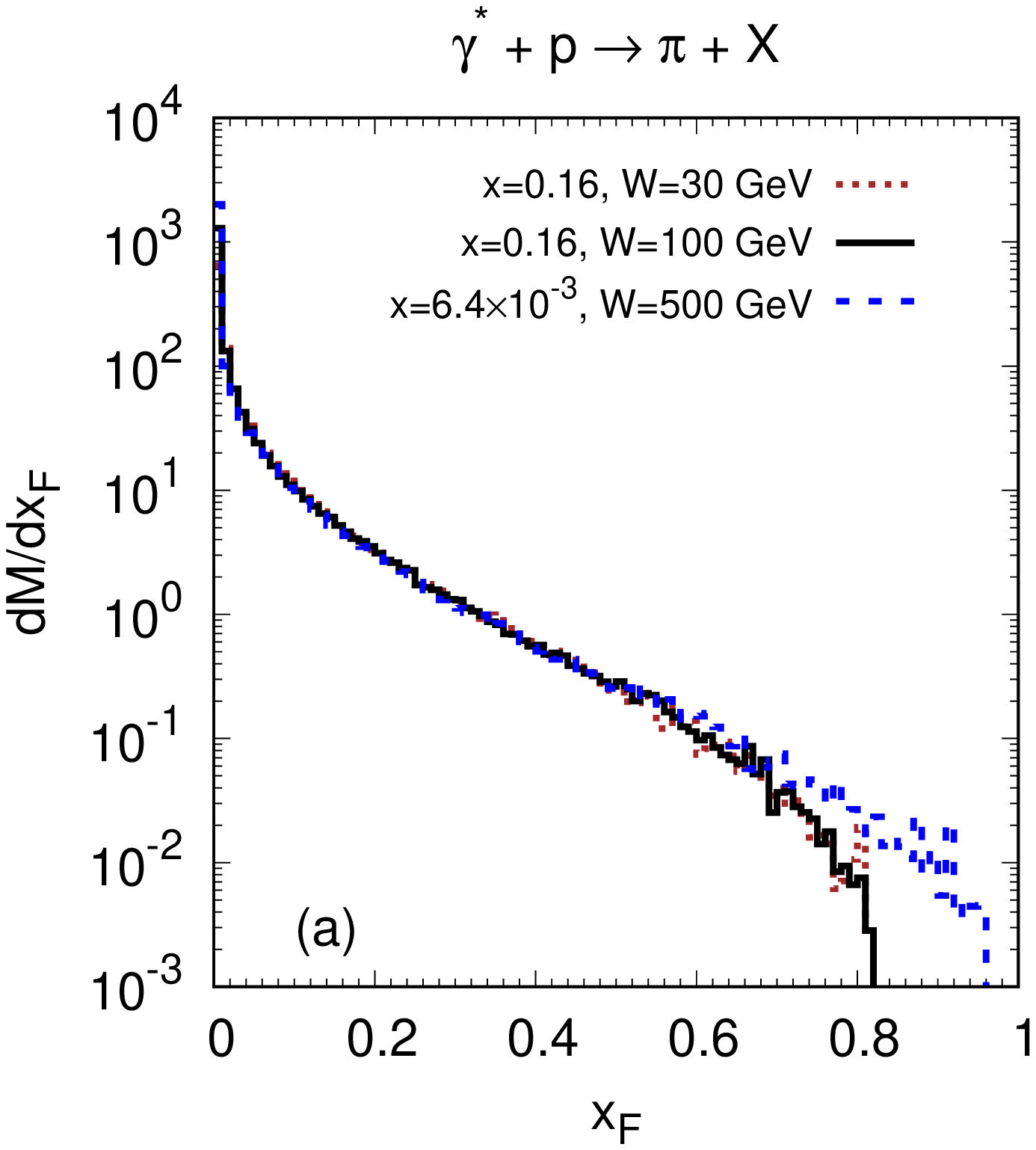}
  \includegraphics[scale = 0.40]{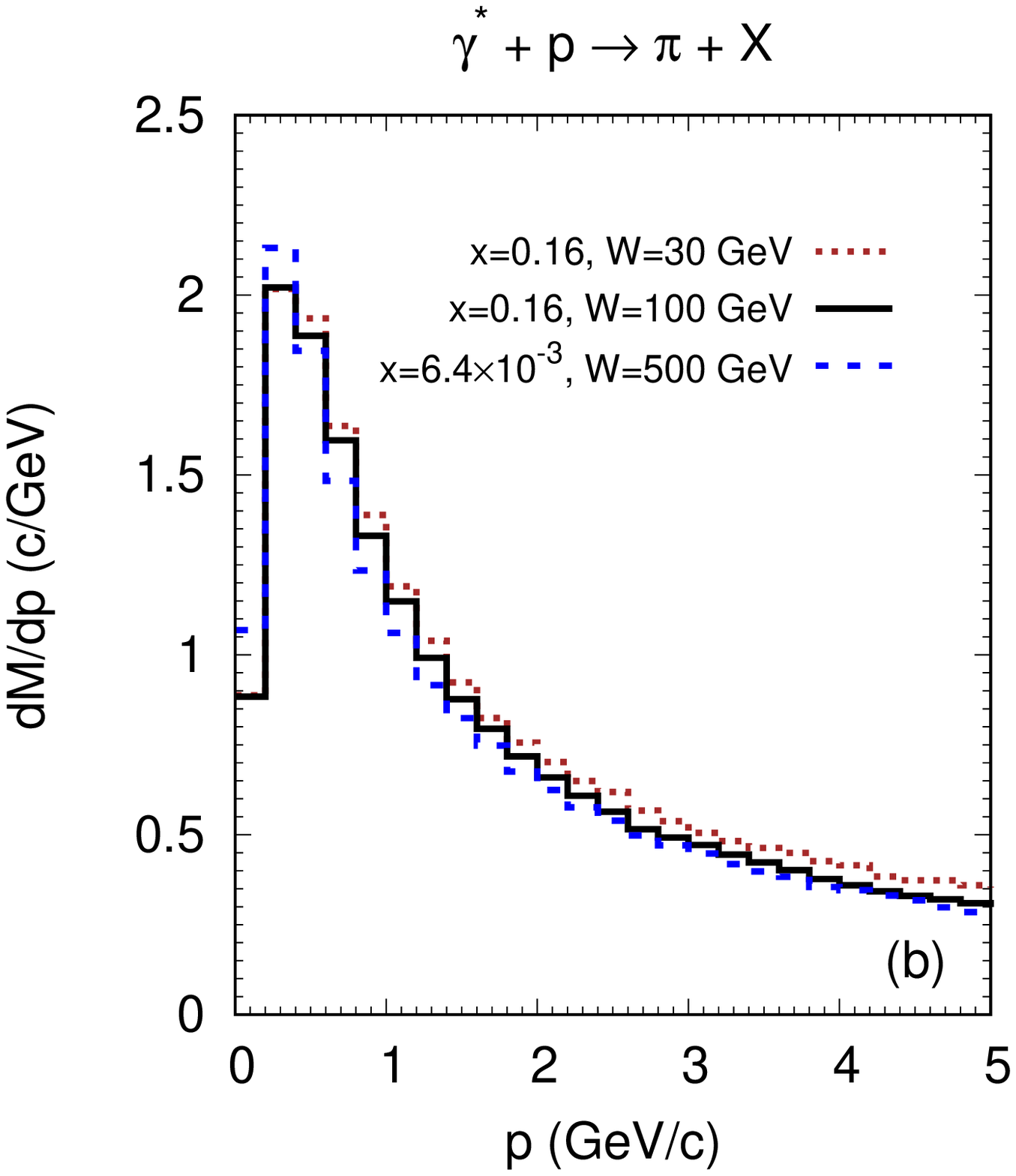}
  \includegraphics[scale = 0.40]{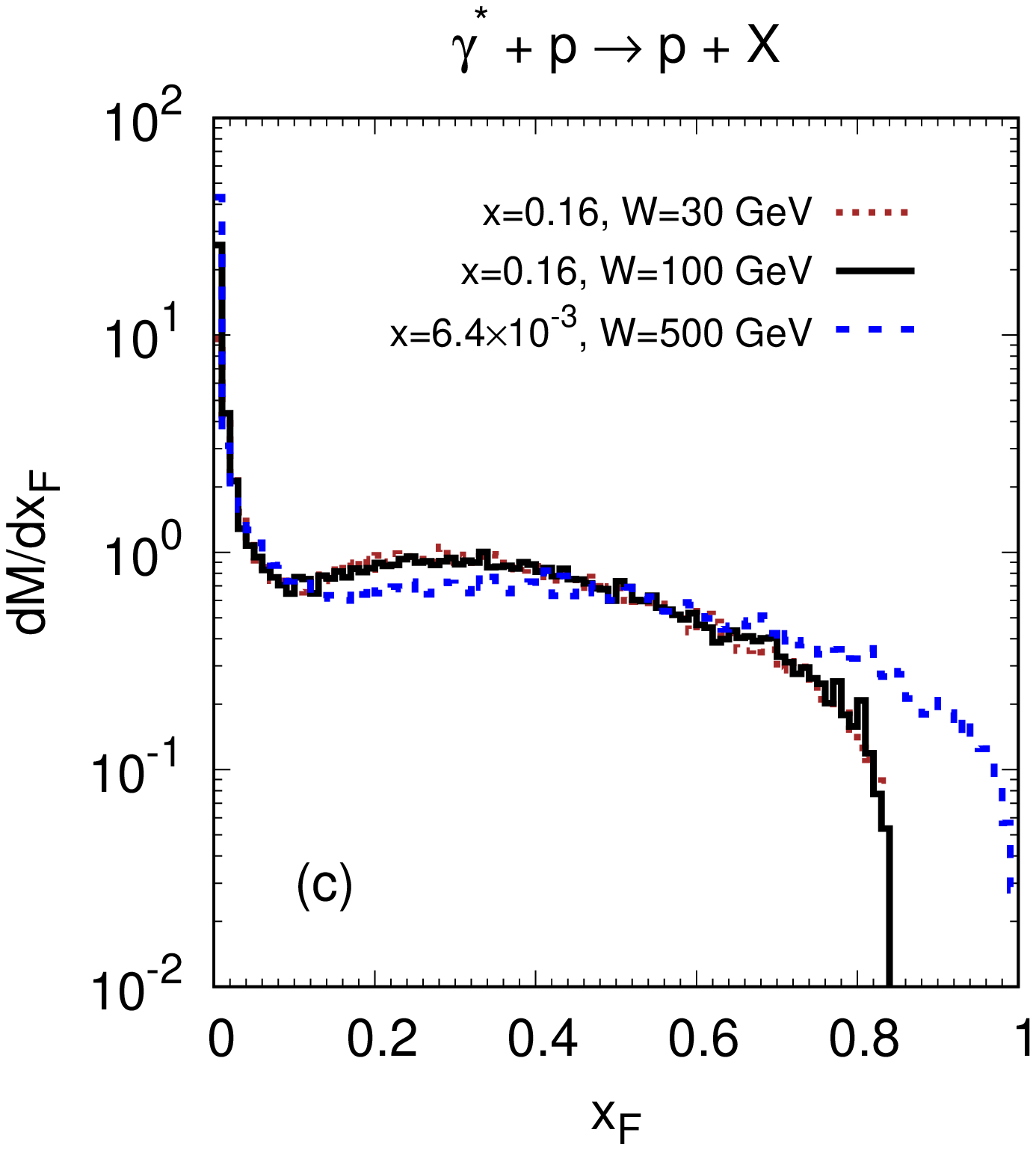}
  \includegraphics[scale = 0.40]{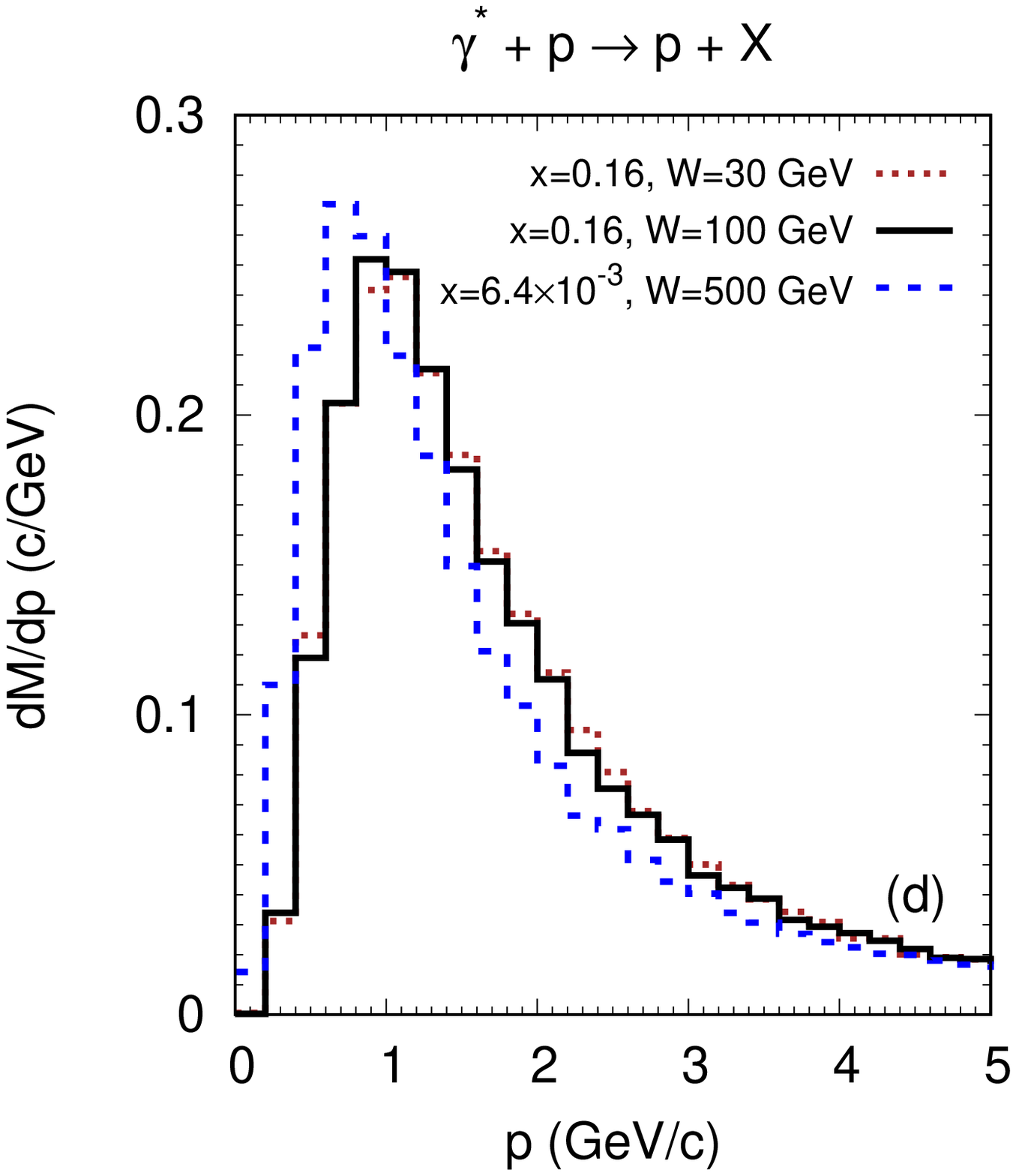}
  \includegraphics[scale = 0.40]{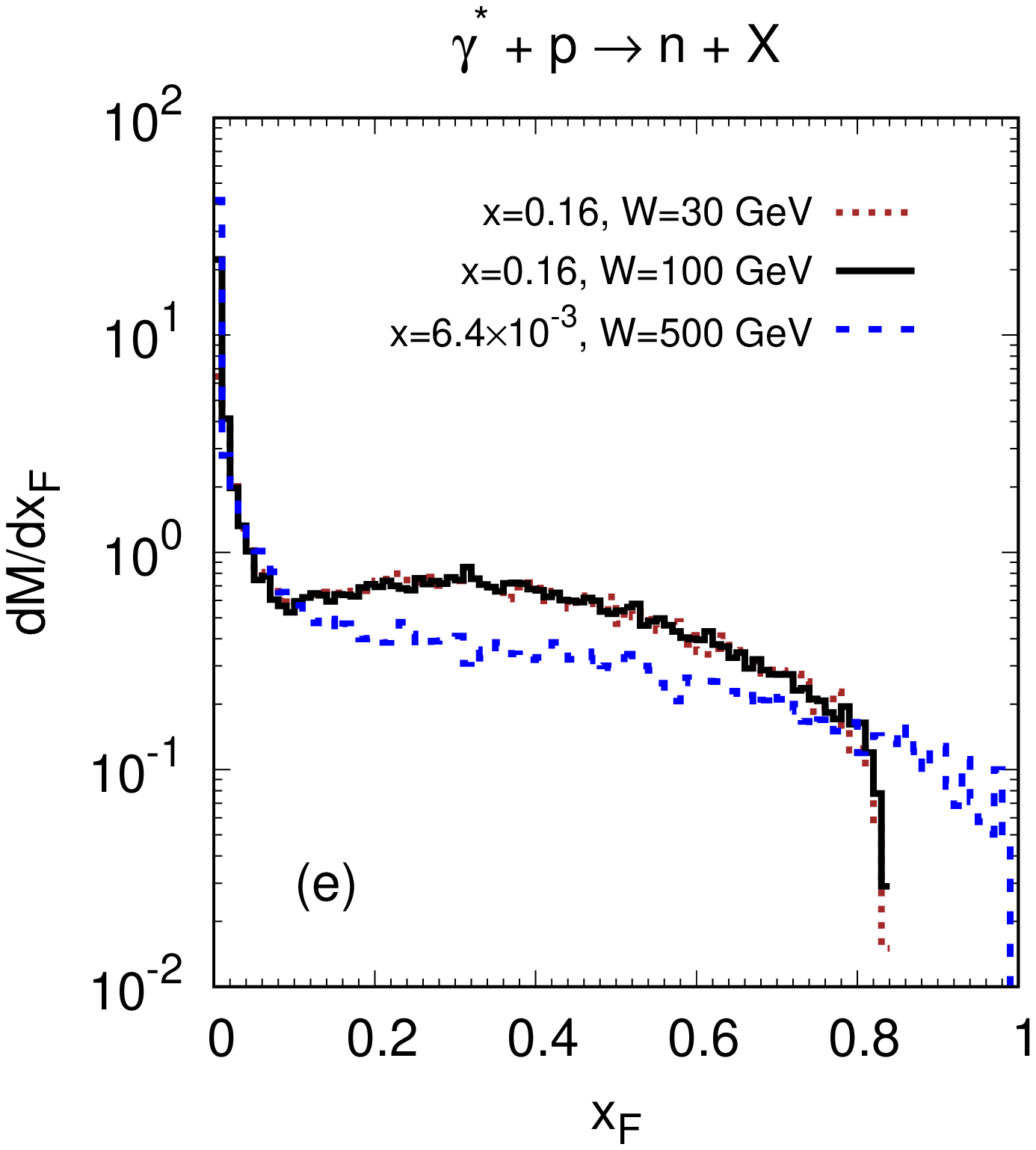}
  \includegraphics[scale = 0.40]{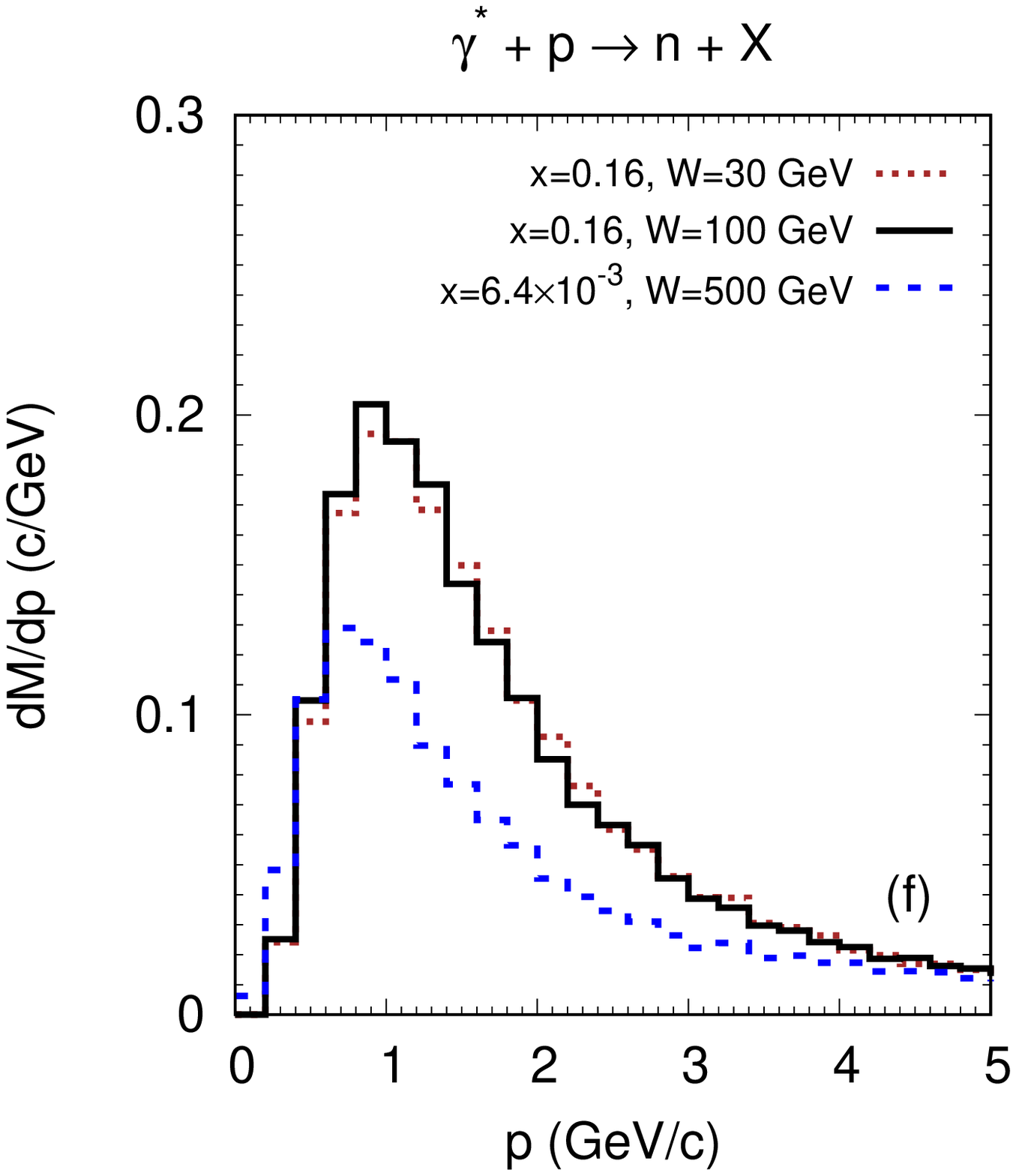}
  \caption{\label{fig:gammap} $x_F$- (left) and momentum (right) spectra of pions (a),(b), protons (c),(d), and neutrons (e),(f)
    produced in $\gamma^* + p$ collisions.
    Dotted (brown) line -- the kinematics with $x=0.16$, $W=30$ GeV,
    solid (black) line -- $x=0.16$, $W=100$ GeV,
    dashed (blue) line -- $x=6.4\times10^{-3}$, $W=500$ GeV.
    The spectra are normalized to the number of respective particles per deep inelastic event.}
\end{figure}
Let us finally take a more detailed look at the primordial particles produced in the elementary $\gamma^* + p$ deep inelastic collision.
Fig.~\ref{fig:gammap} shows the inclusive $x_F$- and momentum-distributions of pions, protons, and neutrons.
The condition $x_F < 1-x$ is fulfilled, which follows from the kinematics of the target fragmentation. Overall, this leads to the enhancement
of the slow particle production with decreasing $x$, which is most  visible for protons (Fig.~\ref{fig:gammap}(c),(d)).

The directly produced slow neutrons are even more sensitive to the photon kinematics (Fig.~\ref{fig:gammap}(e),(f)).
Like in the proton case, the neutron spectrum also shifts to smaller momenta with decreasing $x$, but in addition gets
significantly suppressed because the $p \to n$ transition involves the valence quark.
We can see from Fig.~\ref{fig:dM_d2pt_neut}b that primordial protons and neutrons with momenta $\ltsim 1$ GeV/c
govern the production of low-$p_t$ neutrons.
While the momentum spectra of primordial protons and neutrons are influenced
by the photon kinematics, it is not the case for production of  the low-$p_t$ neutrons.
This nontrivial behaviour can be explained by the fact that the
sum of the proton and neutron primordial spectra above Fermi momentum and below 1 GeV/c is weakly
influenced by the photon kinematics.

We see that pions  are produced most abundantly (Fig.~\ref{fig:gammap}(a),(b)).
The yields of protons and neutrons are suppressed by an order of magnitude relative to the pion yield.
In the case of nuclear target, pions with momentum $\sim0.3$ GeV/c
are strongly absorbed in the processes $\pi N \to \Delta$. This explains large $\Delta$-decay
and $N\Delta \to NN$ absorption components in the neutron transverse momentum spectrum
below 0.5 GeV/c in Fig.~\ref{fig:dM_d2pt_neut}b.

\section{Directions for further studies}
\label{directions}

We have considered the elementary process most close to the leading twist DIS
-- the production of dijets in the direct photon -- gluon interaction in the kinematics where nuclear shadowing effects are small.
Here we outline several  other interesting processes.

(a) In the model we ignored possible differences between the final state interactions for the case of the gluon and quark removals.
One possible difference arises in the Lund string model where effectively two strings are approximating an octet residual system.
Other effect is  a larger energy loss in the propagation of a color octet dijet through the media
leading to a higher nuclear excitation energies due to larger number of wounded nucleons.
Although the production of slow neutrons is expected to be dominated by the interactions of products of the nucleon fragmentation
with surrounding nucleons, the jet propagation may still provide significant contribution since there should be no rapidity gap between
the products of the dijet- and nucleus fragmentation \cite{Angerami:2319206}.

(b) New effects are expected in the kinematics where a dijet is produced in the resolved photon -- nucleus interaction.
In this case, the active parton belonging to the photon wave function carries the light cone fraction 
$x_\gamma$ of the photon momentum which is significantly smaller than one.
The selection of a parton with a smaller $x_\gamma$ in the photon corresponds to the
selection of configurations with a larger number of partons in the photon wave function.
For $x_\gamma \ltsim 0.1$, these configurations should interact with the target with the strength comparable to  
the strength of interaction of configurations responsible for nuclear shadowing in coherent 
diffractive photoproduction of $\rho$-mesons off nuclei \cite{Frankfurt:2015cwa}.
Based on this analysis we expect that the average number of wounded nucleons should be $n_{\rm wound} \sim 3$.
Correspondingly, the average number of emitted neutrons should increase by a factor of 
three, since, to a first approximation, wounded nucleons fragment independently. 

(c) One can consider minimum bias $\gamma A$ collisions at $W \sim 100-500$ GeV.
 Based on the analysis  of the color fluctuations in photon in ref. \cite{Alvioli:2016gfo} 
we  expect that in these collisions the average number of wounded nucleons would be large and fluctuating strongly.
The number of emitted neutrons should be correlated with $p_t$ of the leading particles produced
in the photon fragmentation region (smaller $p_t$ -- larger neutron number).

(d) It appears  feasible to detect charm particles in the UPCs.
At large $W \ge 100$ GeV and small enough $p_t \ltsim \mbox{few GeV/c}$, one reaches the kinematics
where the nuclear shadowing becomes significant (a factor of two reduction of the cross section as compared to the impulse approximation)
resulting in wounding in average two nucleons and hence increasing the number of neutrons by a factor of two.

\section{Summary and conclusions}
\label{summary}

We have performed hybrid, i.e. dynamical plus statistical, calculations of slow neutron production
in hard virtual photon induced inelastic collisions with nuclei.
For the dynamical stage, the quantum-kinetic GiBUU transport model \cite{Buss:2011mx}
has been used. The statistical stage has been described by using the SMM model \cite{Botvina:1987jp,Bondorf:1995ua}.
For the calculation of the characteristics ($A_{\rm res}, Z_{\rm res}, E^*_{\rm res}, \bvec{p}_{\rm res}$) of the nuclear residue,
a procedure based on the particle-hole excitations in the Fermi sea has been applied. We have tested these hybrid calculations
by comparison with available data on the neutron production in muon- and proton-nucleus interactions.

The multiplicity of slow ($E < 10$ MeV) neutrons in $\mu^-$ DIS on Pb measured by the E665 collaboration \cite{Adams:1995nu}
is found to be a factor of two smaller than what is expected from GiBUU+SMM calculations both with the default treatment of hadron formation
(based on hadron production and formation space-time points from JETSET) and with the quantum diffusion model of 
the color transparency. The neutron multiplicity and the energy spectrum in $\mu^- + ^{208}$Pb DIS can be reproduced
only if produced hadrons with momenta above 1 GeV/c are not allowed to interact with the rest of the nucleus. 

This indicates the  presence of a novel dynamics in the production of hadrons in the nuclear fragmentation
region which cannot be  reproduced  by adjusting the parameters of standard models.
In hindsight this may be not too surprising as we are dealing with particles produced in the target fragmentation region,
while the models were designed to describe the space-time picture of hadron formation in  the current fragmentation region.
Hence, further studies are necessary to get a better understanding of the hadron formation in the target fragmentation region.

In the long run, it will be possible to perform such studies at the EIC by measuring the $A$-dependence
of hadron spectra in the nucleus fragmentation region. However, already now the study of hard $\gamma A$ collisions in heavy ion UPCs at the LHC and RHIC with
detection of neutrons in the ZDC would allow to  clarify the space-time picture of hadron formation in the nucleus fragmentation region.
To this end, we have performed  exploratory studies of $\gamma$ + Pb(Au) interactions in the direct photon regime in the LHC
(RHIC) kinematics. We have found a strong dependence of the neutron multiplicities and $p_t$-spectra on the 
model of hadron formation indicating that the slow neutron production could be a sensitive tool for 
studying  the space-time picture of formation of slow hadrons in hard processes.
We have also listed several directions for further studies of the interplay of hard and soft dynamics in the UPCs. 
In particular,  we predict a strong increase  of the neutron multiplicity in the resolved photon interactions as compared to the direct photon interactions.

All modifications of the GiBUU code done in the course of this work concerning the hadron formation time and determination
of the nuclear residue are included in the public release GiBUU 2019. Our hybrid GiBUU+SMM calculations can be readily adopted for
the geometry of the ZDC in particular experiments.

\begin{acknowledgments}
  The support by the Frankfurt Center for Scientific Computing is gratefully acknowledged.
  We thank Marcus Bleicher, Alexander Botvina, Leonid Frankfurt, Kai Gallmeister, and Ulrich Mosel for
  illuminating discussions. 
  We thank A. Angerami for discussion of the experimental feasibility of the studies proposed
  in this paper and Yu.~Dokshitzer for discussion of various mechanisms of the final state interactions. 
  The authors are especially grateful to Ulrich Mosel who read the manuscript before publication and suggested substantial modifications
  of the abstract, introduction and conclusions.
  AL is grateful to the Department of Physics at the Pennsylvania State University, where this work was started,
  for the hospitality and financial support. AL also acknowledges partial financial support of HIC for FAIR within the framework of
  the Hessian LOEWE program. The research of MS was supported by the U.S. Department of Energy, Office of Science, Office of Nuclear Physics,
  under Award No. DE-FG02-93ER40771.
\end{acknowledgments}

\newpage

\appendix

\section{$pA$ collisions}
\label{results_pA}

The purpose of this Appendix is to test our model set-up for the hadron propagation and statistical decay of the residual nuclei
by comparison with experimental data on neutron production in proton-nucleus interactions at the beam energy about 1 GeV.
In this intermediate energy range, the effects of hadron formation are unimportant. The proton-nucleus reaction proceeds
as a cascade of elastic and inelastic (mainly, $\Delta$-resonance production and decay) processes and pion production
and reabsorption. The late stage of the reaction, when the fast cascade particles have left the nucleus and
the slow deexcitation of the nuclear residue starts -- should be, however, similar to the reactions at high energies.
We will consider the cases of the heavy targets, i.e. gold and lead. The proton was initialized at the distance of 0.5 fm
from nuclear surface at the impact parameters $b=0.5,1.0,\ldots,7$ fm with 2000 events for every impact parameter.
The calculated observables are weighted with impact parameter.

\begin{figure}
  \includegraphics[scale = 0.50]{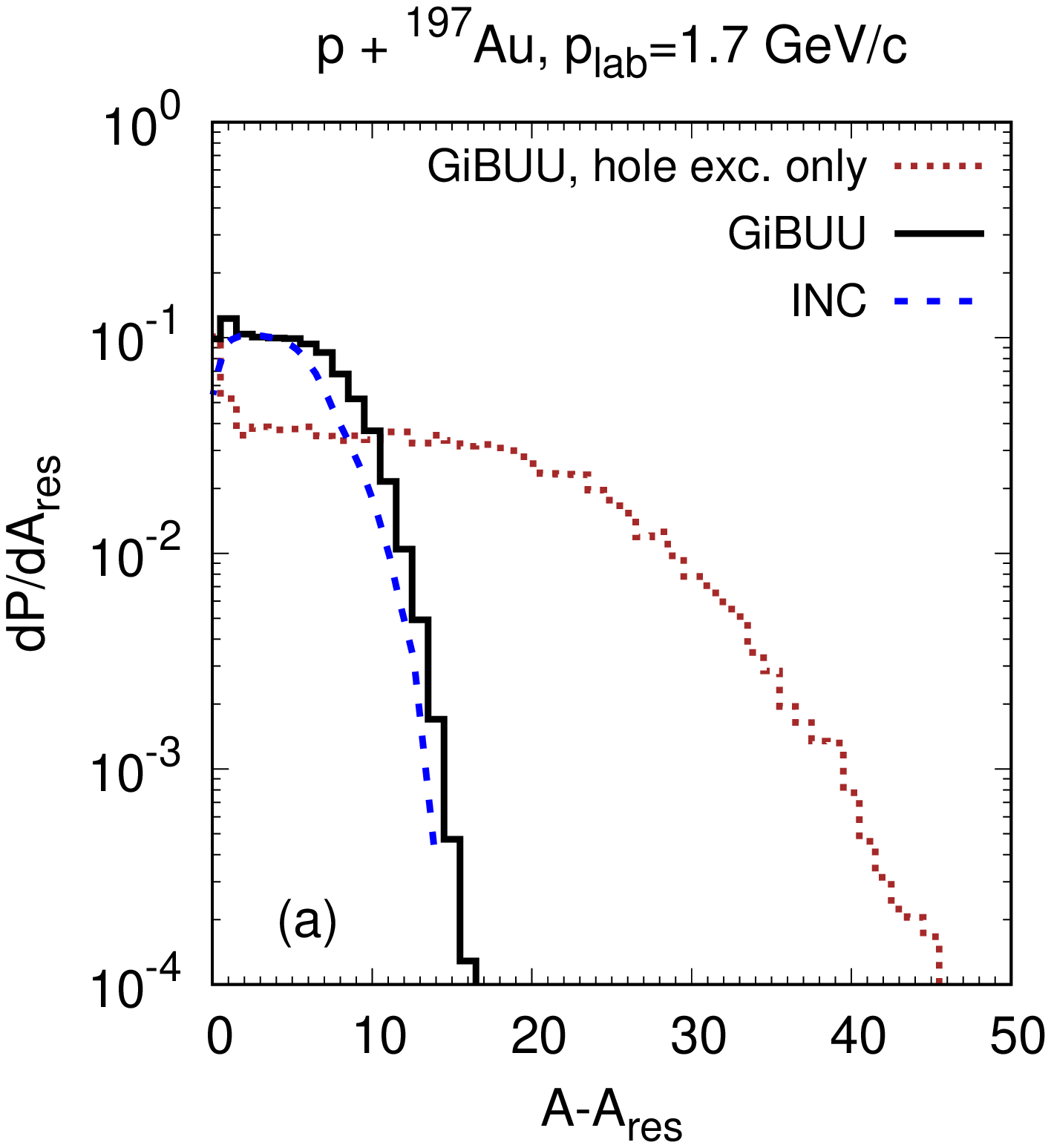}
  \includegraphics[scale = 0.50]{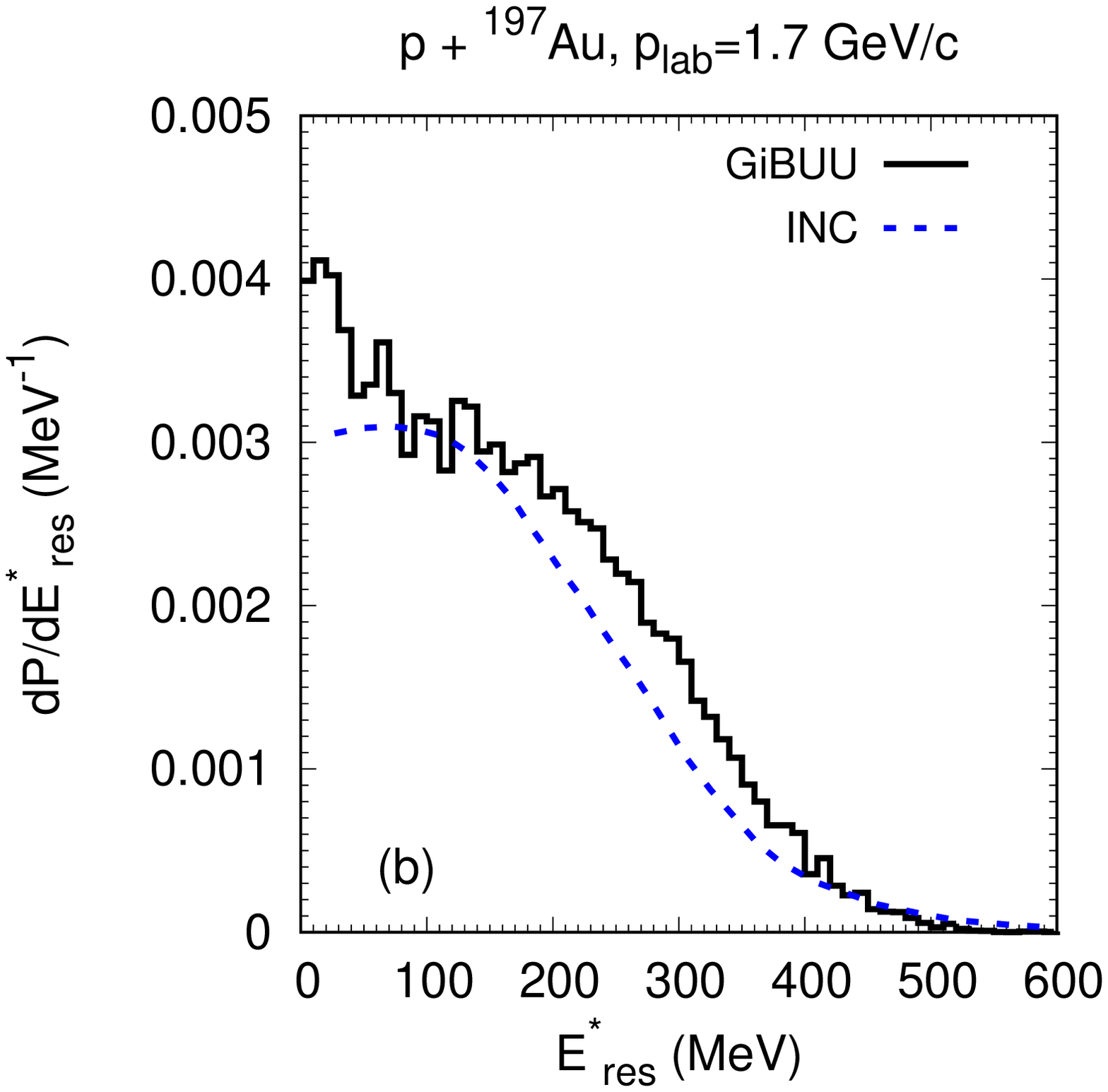}
  \caption{\label{fig:pAu} Probability distribution of the mass number loss (a) and of the excitation energy (b)
    of the residual nucleus produced in $p+^{197}$Au collisions at $E_{\rm lab}=1$ GeV. Dotted line -- GiBUU calculation   
    with hole excitations only (i.e. according to Eq.(\ref{A_res})).
    Solid line -- GiBUU calculation taking into account correction due to particle excitations.
    Average values for GiBUU calculations: $\langle A_{\rm res} \rangle =192~(185)$ with (without) particle excitations,
    $\langle E^*_{\rm res} \rangle = 161$ MeV.
    Dashed line -- INC model results from ref. \cite{Barashenkov:1974qj} in arbitrary units.}
\end{figure}  
It is instructive, first, to test our model setup by comparison with previous calculations. We selected the
INC calculations of ref. \cite{Barashenkov:1974qj}.\footnote{See line ``b'' in Figs. 12 and 13 of ref. \cite{Barashenkov:1974qj},
  corresponding to events without fission. Note that the fission cross section is $\ltsim10\%$ of the inelastic cross section
  for $p+$Au collisions at $E_{\rm lab}=1$ GeV.}
Fig.~\ref{fig:pAu} shows the mass number loss-- and excitation energy distributions of the nuclear residue for $p+^{197}$Au collisions
at the beam energy of 1 GeV. It is evident that the correction of the residue mass number due to the particle excitations is quite large.
Our corrected results are in a good agreement with INC results. 
Thus, we have included the correction for the particle excitations in all calculations discussed
in sec. \ref{results} and below in this Appendix.
However, as we checked, this correction has practically no effect, since the statistical evaporation of neutrons is mostly governed
by the excitation energy of the nucleus and not by its mass and charge numbers. 

\begin{figure}
  \includegraphics[scale = 0.49]{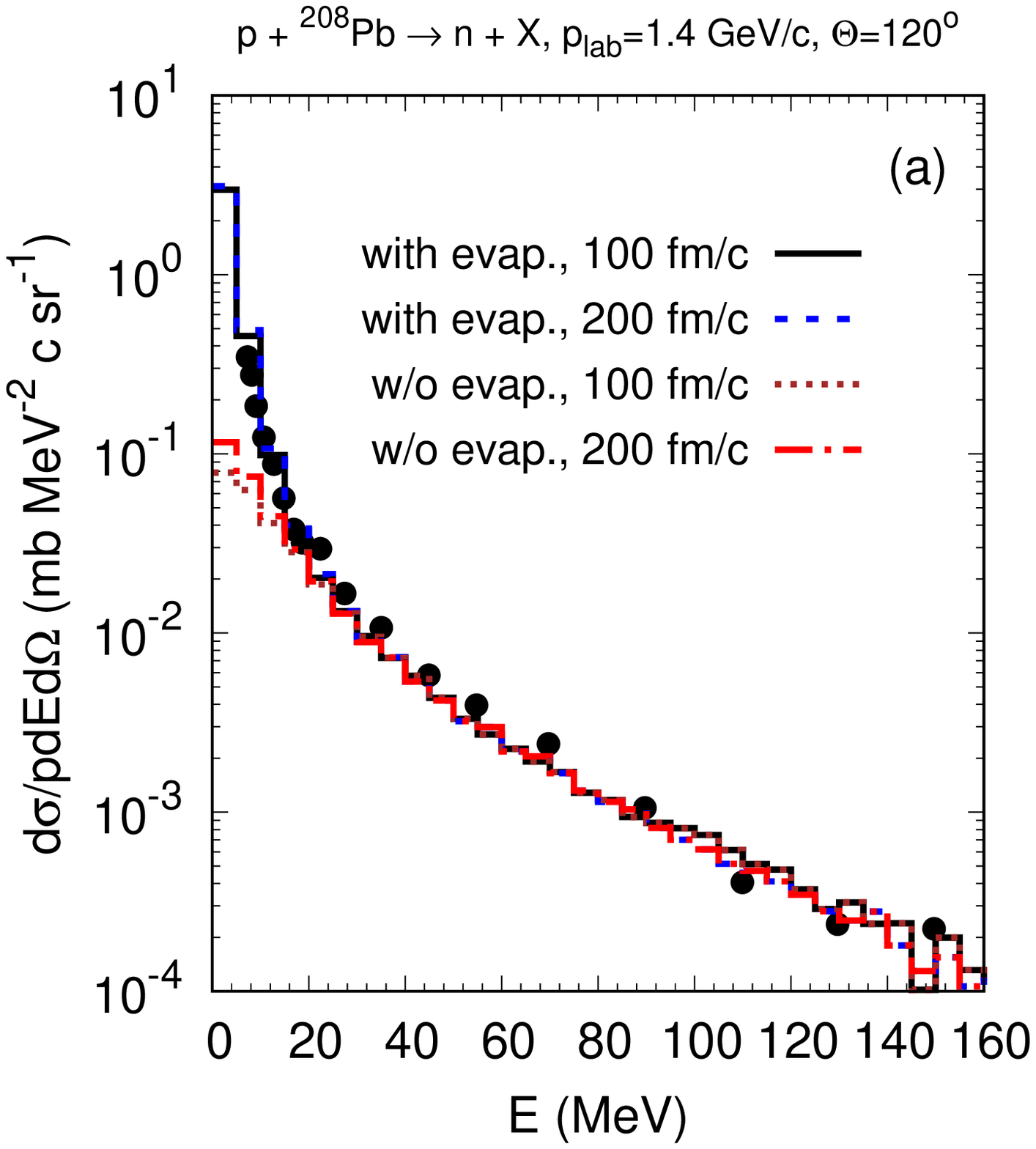}
  \includegraphics[scale = 0.49]{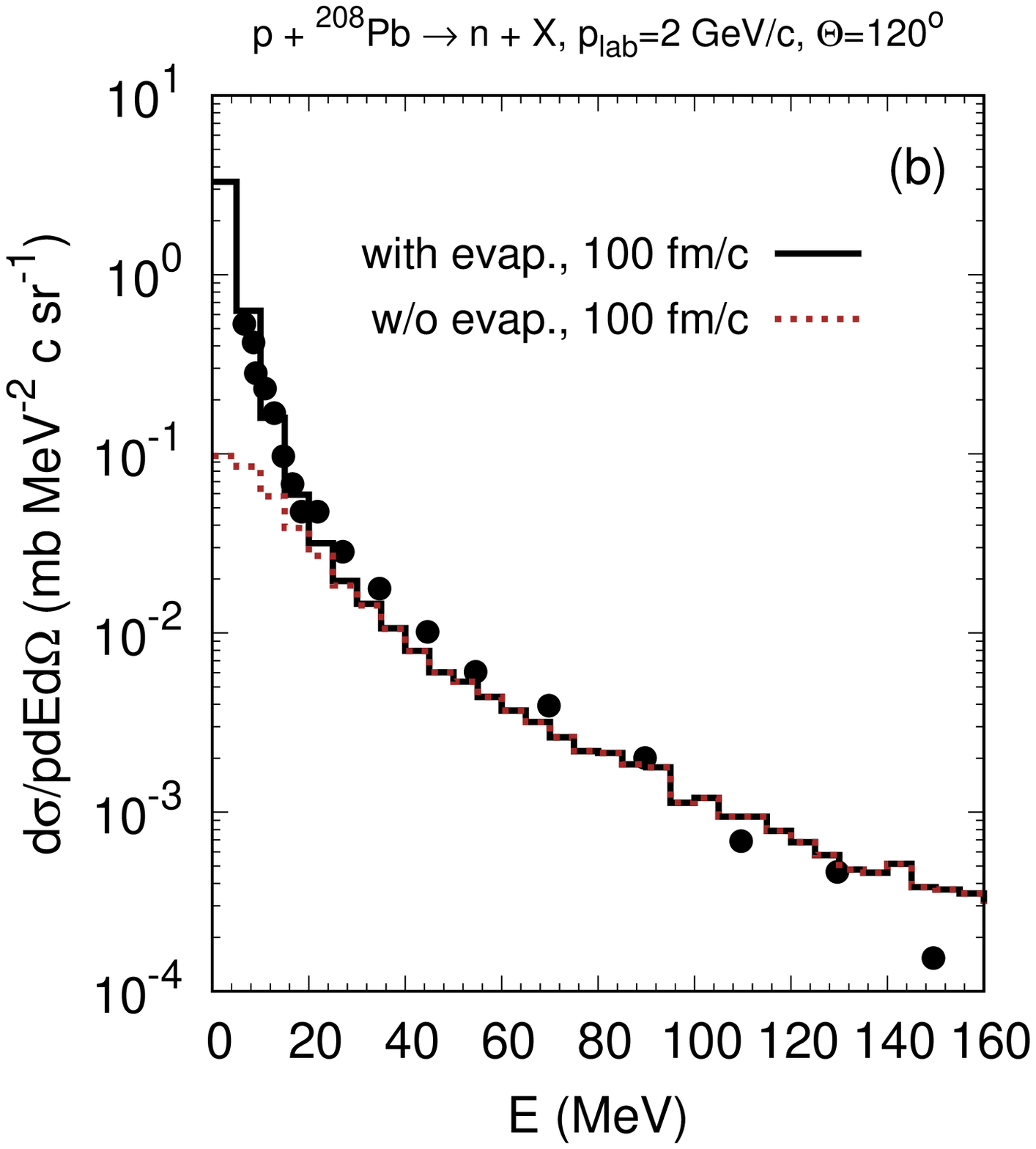}
  \includegraphics[scale = 0.49]{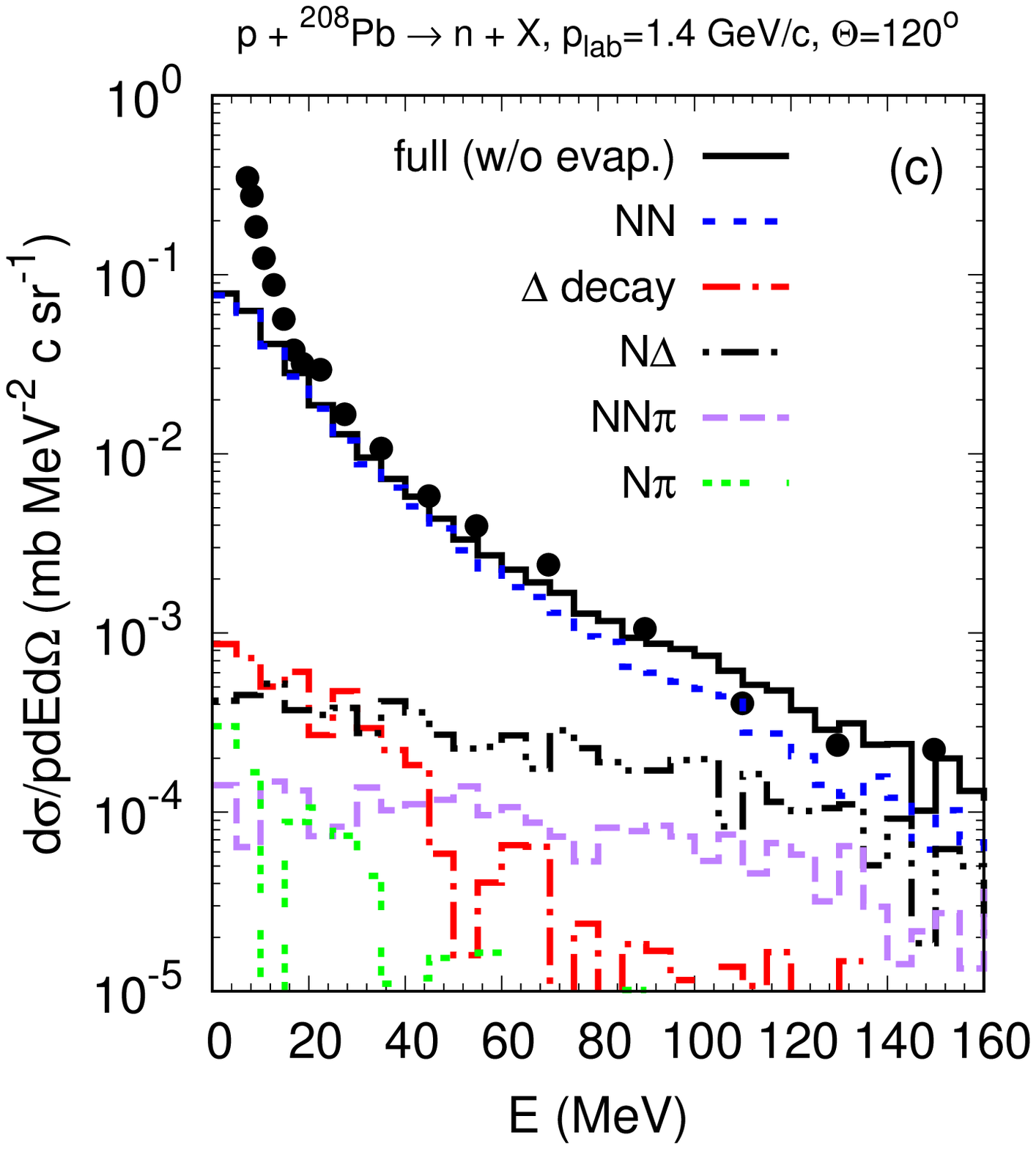}
  \includegraphics[scale = 0.49]{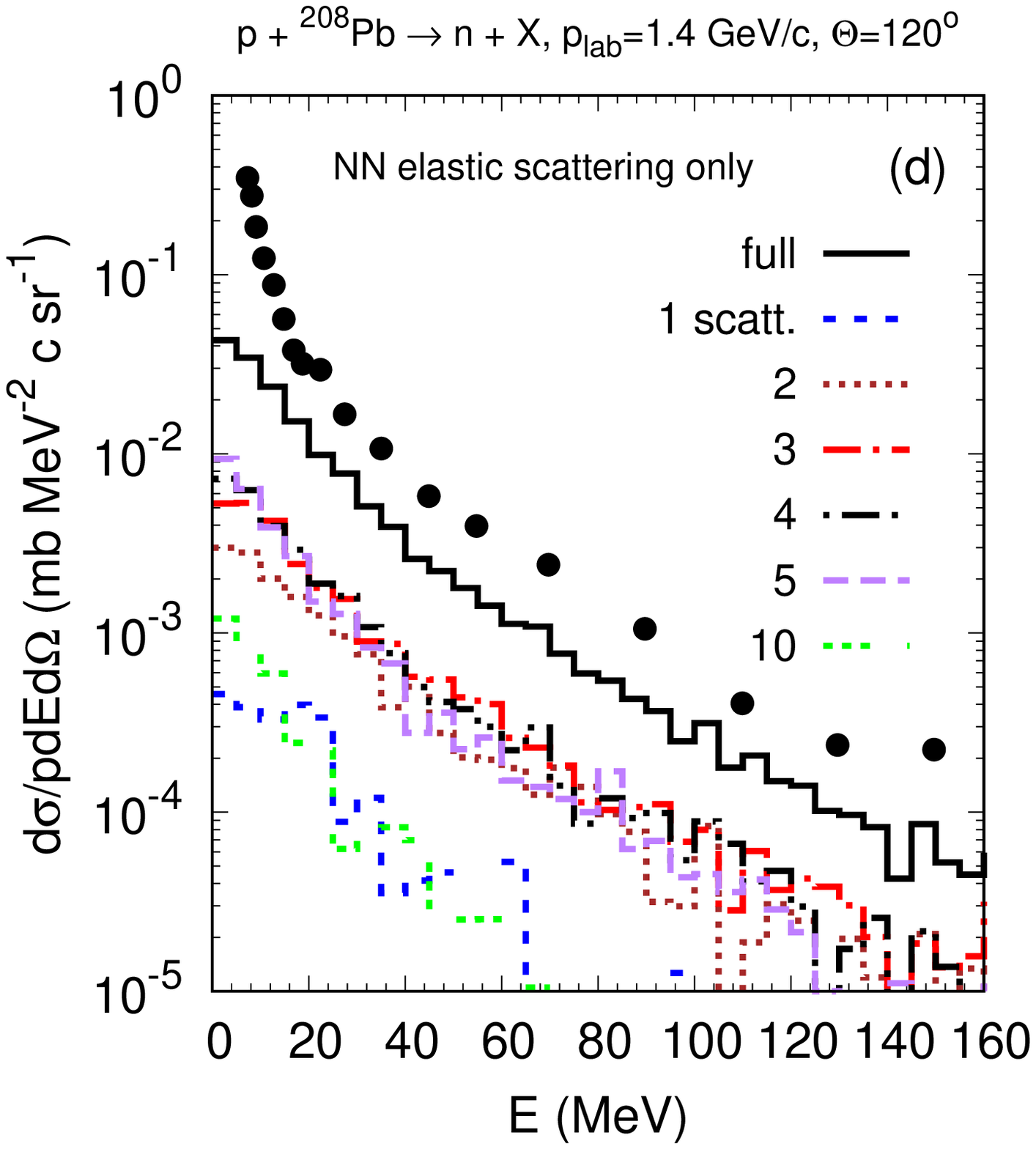}  
  \caption{\label{fig:pPb}
    Energy differential cross section of neutron production at $\Theta=120\degree$
    from $p+^{208}$Pb collisions at 1.4 GeV/c (a),(c),(d) and 2 GeV/c (b).
    Different histograms on the panels (a),(b) show the GiBUU calculations with and without 
    neutron evaporation with the maximum time 100 and 200 fm/c as indicated.
    Solid (black) histogram on the panel (c) displays the full GiBUU calculation stopped
    at 100 fm/c without evaporation. Other histograms show the contributions of various neutron production channels as indicated.
    Solid (black) histogram on the panel (d) displays the GiBUU calculation with $NN$ elastic scattering only.
    Other histograms show the contributions of the neutrons produced in a chain of
    the indicated number of elastic scatterings.
    Experimental data are from \cite{Bayukov:1983wi}.}
\end{figure}
Fig.~\ref{fig:pPb}a,b show the neutron energy spectra at the laboratory angle of $120\degree$ for
the proton-lead collisions at 1.4 and 2 GeV/c.
The GiBUU calculation was run until 100 and 200 fm/c. Then the target residues were 
determined and used as an input to the SMM calculation which was run 10 times for every GiBUU event.
Evaporated neutrons were included in the total spectrum of emitted neutrons. As we see the evaporation substantially
increases the yield of low-energy ($E \ltsim 30$ MeV) neutrons and explains the evident two-slope structure of the
experimental neutron spectra. This is not  surprising since the same data were explained earlier with similar approaches
in ref. \cite{Strikman:1998cc} and recently in ref. \cite{Galoyan:2015wqv}.

Fig.~\ref{fig:pPb}c shows the neutron energy spectrum from 1.4 GeV/c $p+^{208}$Pb collisions with partial
contributions of the $NN$ 
collisions, $N\Delta$ collisions, $NN\pi$ (pion absorption by nucleon pair)
and $N\pi$ collisions (non-resonant channels), and $\Delta$ resonance decays. 
The partial contributions are selected according to the last process in which the neutron was finally produced.
It means that, for example, a neutron may appear in the final state of an inelastic scattering or a resonance decay
and afterwards scatter elastically. Then this neutron belongs to the $NN$ partial contribution. We estimated that such neutrons constitute 
$\sim15\%$ of the $NN$ component. Thus, the main contribution to the spectrum of preequilibrium
neutrons is given by $NN$ elastic collisions. The $N\Delta$ and $NN\pi$ collisions make comparable in the magnitude contribution
above 100 MeV neutron energy. Other neutron production channels are less important.

By performing an additional calculation
keeping only elastic scatterings we have established that the main contribution to the neutron spectrum comes from multiple scattering
(from two- to five-step) processes as displayed in Fig.~\ref{fig:pPb}d. This suggests that the residual nuclear system
is approaching thermally equilibrated state while dynamically emitting neutrons. This is a type of the situation where it is natural 
to apply statistical approaches.

\bibliography{gammaAn}

\begin{thebibliography}{51}%
\makeatletter
\providecommand \@ifxundefined [1]{%
 \@ifx{#1\undefined}
}%
\providecommand \@ifnum [1]{%
 \ifnum #1\expandafter \@firstoftwo
 \else \expandafter \@secondoftwo
 \fi
}%
\providecommand \@ifx [1]{%
 \ifx #1\expandafter \@firstoftwo
 \else \expandafter \@secondoftwo
 \fi
}%
\providecommand \natexlab [1]{#1}%
\providecommand \enquote  [1]{``#1''}%
\providecommand \bibnamefont  [1]{#1}%
\providecommand \bibfnamefont [1]{#1}%
\providecommand \citenamefont [1]{#1}%
\providecommand \href@noop [0]{\@secondoftwo}%
\providecommand \href [0]{\begingroup \@sanitize@url \@href}%
\providecommand \@href[1]{\@@startlink{#1}\@@href}%
\providecommand \@@href[1]{\endgroup#1\@@endlink}%
\providecommand \@sanitize@url [0]{\catcode `\\12\catcode `\$12\catcode
  `\&12\catcode `\#12\catcode `\^12\catcode `\_12\catcode `\%12\relax}%
\providecommand \@@startlink[1]{}%
\providecommand \@@endlink[0]{}%
\providecommand \url  [0]{\begingroup\@sanitize@url \@url }%
\providecommand \@url [1]{\endgroup\@href {#1}{\urlprefix }}%
\providecommand \urlprefix  [0]{URL }%
\providecommand \Eprint [0]{\href }%
\providecommand \doibase [0]{http://dx.doi.org/}%
\providecommand \selectlanguage [0]{\@gobble}%
\providecommand \bibinfo  [0]{\@secondoftwo}%
\providecommand \bibfield  [0]{\@secondoftwo}%
\providecommand \translation [1]{[#1]}%
\providecommand \BibitemOpen [0]{}%
\providecommand \bibitemStop [0]{}%
\providecommand \bibitemNoStop [0]{.\EOS\space}%
\providecommand \EOS [0]{\spacefactor3000\relax}%
\providecommand \BibitemShut  [1]{\csname bibitem#1\endcsname}%
\let\auto@bib@innerbib\@empty
%</preamble>
\bibitem [{\citenamefont {Accardi}\ \emph {et~al.}(2016)\citenamefont {Accardi}
  \emph {et~al.}}]{Accardi:2012qut}%
  \BibitemOpen
  \bibfield  {author} {\bibinfo {author} {\bibfnamefont {A.}~\bibnamefont
  {Accardi}} \emph {et~al.},\ }\href {\doibase 10.1140/epja/i2016-16268-9}
  {\bibfield  {journal} {\bibinfo  {journal} {Eur. Phys. J.}\ }\textbf
  {\bibinfo {volume} {A52}},\ \bibinfo {pages} {268} (\bibinfo {year}
  {2016})},\ \Eprint {http://arxiv.org/abs/1212.1701} {arXiv:1212.1701
  [nucl-ex]} \BibitemShut {NoStop}%
%%CITATION = ARXIV:1212.1701;%%
\bibitem [{\citenamefont {Ashman}\ \emph {et~al.}(1991)\citenamefont {Ashman}
  \emph {et~al.}}]{Ashman:1991cx}%
  \BibitemOpen
  \bibfield  {author} {\bibinfo {author} {\bibfnamefont {J.}~\bibnamefont
  {Ashman}} \emph {et~al.} (\bibinfo {collaboration} {European Muon}),\ }\href
  {\doibase 10.1007/BF01412322} {\bibfield  {journal} {\bibinfo  {journal} {Z.
  Phys.}\ }\textbf {\bibinfo {volume} {C52}},\ \bibinfo {pages} {1} (\bibinfo
  {year} {1991})}\BibitemShut {NoStop}%
%%CITATION = ZEPYA,C52,1;%%
\bibitem [{\citenamefont {Airapetian}\ \emph {et~al.}(2001)\citenamefont
  {Airapetian} \emph {et~al.}}]{Airapetian:2000ks}%
  \BibitemOpen
  \bibfield  {author} {\bibinfo {author} {\bibfnamefont {A.}~\bibnamefont
  {Airapetian}} \emph {et~al.} (\bibinfo {collaboration} {HERMES}),\ }\href
  {\doibase 10.1007/s100520100697} {\bibfield  {journal} {\bibinfo  {journal}
  {Eur. Phys. J.}\ }\textbf {\bibinfo {volume} {C20}},\ \bibinfo {pages} {479}
  (\bibinfo {year} {2001})},\ \Eprint {http://arxiv.org/abs/hep-ex/0012049}
  {arXiv:hep-ex/0012049 [hep-ex]} \BibitemShut {NoStop}%
%%CITATION = HEP-EX/0012049;%%
\bibitem [{\citenamefont {Accardi}\ \emph {et~al.}(2003)\citenamefont
  {Accardi}, \citenamefont {Muccifora},\ and\ \citenamefont
  {Pirner}}]{Accardi:2002tv}%
  \BibitemOpen
  \bibfield  {author} {\bibinfo {author} {\bibfnamefont {A.}~\bibnamefont
  {Accardi}}, \bibinfo {author} {\bibfnamefont {V.}~\bibnamefont {Muccifora}},
  \ and\ \bibinfo {author} {\bibfnamefont {H.-J.}\ \bibnamefont {Pirner}},\
  }\href {\doibase 10.1016/S0375-9474(03)00670-5} {\bibfield  {journal}
  {\bibinfo  {journal} {Nucl. Phys.}\ }\textbf {\bibinfo {volume} {A720}},\
  \bibinfo {pages} {131} (\bibinfo {year} {2003})},\ \Eprint
  {http://arxiv.org/abs/nucl-th/0211011} {arXiv:nucl-th/0211011 [nucl-th]}
  \BibitemShut {NoStop}%
%%CITATION = NUCL-TH/0211011;%%
\bibitem [{\citenamefont {Akopov}\ \emph {et~al.}(2002)\citenamefont {Akopov},
  \citenamefont {Elbakian},\ and\ \citenamefont {Grigoryan}}]{Akopov:2002nc}%
  \BibitemOpen
  \bibfield  {author} {\bibinfo {author} {\bibfnamefont {N.~Z.}\ \bibnamefont
  {Akopov}}, \bibinfo {author} {\bibfnamefont {G.~M.}\ \bibnamefont
  {Elbakian}}, \ and\ \bibinfo {author} {\bibfnamefont {L.~A.}\ \bibnamefont
  {Grigoryan}},\ }\href@noop {} {\  (\bibinfo {year} {2002})},\ \Eprint
  {http://arxiv.org/abs/hep-ph/0205123} {arXiv:hep-ph/0205123 [hep-ph]}
  \BibitemShut {NoStop}%
%%CITATION = HEP-PH/0205123;%%
\bibitem [{\citenamefont {Gallmeister}\ and\ \citenamefont
  {Falter}(2005)}]{Gallmeister:2005ad}%
  \BibitemOpen
  \bibfield  {author} {\bibinfo {author} {\bibfnamefont {K.}~\bibnamefont
  {Gallmeister}}\ and\ \bibinfo {author} {\bibfnamefont {T.}~\bibnamefont
  {Falter}},\ }\href {\doibase 10.1016/j.physletb.2005.08.135} {\bibfield
  {journal} {\bibinfo  {journal} {Phys. Lett.}\ }\textbf {\bibinfo {volume}
  {B630}},\ \bibinfo {pages} {40} (\bibinfo {year} {2005})},\ \Eprint
  {http://arxiv.org/abs/nucl-th/0502015} {arXiv:nucl-th/0502015 [nucl-th]}
  \BibitemShut {NoStop}%
%%CITATION = NUCL-TH/0502015;%%
\bibitem [{\citenamefont {Gallmeister}\ and\ \citenamefont
  {Mosel}(2008)}]{Gallmeister:2007an}%
  \BibitemOpen
  \bibfield  {author} {\bibinfo {author} {\bibfnamefont {K.}~\bibnamefont
  {Gallmeister}}\ and\ \bibinfo {author} {\bibfnamefont {U.}~\bibnamefont
  {Mosel}},\ }\href {\doibase 10.1016/j.nuclphysa.2007.12.009} {\bibfield
  {journal} {\bibinfo  {journal} {Nucl. Phys.}\ }\textbf {\bibinfo {volume}
  {A801}},\ \bibinfo {pages} {68} (\bibinfo {year} {2008})},\ \Eprint
  {http://arxiv.org/abs/nucl-th/0701064} {arXiv:nucl-th/0701064 [nucl-th]}
  \BibitemShut {NoStop}%
%%CITATION = NUCL-TH/0701064;%%
\bibitem [{\citenamefont {Kopeliovich}\ \emph {et~al.}(2004)\citenamefont
  {Kopeliovich}, \citenamefont {Nemchik}, \citenamefont {Predazzi},\ and\
  \citenamefont {Hayashigaki}}]{Kopeliovich:2003py}%
  \BibitemOpen
  \bibfield  {author} {\bibinfo {author} {\bibfnamefont {B.~Z.}\ \bibnamefont
  {Kopeliovich}}, \bibinfo {author} {\bibfnamefont {J.}~\bibnamefont
  {Nemchik}}, \bibinfo {author} {\bibfnamefont {E.}~\bibnamefont {Predazzi}}, \
  and\ \bibinfo {author} {\bibfnamefont {A.}~\bibnamefont {Hayashigaki}},\
  }\bibfield  {booktitle} {\emph {\bibinfo {booktitle} {{EURESCO Conference on
  Hadron Structure Viewed with Electromagnetic Probes Santorini, Greece,
  October 7-12, 2003}}},\ }\href {\doibase 10.1016/j.nuclphysa.2004.04.110}
  {\bibfield  {journal} {\bibinfo  {journal} {Nucl. Phys.}\ }\textbf {\bibinfo
  {volume} {A740}},\ \bibinfo {pages} {211} (\bibinfo {year} {2004})},\ \Eprint
  {http://arxiv.org/abs/hep-ph/0311220} {arXiv:hep-ph/0311220 [hep-ph]}
  \BibitemShut {NoStop}%
%%CITATION = HEP-PH/0311220;%%
\bibitem [{\citenamefont {Wang}\ and\ \citenamefont
  {Wang}(2002)}]{Wang:2002ri}%
  \BibitemOpen
  \bibfield  {author} {\bibinfo {author} {\bibfnamefont {E.}~\bibnamefont
  {Wang}}\ and\ \bibinfo {author} {\bibfnamefont {X.-N.}\ \bibnamefont
  {Wang}},\ }\href {\doibase 10.1103/PhysRevLett.89.162301} {\bibfield
  {journal} {\bibinfo  {journal} {Phys. Rev. Lett.}\ }\textbf {\bibinfo
  {volume} {89}},\ \bibinfo {pages} {162301} (\bibinfo {year} {2002})},\
  \Eprint {http://arxiv.org/abs/hep-ph/0202105} {arXiv:hep-ph/0202105 [hep-ph]}
  \BibitemShut {NoStop}%
%%CITATION = HEP-PH/0202105;%%
\bibitem [{\citenamefont {Arleo}(2003)}]{Arleo:2003jz}%
  \BibitemOpen
  \bibfield  {author} {\bibinfo {author} {\bibfnamefont {F.}~\bibnamefont
  {Arleo}},\ }\href {\doibase 10.1140/epjc/s2003-01289-x} {\bibfield  {journal}
  {\bibinfo  {journal} {Eur. Phys. J.}\ }\textbf {\bibinfo {volume} {C30}},\
  \bibinfo {pages} {213} (\bibinfo {year} {2003})},\ \Eprint
  {http://arxiv.org/abs/hep-ph/0306235} {arXiv:hep-ph/0306235 [hep-ph]}
  \BibitemShut {NoStop}%
%%CITATION = HEP-PH/0306235;%%
\bibitem [{\citenamefont {Falter}\ \emph {et~al.}(2004)\citenamefont {Falter},
  \citenamefont {Cassing}, \citenamefont {Gallmeister},\ and\ \citenamefont
  {Mosel}}]{Falter:2004uc}%
  \BibitemOpen
  \bibfield  {author} {\bibinfo {author} {\bibfnamefont {T.}~\bibnamefont
  {Falter}}, \bibinfo {author} {\bibfnamefont {W.}~\bibnamefont {Cassing}},
  \bibinfo {author} {\bibfnamefont {K.}~\bibnamefont {Gallmeister}}, \ and\
  \bibinfo {author} {\bibfnamefont {U.}~\bibnamefont {Mosel}},\ }\href
  {\doibase 10.1103/PhysRevC.70.054609} {\bibfield  {journal} {\bibinfo
  {journal} {Phys. Rev.}\ }\textbf {\bibinfo {volume} {C70}},\ \bibinfo {pages}
  {054609} (\bibinfo {year} {2004})},\ \Eprint
  {http://arxiv.org/abs/nucl-th/0406023} {arXiv:nucl-th/0406023 [nucl-th]}
  \BibitemShut {NoStop}%
%%CITATION = NUCL-TH/0406023;%%
\bibitem [{\citenamefont {Adams}\ \emph
  {et~al.}(1995{\natexlab{a}})\citenamefont {Adams} \emph
  {et~al.}}]{Adams:1995nu}%
  \BibitemOpen
  \bibfield  {author} {\bibinfo {author} {\bibfnamefont {M.~R.}\ \bibnamefont
  {Adams}} \emph {et~al.} (\bibinfo {collaboration} {E665}),\ }\href {\doibase
  10.1103/PhysRevLett.74.5198, 10.1103/PhysRevLett.80.2020} {\bibfield
  {journal} {\bibinfo  {journal} {Phys. Rev. Lett.}\ }\textbf {\bibinfo
  {volume} {74}},\ \bibinfo {pages} {5198} (\bibinfo {year}
  {1995}{\natexlab{a}})},\ \bibinfo {note} {[Erratum: Phys. Rev.
  Lett.80,2020(1998)]}\BibitemShut {NoStop}%
%%CITATION = PRLTA,74,5198;%%
\bibitem [{\citenamefont {Strikman}\ \emph {et~al.}(1999)\citenamefont
  {Strikman}, \citenamefont {Tverskoy},\ and\ \citenamefont
  {Zhalov}}]{Strikman:1998cc}%
  \BibitemOpen
  \bibfield  {author} {\bibinfo {author} {\bibfnamefont {M.}~\bibnamefont
  {Strikman}}, \bibinfo {author} {\bibfnamefont {M.~G.}\ \bibnamefont
  {Tverskoy}}, \ and\ \bibinfo {author} {\bibfnamefont {M.~B.}\ \bibnamefont
  {Zhalov}},\ }\href {\doibase 10.1016/S0370-2693(99)00627-9} {\bibfield
  {journal} {\bibinfo  {journal} {Phys. Lett.}\ }\textbf {\bibinfo {volume}
  {B459}},\ \bibinfo {pages} {37} (\bibinfo {year} {1999})},\ \Eprint
  {http://arxiv.org/abs/nucl-th/9806099} {arXiv:nucl-th/9806099 [nucl-th]}
  \BibitemShut {NoStop}%
%%CITATION = NUCL-TH/9806099;%%
\bibitem [{\citenamefont {Adams}\ \emph
  {et~al.}(1995{\natexlab{b}})\citenamefont {Adams} \emph
  {et~al.}}]{Adams:1994ri}%
  \BibitemOpen
  \bibfield  {author} {\bibinfo {author} {\bibfnamefont {M.~R.}\ \bibnamefont
  {Adams}} \emph {et~al.} (\bibinfo {collaboration} {E665}),\ }\href {\doibase
  10.1007/BF01571879} {\bibfield  {journal} {\bibinfo  {journal} {Z. Phys.}\
  }\textbf {\bibinfo {volume} {C65}},\ \bibinfo {pages} {225} (\bibinfo {year}
  {1995}{\natexlab{b}})}\BibitemShut {NoStop}%
%%CITATION = ZEPYA,C65,225;%%
\bibitem [{\citenamefont {Ciofi~degli Atti}\ and\ \citenamefont
  {Kopeliovich}(2005)}]{CiofidegliAtti:2004pv}%
  \BibitemOpen
  \bibfield  {author} {\bibinfo {author} {\bibfnamefont {C.}~\bibnamefont
  {Ciofi~degli Atti}}\ and\ \bibinfo {author} {\bibfnamefont {B.~Z.}\
  \bibnamefont {Kopeliovich}},\ }\href {\doibase
  10.1016/j.physletb.2004.12.021} {\bibfield  {journal} {\bibinfo  {journal}
  {Phys. Lett.}\ }\textbf {\bibinfo {volume} {B606}},\ \bibinfo {pages} {281}
  (\bibinfo {year} {2005})},\ \Eprint {http://arxiv.org/abs/hep-ph/0409077}
  {arXiv:hep-ph/0409077 [hep-ph]} \BibitemShut {NoStop}%
%%CITATION = HEP-PH/0409077;%%
\bibitem [{\citenamefont {Angerami}(2018)}]{Angerami:2319206}%
  \BibitemOpen
  \bibfield  {author} {\bibinfo {author} {\bibfnamefont {A.}~\bibnamefont
  {Angerami}} (\bibinfo {collaboration} {ATLAS Collaboration}),\ }\href
  {http://cds.cern.ch/record/2319206} {\bibfield  {journal} {\bibinfo
  {journal} {Talk at QM 2018, ATL-PHYS-SLIDE-2018-270}\ } (\bibinfo {year}
  {2018})}\BibitemShut {NoStop}%
\bibitem [{\citenamefont {Buss}\ \emph {et~al.}(2012)\citenamefont {Buss},
  \citenamefont {Gaitanos}, \citenamefont {Gallmeister}, \citenamefont {van
  Hees}, \citenamefont {Kaskulov}, \citenamefont {Lalakulich}, \citenamefont
  {Larionov}, \citenamefont {Leitner}, \citenamefont {Weil},\ and\
  \citenamefont {Mosel}}]{Buss:2011mx}%
  \BibitemOpen
  \bibfield  {author} {\bibinfo {author} {\bibfnamefont {O.}~\bibnamefont
  {Buss}}, \bibinfo {author} {\bibfnamefont {T.}~\bibnamefont {Gaitanos}},
  \bibinfo {author} {\bibfnamefont {K.}~\bibnamefont {Gallmeister}}, \bibinfo
  {author} {\bibfnamefont {H.}~\bibnamefont {van Hees}}, \bibinfo {author}
  {\bibfnamefont {M.}~\bibnamefont {Kaskulov}}, \bibinfo {author}
  {\bibfnamefont {O.}~\bibnamefont {Lalakulich}}, \bibinfo {author}
  {\bibfnamefont {A.~B.}\ \bibnamefont {Larionov}}, \bibinfo {author}
  {\bibfnamefont {T.}~\bibnamefont {Leitner}}, \bibinfo {author} {\bibfnamefont
  {J.}~\bibnamefont {Weil}}, \ and\ \bibinfo {author} {\bibfnamefont
  {U.}~\bibnamefont {Mosel}},\ }\href {\doibase 10.1016/j.physrep.2011.12.001}
  {\bibfield  {journal} {\bibinfo  {journal} {Phys. Rept.}\ }\textbf {\bibinfo
  {volume} {512}},\ \bibinfo {pages} {1} (\bibinfo {year} {2012})},\ \Eprint
  {http://arxiv.org/abs/1106.1344} {arXiv:1106.1344 [hep-ph]} \BibitemShut
  {NoStop}%
%%CITATION = ARXIV:1106.1344;%%
\bibitem [{\citenamefont {Botvina}\ \emph {et~al.}(1987)\citenamefont
  {Botvina}, \citenamefont {Iljinov}, \citenamefont {Mishustin}, \citenamefont
  {Bondorf}, \citenamefont {Donangelo},\ and\ \citenamefont
  {Sneppen}}]{Botvina:1987jp}%
  \BibitemOpen
  \bibfield  {author} {\bibinfo {author} {\bibfnamefont {A.~S.}\ \bibnamefont
  {Botvina}}, \bibinfo {author} {\bibfnamefont {A.~S.}\ \bibnamefont
  {Iljinov}}, \bibinfo {author} {\bibfnamefont {I.~N.}\ \bibnamefont
  {Mishustin}}, \bibinfo {author} {\bibfnamefont {J.~P.}\ \bibnamefont
  {Bondorf}}, \bibinfo {author} {\bibfnamefont {R.}~\bibnamefont {Donangelo}},
  \ and\ \bibinfo {author} {\bibfnamefont {K.}~\bibnamefont {Sneppen}},\ }\href
  {\doibase 10.1016/0375-9474(87)90232-6} {\bibfield  {journal} {\bibinfo
  {journal} {Nucl. Phys.}\ }\textbf {\bibinfo {volume} {A475}},\ \bibinfo
  {pages} {663} (\bibinfo {year} {1987})}\BibitemShut {NoStop}%
%%CITATION = NUPHA,A475,663;%%
\bibitem [{\citenamefont {Bondorf}\ \emph {et~al.}(1995)\citenamefont
  {Bondorf}, \citenamefont {Botvina}, \citenamefont {Ilinov}, \citenamefont
  {Mishustin},\ and\ \citenamefont {Sneppen}}]{Bondorf:1995ua}%
  \BibitemOpen
  \bibfield  {author} {\bibinfo {author} {\bibfnamefont {J.~P.}\ \bibnamefont
  {Bondorf}}, \bibinfo {author} {\bibfnamefont {A.~S.}\ \bibnamefont
  {Botvina}}, \bibinfo {author} {\bibfnamefont {A.~S.}\ \bibnamefont {Ilinov}},
  \bibinfo {author} {\bibfnamefont {I.~N.}\ \bibnamefont {Mishustin}}, \ and\
  \bibinfo {author} {\bibfnamefont {K.}~\bibnamefont {Sneppen}},\ }\href
  {\doibase 10.1016/0370-1573(94)00097-M} {\bibfield  {journal} {\bibinfo
  {journal} {Phys. Rept.}\ }\textbf {\bibinfo {volume} {257}},\ \bibinfo
  {pages} {133} (\bibinfo {year} {1995})}\BibitemShut {NoStop}%
%%CITATION = PRPLC,257,133;%%
\bibitem [{\citenamefont {Barashenkov}\ \emph {et~al.}(1974)\citenamefont
  {Barashenkov}, \citenamefont {Gereghi}, \citenamefont {Iljinov},\ and\
  \citenamefont {Toneev}}]{Barashenkov:1974qj}%
  \BibitemOpen
  \bibfield  {author} {\bibinfo {author} {\bibfnamefont {V.~S.}\ \bibnamefont
  {Barashenkov}}, \bibinfo {author} {\bibfnamefont {F.~G.}\ \bibnamefont
  {Gereghi}}, \bibinfo {author} {\bibfnamefont {A.~S.}\ \bibnamefont
  {Iljinov}}, \ and\ \bibinfo {author} {\bibfnamefont {V.~D.}\ \bibnamefont
  {Toneev}},\ }\href {\doibase 10.1016/0375-9474(74)90594-6} {\bibfield
  {journal} {\bibinfo  {journal} {Nucl. Phys.}\ }\textbf {\bibinfo {volume}
  {A222}},\ \bibinfo {pages} {204} (\bibinfo {year} {1974})}\BibitemShut
  {NoStop}%
%%CITATION = NUPHA,A222,204;%%
\bibitem [{\citenamefont {Botvina}\ \emph {et~al.}(1995)\citenamefont
  {Botvina}, \citenamefont {Larionov},\ and\ \citenamefont
  {Mishustin}}]{Botvina:1995}%
  \BibitemOpen
  \bibfield  {author} {\bibinfo {author} {\bibfnamefont {A.~S.}\ \bibnamefont
  {Botvina}}, \bibinfo {author} {\bibfnamefont {A.~B.}\ \bibnamefont
  {Larionov}}, \ and\ \bibinfo {author} {\bibfnamefont {I.~N.}\ \bibnamefont
  {Mishustin}},\ }\href@noop {} {\bibfield  {journal} {\bibinfo  {journal}
  {Phys. At. Nucl.}\ }\textbf {\bibinfo {volume} {58}},\ \bibinfo {pages}
  {1703} (\bibinfo {year} {1995})}\BibitemShut {NoStop}%
\bibitem [{\citenamefont {Sjostrand}\ \emph {et~al.}(2006)\citenamefont
  {Sjostrand}, \citenamefont {Mrenna},\ and\ \citenamefont
  {Skands}}]{Sjostrand:2006za}%
  \BibitemOpen
  \bibfield  {author} {\bibinfo {author} {\bibfnamefont {T.}~\bibnamefont
  {Sjostrand}}, \bibinfo {author} {\bibfnamefont {S.}~\bibnamefont {Mrenna}}, \
  and\ \bibinfo {author} {\bibfnamefont {P.~Z.}\ \bibnamefont {Skands}},\
  }\href {\doibase 10.1088/1126-6708/2006/05/026} {\bibfield  {journal}
  {\bibinfo  {journal} {JHEP}\ }\textbf {\bibinfo {volume} {05}},\ \bibinfo
  {pages} {026} (\bibinfo {year} {2006})},\ \Eprint
  {http://arxiv.org/abs/hep-ph/0603175} {arXiv:hep-ph/0603175 [hep-ph]}
  \BibitemShut {NoStop}%
%%CITATION = HEP-PH/0603175;%%
\bibitem [{\citenamefont {Christy}\ and\ \citenamefont
  {Bosted}(2010)}]{Christy:2007ve}%
  \BibitemOpen
  \bibfield  {author} {\bibinfo {author} {\bibfnamefont {M.~E.}\ \bibnamefont
  {Christy}}\ and\ \bibinfo {author} {\bibfnamefont {P.~E.}\ \bibnamefont
  {Bosted}},\ }\href {\doibase 10.1103/PhysRevC.81.055213} {\bibfield
  {journal} {\bibinfo  {journal} {Phys. Rev.}\ }\textbf {\bibinfo {volume}
  {C81}},\ \bibinfo {pages} {055213} (\bibinfo {year} {2010})},\ \Eprint
  {http://arxiv.org/abs/0712.3731} {arXiv:0712.3731 [hep-ph]} \BibitemShut
  {NoStop}%
%%CITATION = ARXIV:0712.3731;%%
\bibitem [{\citenamefont {Andersson}\ \emph {et~al.}(1983)\citenamefont
  {Andersson}, \citenamefont {Gustafson}, \citenamefont {Ingelman},\ and\
  \citenamefont {Sjostrand}}]{Andersson:1983ia}%
  \BibitemOpen
  \bibfield  {author} {\bibinfo {author} {\bibfnamefont {B.}~\bibnamefont
  {Andersson}}, \bibinfo {author} {\bibfnamefont {G.}~\bibnamefont
  {Gustafson}}, \bibinfo {author} {\bibfnamefont {G.}~\bibnamefont {Ingelman}},
  \ and\ \bibinfo {author} {\bibfnamefont {T.}~\bibnamefont {Sjostrand}},\
  }\href {\doibase 10.1016/0370-1573(83)90080-7} {\bibfield  {journal}
  {\bibinfo  {journal} {Phys. Rept.}\ }\textbf {\bibinfo {volume} {97}},\
  \bibinfo {pages} {31} (\bibinfo {year} {1983})}\BibitemShut {NoStop}%
%%CITATION = PRPLC,97,31;%%
\bibitem [{\citenamefont {Dutta}\ \emph {et~al.}(2013)\citenamefont {Dutta},
  \citenamefont {Hafidi},\ and\ \citenamefont {Strikman}}]{Dutta:2012ii}%
  \BibitemOpen
  \bibfield  {author} {\bibinfo {author} {\bibfnamefont {D.}~\bibnamefont
  {Dutta}}, \bibinfo {author} {\bibfnamefont {K.}~\bibnamefont {Hafidi}}, \
  and\ \bibinfo {author} {\bibfnamefont {M.}~\bibnamefont {Strikman}},\ }\href
  {\doibase 10.1016/j.ppnp.2012.11.001} {\bibfield  {journal} {\bibinfo
  {journal} {Prog.Part.Nucl.Phys.}\ }\textbf {\bibinfo {volume} {69}},\
  \bibinfo {pages} {1} (\bibinfo {year} {2013})}\BibitemShut {NoStop}%
%%CITATION = ARXIV:1211.2826;%%
\bibitem [{\citenamefont {Larson}\ \emph {et~al.}(2006)\citenamefont {Larson},
  \citenamefont {Miller},\ and\ \citenamefont {Strikman}}]{Larson:2006ge}%
  \BibitemOpen
  \bibfield  {author} {\bibinfo {author} {\bibfnamefont {A.}~\bibnamefont
  {Larson}}, \bibinfo {author} {\bibfnamefont {G.~A.}\ \bibnamefont {Miller}},
  \ and\ \bibinfo {author} {\bibfnamefont {M.}~\bibnamefont {Strikman}},\
  }\href {\doibase 10.1103/PhysRevC.74.018201} {\bibfield  {journal} {\bibinfo
  {journal} {Phys. Rev.}\ }\textbf {\bibinfo {volume} {C74}},\ \bibinfo {pages}
  {018201} (\bibinfo {year} {2006})}\BibitemShut {NoStop}%
\bibitem [{\citenamefont {Larionov}\ \emph {et~al.}(2016)\citenamefont
  {Larionov}, \citenamefont {Strikman},\ and\ \citenamefont
  {Bleicher}}]{Larionov:2016phv}%
  \BibitemOpen
  \bibfield  {author} {\bibinfo {author} {\bibfnamefont {A.~B.}\ \bibnamefont
  {Larionov}}, \bibinfo {author} {\bibfnamefont {M.}~\bibnamefont {Strikman}},
  \ and\ \bibinfo {author} {\bibfnamefont {M.}~\bibnamefont {Bleicher}},\
  }\href {\doibase 10.1103/PhysRevC.93.034618} {\bibfield  {journal} {\bibinfo
  {journal} {Phys. Rev.}\ }\textbf {\bibinfo {volume} {C93}},\ \bibinfo {pages}
  {034618} (\bibinfo {year} {2016})},\ \Eprint
  {http://arxiv.org/abs/1601.00189} {arXiv:1601.00189 [nucl-th]} \BibitemShut
  {NoStop}%
%%CITATION = ARXIV:1601.00189;%%
\bibitem [{\citenamefont {Farrar}\ \emph {et~al.}(1988)\citenamefont {Farrar},
  \citenamefont {Liu}, \citenamefont {Frankfurt},\ and\ \citenamefont
  {Strikman}}]{Farrar:1988me}%
  \BibitemOpen
  \bibfield  {author} {\bibinfo {author} {\bibfnamefont {G.}~\bibnamefont
  {Farrar}}, \bibinfo {author} {\bibfnamefont {H.}~\bibnamefont {Liu}},
  \bibinfo {author} {\bibfnamefont {L.}~\bibnamefont {Frankfurt}}, \ and\
  \bibinfo {author} {\bibfnamefont {M.}~\bibnamefont {Strikman}},\ }\href
  {\doibase 10.1103/PhysRevLett.61.686} {\bibfield  {journal} {\bibinfo
  {journal} {Phys. Rev. Lett.}\ }\textbf {\bibinfo {volume} {61}},\ \bibinfo
  {pages} {686} (\bibinfo {year} {1988})}\BibitemShut {NoStop}%
%%CITATION = PRLTA,61,686;%%
\bibitem [{\citenamefont {Lalazissis}\ \emph {et~al.}(1997)\citenamefont
  {Lalazissis}, \citenamefont {K\"onig},\ and\ \citenamefont
  {Ring}}]{Lalazissis:1996rd}%
  \BibitemOpen
  \bibfield  {author} {\bibinfo {author} {\bibfnamefont {G.~A.}\ \bibnamefont
  {Lalazissis}}, \bibinfo {author} {\bibfnamefont {J.}~\bibnamefont {K\"onig}},
  \ and\ \bibinfo {author} {\bibfnamefont {P.}~\bibnamefont {Ring}},\ }\href
  {\doibase 10.1103/PhysRevC.55.540} {\bibfield  {journal} {\bibinfo  {journal}
  {Phys. Rev.}\ }\textbf {\bibinfo {volume} {C55}},\ \bibinfo {pages} {540}
  (\bibinfo {year} {1997})}\BibitemShut {NoStop}%
%%CITATION = NUCL-TH/9607039;%%
\bibitem [{\citenamefont {Andersson}\ \emph {et~al.}(1993)\citenamefont
  {Andersson}, \citenamefont {Gustafson},\ and\ \citenamefont
  {Pi}}]{Andersson:1992iq}%
  \BibitemOpen
  \bibfield  {author} {\bibinfo {author} {\bibfnamefont {B.}~\bibnamefont
  {Andersson}}, \bibinfo {author} {\bibfnamefont {G.}~\bibnamefont
  {Gustafson}}, \ and\ \bibinfo {author} {\bibfnamefont {H.}~\bibnamefont
  {Pi}},\ }\href {\doibase 10.1007/BF01474343} {\bibfield  {journal} {\bibinfo
  {journal} {Z. Phys.}\ }\textbf {\bibinfo {volume} {C57}},\ \bibinfo {pages}
  {485} (\bibinfo {year} {1993})}\BibitemShut {NoStop}%
%%CITATION = ZEPYA,C57,485;%%
\bibitem [{\citenamefont {Larionov}\ \emph {et~al.}(2012)\citenamefont
  {Larionov}, \citenamefont {Gaitanos},\ and\ \citenamefont
  {Mosel}}]{Larionov:2011fs}%
  \BibitemOpen
  \bibfield  {author} {\bibinfo {author} {\bibfnamefont {A.~B.}\ \bibnamefont
  {Larionov}}, \bibinfo {author} {\bibfnamefont {T.}~\bibnamefont {Gaitanos}},
  \ and\ \bibinfo {author} {\bibfnamefont {U.}~\bibnamefont {Mosel}},\ }\href
  {\doibase 10.1103/PhysRevC.85.024614} {\bibfield  {journal} {\bibinfo
  {journal} {Phys. Rev.}\ }\textbf {\bibinfo {volume} {C85}},\ \bibinfo {pages}
  {024614} (\bibinfo {year} {2012})},\ \Eprint {http://arxiv.org/abs/1107.2326}
  {arXiv:1107.2326 [nucl-th]} \BibitemShut {NoStop}%
%%CITATION = ARXIV:1107.2326;%%
\bibitem [{\citenamefont {Weil}(2013)}]{WeilPhD}%
  \BibitemOpen
  \bibfield  {author} {\bibinfo {author} {\bibfnamefont {J.}~\bibnamefont
  {Weil}},\ }\emph {\bibinfo {title} {Vector Mesons in Medium in a Transport
  Approach}},\ \href@noop {} {Ph.D. thesis},\ \bibinfo  {school} {University of
  Giessen} (\bibinfo {year} {2013})\BibitemShut {NoStop}%
\bibitem [{\citenamefont {Lenske}()}]{Lenske_priv}%
  \BibitemOpen
  \bibfield  {author} {\bibinfo {author} {\bibfnamefont {H.}~\bibnamefont
  {Lenske}},\ }\href@noop {} {}\bibinfo {note} {{private
  communication}}\BibitemShut {NoStop}%
\bibitem [{\citenamefont {Bertini}(1963)}]{Bertini:1963zzc}%
  \BibitemOpen
  \bibfield  {author} {\bibinfo {author} {\bibfnamefont {H.~W.}\ \bibnamefont
  {Bertini}},\ }\href {\doibase 10.1103/PhysRev.131.1801} {\bibfield  {journal}
  {\bibinfo  {journal} {Phys. Rev.}\ }\textbf {\bibinfo {volume} {131}},\
  \bibinfo {pages} {1801} (\bibinfo {year} {1963})}\BibitemShut {NoStop}%
%%CITATION = PHRVA,131,1801;%%
\bibitem [{\citenamefont {Sangster}\ \emph {et~al.}(1992)\citenamefont
  {Sangster} \emph {et~al.}}]{Sangster:1992qr}%
  \BibitemOpen
  \bibfield  {author} {\bibinfo {author} {\bibfnamefont {T.~C.}\ \bibnamefont
  {Sangster}} \emph {et~al.},\ }\href {\doibase 10.1103/PhysRevC.46.1404}
  {\bibfield  {journal} {\bibinfo  {journal} {Phys. Rev.}\ }\textbf {\bibinfo
  {volume} {C46}},\ \bibinfo {pages} {1404} (\bibinfo {year}
  {1992})}\BibitemShut {NoStop}%
%%CITATION = PHRVA,C46,1404;%%
\bibitem [{\citenamefont {Gaitanos}\ \emph {et~al.}(2008)\citenamefont
  {Gaitanos}, \citenamefont {Lenske},\ and\ \citenamefont
  {Mosel}}]{Gaitanos:2007mm}%
  \BibitemOpen
  \bibfield  {author} {\bibinfo {author} {\bibfnamefont {T.}~\bibnamefont
  {Gaitanos}}, \bibinfo {author} {\bibfnamefont {H.}~\bibnamefont {Lenske}}, \
  and\ \bibinfo {author} {\bibfnamefont {U.}~\bibnamefont {Mosel}},\ }\href
  {\doibase 10.1016/j.physletb.2008.04.011} {\bibfield  {journal} {\bibinfo
  {journal} {Phys. Lett.}\ }\textbf {\bibinfo {volume} {B663}},\ \bibinfo
  {pages} {197} (\bibinfo {year} {2008})},\ \Eprint
  {http://arxiv.org/abs/0712.3292} {arXiv:0712.3292 [nucl-th]} \BibitemShut
  {NoStop}%
%%CITATION = ARXIV:0712.3292;%%
\bibitem [{\citenamefont {Gaitanos}\ \emph {et~al.}(2010)\citenamefont
  {Gaitanos}, \citenamefont {Larionov}, \citenamefont {Lenske},\ and\
  \citenamefont {Mosel}}]{Gaitanos:2010fd}%
  \BibitemOpen
  \bibfield  {author} {\bibinfo {author} {\bibfnamefont {T.}~\bibnamefont
  {Gaitanos}}, \bibinfo {author} {\bibfnamefont {A.~B.}\ \bibnamefont
  {Larionov}}, \bibinfo {author} {\bibfnamefont {H.}~\bibnamefont {Lenske}}, \
  and\ \bibinfo {author} {\bibfnamefont {U.}~\bibnamefont {Mosel}},\ }\href
  {\doibase 10.1103/PhysRevC.81.054316} {\bibfield  {journal} {\bibinfo
  {journal} {Phys. Rev.}\ }\textbf {\bibinfo {volume} {C81}},\ \bibinfo {pages}
  {054316} (\bibinfo {year} {2010})},\ \Eprint {http://arxiv.org/abs/1003.4863}
  {arXiv:1003.4863 [nucl-th]} \BibitemShut {NoStop}%
%%CITATION = ARXIV:1003.4863;%%
\bibitem [{\citenamefont {Barashenkov}\ \emph {et~al.}(1972)\citenamefont
  {Barashenkov}, \citenamefont {Bertini}, \citenamefont {Chen}, \citenamefont
  {Friedlander}, \citenamefont {Harp}, \citenamefont {Iljinov}, \citenamefont
  {Miller},\ and\ \citenamefont {Toneev}}]{Barashenkov:1972id}%
  \BibitemOpen
  \bibfield  {author} {\bibinfo {author} {\bibfnamefont {V.~S.}\ \bibnamefont
  {Barashenkov}}, \bibinfo {author} {\bibfnamefont {H.~W.}\ \bibnamefont
  {Bertini}}, \bibinfo {author} {\bibfnamefont {K.}~\bibnamefont {Chen}},
  \bibinfo {author} {\bibfnamefont {G.}~\bibnamefont {Friedlander}}, \bibinfo
  {author} {\bibfnamefont {G.~D.}\ \bibnamefont {Harp}}, \bibinfo {author}
  {\bibfnamefont {A.~S.}\ \bibnamefont {Iljinov}}, \bibinfo {author}
  {\bibfnamefont {J.~M.}\ \bibnamefont {Miller}}, \ and\ \bibinfo {author}
  {\bibfnamefont {V.~D.}\ \bibnamefont {Toneev}},\ }\href {\doibase
  10.1016/0375-9474(72)90678-1} {\bibfield  {journal} {\bibinfo  {journal}
  {Nucl. Phys.}\ }\textbf {\bibinfo {volume} {A187}},\ \bibinfo {pages} {531}
  (\bibinfo {year} {1972})}\BibitemShut {NoStop}%
%%CITATION = NUPHA,A187,531;%%
\bibitem [{\citenamefont {Weisskopf}(1937)}]{Weisskopf:1937zz}%
  \BibitemOpen
  \bibfield  {author} {\bibinfo {author} {\bibfnamefont {V.}~\bibnamefont
  {Weisskopf}},\ }\href {\doibase 10.1103/PhysRev.52.295} {\bibfield  {journal}
  {\bibinfo  {journal} {Phys. Rev.}\ }\textbf {\bibinfo {volume} {52}},\
  \bibinfo {pages} {295} (\bibinfo {year} {1937})}\BibitemShut {NoStop}%
%%CITATION = PHRVA,52,295;%%
\bibitem [{\citenamefont {Abramovsky}\ \emph {et~al.}(1973)\citenamefont
  {Abramovsky}, \citenamefont {Gribov},\ and\ \citenamefont
  {Kancheli}}]{Abramovsky:1973fm}%
  \BibitemOpen
  \bibfield  {author} {\bibinfo {author} {\bibfnamefont {V.~A.}\ \bibnamefont
  {Abramovsky}}, \bibinfo {author} {\bibfnamefont {V.~N.}\ \bibnamefont
  {Gribov}}, \ and\ \bibinfo {author} {\bibfnamefont {O.~V.}\ \bibnamefont
  {Kancheli}},\ }\href@noop {} {\bibfield  {journal} {\bibinfo  {journal} {Yad.
  Fiz.}\ }\textbf {\bibinfo {volume} {18}},\ \bibinfo {pages} {595} (\bibinfo
  {year} {1973})},\ \bibinfo {note} {[Sov. J. Nucl.
  Phys.18,308(1974)]}\BibitemShut {NoStop}%
%%CITATION = YAFIA,18,595;%%
\bibitem [{\citenamefont {Bertocchi}\ and\ \citenamefont
  {Treleani}(1977)}]{Bertocchi:1976bq}%
  \BibitemOpen
  \bibfield  {author} {\bibinfo {author} {\bibfnamefont {L.}~\bibnamefont
  {Bertocchi}}\ and\ \bibinfo {author} {\bibfnamefont {D.}~\bibnamefont
  {Treleani}},\ }\href {\doibase 10.1088/0305-4616/3/2/007} {\bibfield
  {journal} {\bibinfo  {journal} {J. Phys.}\ }\textbf {\bibinfo {volume}
  {G3}},\ \bibinfo {pages} {147} (\bibinfo {year} {1977})}\BibitemShut
  {NoStop}%
%%CITATION = JPAGA,G3,147;%%
\bibitem [{\citenamefont {Li}\ and\ \citenamefont
  {Machleidt}(1993)}]{Li:1993rwa}%
  \BibitemOpen
  \bibfield  {author} {\bibinfo {author} {\bibfnamefont {G.-Q.}\ \bibnamefont
  {Li}}\ and\ \bibinfo {author} {\bibfnamefont {R.}~\bibnamefont {Machleidt}},\
  }\href {\doibase 10.1103/PhysRevC.48.1702} {\bibfield  {journal} {\bibinfo
  {journal} {Phys. Rev.}\ }\textbf {\bibinfo {volume} {C48}},\ \bibinfo {pages}
  {1702} (\bibinfo {year} {1993})},\ \Eprint
  {http://arxiv.org/abs/nucl-th/9307028} {arXiv:nucl-th/9307028 [nucl-th]}
  \BibitemShut {NoStop}%
%%CITATION = NUCL-TH/9307028;%%
\bibitem [{\citenamefont {Li}\ and\ \citenamefont
  {Machleidt}(1994)}]{Li:1993ef}%
  \BibitemOpen
  \bibfield  {author} {\bibinfo {author} {\bibfnamefont {G.-Q.}\ \bibnamefont
  {Li}}\ and\ \bibinfo {author} {\bibfnamefont {R.}~\bibnamefont {Machleidt}},\
  }\href {\doibase 10.1103/PhysRevC.49.566} {\bibfield  {journal} {\bibinfo
  {journal} {Phys. Rev.}\ }\textbf {\bibinfo {volume} {C49}},\ \bibinfo {pages}
  {566} (\bibinfo {year} {1994})},\ \Eprint
  {http://arxiv.org/abs/nucl-th/9308016} {arXiv:nucl-th/9308016 [nucl-th]}
  \BibitemShut {NoStop}%
%%CITATION = NUCL-TH/9308016;%%
\bibitem [{\citenamefont {Baltz}(2008)}]{Baltz:2007kq}%
  \BibitemOpen
  \bibfield  {author} {\bibinfo {author} {\bibfnamefont {A.~J.}\ \bibnamefont
  {Baltz}},\ }\href {\doibase 10.1016/j.physrep.2007.12.001} {\bibfield
  {journal} {\bibinfo  {journal} {Phys. Rept.}\ }\textbf {\bibinfo {volume}
  {458}},\ \bibinfo {pages} {1} (\bibinfo {year} {2008})},\ \Eprint
  {http://arxiv.org/abs/0706.3356} {arXiv:0706.3356 [nucl-ex]} \BibitemShut
  {NoStop}%
%%CITATION = ARXIV:0706.3356;%%
\bibitem [{\citenamefont {Aad}\ \emph {et~al.}(2016)\citenamefont {Aad} \emph
  {et~al.}}]{Aad:2015xis}%
  \BibitemOpen
  \bibfield  {author} {\bibinfo {author} {\bibfnamefont {G.}~\bibnamefont
  {Aad}} \emph {et~al.} (\bibinfo {collaboration} {ATLAS}),\ }\href {\doibase
  10.1016/j.physletb.2016.01.028} {\bibfield  {journal} {\bibinfo  {journal}
  {Phys. Lett.}\ }\textbf {\bibinfo {volume} {B754}},\ \bibinfo {pages} {214}
  (\bibinfo {year} {2016})},\ \Eprint {http://arxiv.org/abs/1511.00502}
  {arXiv:1511.00502 [hep-ex]} \BibitemShut {NoStop}%
%%CITATION = ARXIV:1511.00502;%%
\bibitem [{\citenamefont {White}(2011)}]{White:2011tq}%
  \BibitemOpen
  \bibfield  {author} {\bibinfo {author} {\bibfnamefont {S.~N.}\ \bibnamefont
  {White}},\ }\bibfield  {booktitle} {\emph {\bibinfo {booktitle}
  {{Proceedings, 6th International Workshop on Diffraction in high energy
  physics (Diffraction 2010): Otranto, Italy, September 10-15, 2010}}},\ }\href
  {\doibase 10.1063/1.3601384} {\bibfield  {journal} {\bibinfo  {journal} {AIP
  Conf. Proc.}\ }\textbf {\bibinfo {volume} {1350}},\ \bibinfo {pages} {95}
  (\bibinfo {year} {2011})},\ \Eprint {http://arxiv.org/abs/1101.2889}
  {arXiv:1101.2889 [hep-ex]} \BibitemShut {NoStop}%
%%CITATION = ARXIV:1101.2889;%%
\bibitem [{\citenamefont {Adriani}\ \emph {et~al.}(2018)\citenamefont {Adriani}
  \emph {et~al.}}]{Adriani:2018ess}%
  \BibitemOpen
  \bibfield  {author} {\bibinfo {author} {\bibfnamefont {O.}~\bibnamefont
  {Adriani}} \emph {et~al.} (\bibinfo {collaboration} {LHCf}),\ }\href
  {\doibase 10.1007/JHEP11(2018)073} {\bibfield  {journal} {\bibinfo  {journal}
  {JHEP}\ }\textbf {\bibinfo {volume} {11}},\ \bibinfo {pages} {073} (\bibinfo
  {year} {2018})},\ \Eprint {http://arxiv.org/abs/1808.09877} {arXiv:1808.09877
  [hep-ex]} \BibitemShut {NoStop}%
%%CITATION = ARXIV:1808.09877;%%
\bibitem [{\citenamefont {Frankfurt}\ \emph {et~al.}(2016)\citenamefont
  {Frankfurt}, \citenamefont {Guzey}, \citenamefont {Strikman},\ and\
  \citenamefont {Zhalov}}]{Frankfurt:2015cwa}%
  \BibitemOpen
  \bibfield  {author} {\bibinfo {author} {\bibfnamefont {L.}~\bibnamefont
  {Frankfurt}}, \bibinfo {author} {\bibfnamefont {V.}~\bibnamefont {Guzey}},
  \bibinfo {author} {\bibfnamefont {M.}~\bibnamefont {Strikman}}, \ and\
  \bibinfo {author} {\bibfnamefont {M.}~\bibnamefont {Zhalov}},\ }\href
  {\doibase 10.1016/j.physletb.2015.11.012} {\bibfield  {journal} {\bibinfo
  {journal} {Phys. Lett.}\ }\textbf {\bibinfo {volume} {B752}},\ \bibinfo
  {pages} {51} (\bibinfo {year} {2016})},\ \Eprint
  {http://arxiv.org/abs/1506.07150} {arXiv:1506.07150 [hep-ph]} \BibitemShut
  {NoStop}%
%%CITATION = ARXIV:1506.07150;%%
\bibitem [{\citenamefont {Alvioli}\ \emph {et~al.}(2017)\citenamefont
  {Alvioli}, \citenamefont {Frankfurt}, \citenamefont {Guzey}, \citenamefont
  {Strikman},\ and\ \citenamefont {Zhalov}}]{Alvioli:2016gfo}%
  \BibitemOpen
  \bibfield  {author} {\bibinfo {author} {\bibfnamefont {M.}~\bibnamefont
  {Alvioli}}, \bibinfo {author} {\bibfnamefont {L.}~\bibnamefont {Frankfurt}},
  \bibinfo {author} {\bibfnamefont {V.}~\bibnamefont {Guzey}}, \bibinfo
  {author} {\bibfnamefont {M.}~\bibnamefont {Strikman}}, \ and\ \bibinfo
  {author} {\bibfnamefont {M.}~\bibnamefont {Zhalov}},\ }\href {\doibase
  10.1016/j.physletb.2017.02.034} {\bibfield  {journal} {\bibinfo  {journal}
  {Phys. Lett.}\ }\textbf {\bibinfo {volume} {B767}},\ \bibinfo {pages} {450}
  (\bibinfo {year} {2017})},\ \Eprint {http://arxiv.org/abs/1605.06606}
  {arXiv:1605.06606 [hep-ph]} \BibitemShut {NoStop}%
%%CITATION = ARXIV:1605.06606;%%
\bibitem [{\citenamefont {Bayukov}\ \emph {et~al.}(1983)\citenamefont {Bayukov}
  \emph {et~al.}}]{Bayukov:1983wi}%
  \BibitemOpen
  \bibfield  {author} {\bibinfo {author} {\bibfnamefont {{\relax Yu}.~D.}\
  \bibnamefont {Bayukov}} \emph {et~al.},\ }\href@noop {} {\bibfield  {journal}
  {\bibinfo  {journal} {ITEP-172-1983}\ } (\bibinfo {year} {1983})}\BibitemShut
  {NoStop}%
\bibitem [{\citenamefont {Galoyan}\ \emph {et~al.}(2015)\citenamefont
  {Galoyan}, \citenamefont {Ribon},\ and\ \citenamefont
  {Uzhinsky}}]{Galoyan:2015wqv}%
  \BibitemOpen
  \bibfield  {author} {\bibinfo {author} {\bibfnamefont {A.~S.}\ \bibnamefont
  {Galoyan}}, \bibinfo {author} {\bibfnamefont {A.}~\bibnamefont {Ribon}}, \
  and\ \bibinfo {author} {\bibfnamefont {V.~V.}\ \bibnamefont {Uzhinsky}},\
  }\href {\doibase 10.1134/S0021364015180058} {\bibfield  {journal} {\bibinfo
  {journal} {JETP Lett.}\ }\textbf {\bibinfo {volume} {102}},\ \bibinfo {pages}
  {324} (\bibinfo {year} {2015})},\ \bibinfo {note} {[Pisma Zh. Eksp. Teor.
  Fiz.102,no.6,361(2015)]}\BibitemShut {NoStop}%
%%CITATION = JTPLA,102,324;%%
\end{thebibliography}%

\end{document}